\documentclass[11pt]{article}

\usepackage[margin=1in]{geometry}
\usepackage[utf8]{inputenc}
\usepackage[T1]{fontenc}
\usepackage{amsfonts,amsmath,amssymb,amsthm,mathtools}
\usepackage{booktabs}
\usepackage{enumitem}
\usepackage{nicefrac}
\usepackage{mathrsfs}
\usepackage{algorithm,algpseudocode}
\usepackage[square,sort&compress,numbers]{natbib}
\usepackage[colorlinks=true,linkcolor=blue,citecolor=blue,urlcolor=blue]{hyperref}
\usepackage[capitalise,noabbrev]{cleveref}
\usepackage{tikz}
\newcommand{\tikzxmark}{%
\tikz[scale=0.23] {
    \draw[line width=0.7,line cap=round] (0,0) to [bend left=6] (1,1);
    \draw[line width=0.7,line cap=round] (0.2,0.95) to [bend right=3] (0.8,0.05);
}}
\newcommand{\tikzcmark}{%
\tikz[scale=0.23] {
    \draw[line width=0.7,line cap=round] (0.25,0) to [bend left=10] (1,1);
    \draw[line width=0.8,line cap=round] (0,0.35) to [bend right=1] (0.23,0);
}}

\newtheorem{theorem}{Theorem}
\newtheorem{lemma}{Lemma}
\newtheorem{proposition}{Proposition}
\newtheorem{corollary}{Corollary}
\newtheorem{claim}{Claim}

\theoremstyle{definition}
\newtheorem{definition}{Definition}
\newtheorem{example}{Example}

\theoremstyle{remark}
\newtheorem{remark}{Remark}

\Crefname{theorem}{Theorem}{Theorems}
\Crefname{lemma}{Lemma}{Lemmas}
\Crefname{claim}{Claim}{Claims}
\Crefname{proposition}{Proposition}{Propositions}
\Crefname{corollary}{Corollary}{Corollaries}
\Crefname{definition}{Definition}{Definitions}
\Crefname{example}{Example}{Examples}
\Crefname{table}{Table}{Tables}
\Crefname{algorithm}{Algorithm}{Algorithms}
\Crefname{enumi}{Condition}{Conditions}

\DeclareMathOperator*{\argmax}{arg\,max}

\newenvironment{psketch}{\medskip\noindent\emph{Proof sketch.}}{\hfill$\square$\medskip}

\usepackage[usenames,svgnames,xcdraw,table]{xcolor}
\definecolor{auburn}{rgb}{0.33, 0.51, 0.2}
\definecolor{azure}{rgb}{0.0, 0.5, 1.0}
\allowdisplaybreaks
\newif\ifverbose
\verbosetrue %

\allowdisplaybreaks

\title{Likes, Budgets, and Equilibria:\\Designing Contests for Socially Optimal Advertising}
\author{Sayantika Mandal\\ Indian Institute of Technology Bombay\\ \texttt{22d0379@iitb.ac.in}
\and Swaprava Nath\\ Indian Institute of Technology Bombay\\ \texttt{swaprava@iitb.ac.in}}
\date{}

\begin{document}

\maketitle

\begin{abstract}
We study competitive advertising on a social network. Each of $m$ firms allocates a budget across consumers, and a platform decides, through a contest success function (CSF), what share of each consumer's attention every firm captures from the budgets placed on that consumer; awareness then spreads through the network. We model awareness and budgets on two timescales and bound how many fast steps awareness needs to settle between two budget updates. On the slow timescale, we show that best-response dynamics converge at a geometric rate to a Nash equilibrium whenever the firms' utilities are strongly concave and smooth, and that the standard ratio-form CSFs satisfy these conditions. Their equilibrium, however, need not maximize welfare. A firm's spending lowers its rivals' shares of attention, leading to the individual optima being different from the social optima, and we show that a constant per-unit charge on budget cannot close this gap in general. We then give a sufficient condition for the equilibrium to coincide with the welfare optimum. To meet it, we multiply each firm's ratio share by a function of the total spending at the consumer, a class of CSFs that we call scaled ratio CSFs (SRCSFs), and we show that within this class the condition holds exactly when this function is a power of the total spending; we call the resulting CSF the Power RCSF. The Power RCSF alone still falls short in two ways. Its equilibrium is optimal only on instances where the firms exhaust their budgets, and we remove this restriction with a constant per-unit charge that is budget-balanced. Its curvature also becomes unbounded as spending at a consumer approaches zero, which breaks our convergence guarantee, and we restore the guarantee by requiring every firm to spend a small minimum amount on each consumer. With these two additions, best-response dynamics converge to a unique, symmetric equilibrium that maximizes $p$-mean social welfare for every $p\leqslant 1$ at the same time, so the Price of Anarchy is unity.
\end{abstract}

\section{Introduction}
\label{sec:intro}

The rise of AI-powered platforms has created a new class of competitive markets in which several providers compete for user attention through a common intermediary. Consider an AI marketplace where providers of competing coding assistants, search tools, or recommendation engines each invest resources to be visible to users. For each user, the platform decides whose content to surface and how prominently, based on how much each provider has invested in reaching that user. A user who adopts a product then recommends it to friends and contacts, so awareness spreads further through the social network. The same structure appears in any market where an intermediary controls initial exposure and peer influence shapes further adoption.

The platform's allocation rule therefore plays a dual role. For a provider, it determines how much of an investment turns into user attention. For a regulator or a platform designer, it is a design instrument, and a natural question is whether the rule can be chosen so that competition among providers produces collectively desirable outcomes instead of a wasteful race for visibility. We study this question formally. Each user is a vertex of a social network, and each provider's share of a user's attention is an \emph{awareness level} in $[0,1]$. Awareness comes from direct investment by the provider and from peer influence through the network. Following \citet{skaperdas1996contest}, we model the platform's rule as a \emph{contest success function} (CSF), which maps the providers' investments at a user into the shares of that user's attention they capture. Providers play a budget allocation game, and a planner chooses the CSF. Our central question is:
\begin{quote}
    \emph{can a planner choose the CSF so that the competitive game has a Nash equilibrium that is socially optimal, and that the providers actually reach?}
\end{quote}
In practice, AI-driven ad systems such as Google's Performance Max, Meta's Advantage+, and the sponsored answers now shown in ChatGPT and Copilot decide what share of a user's attention each advertiser receives as a function of competing budgets, which we model as a CSF.

Our answer is a qualified yes, and the qualifications are the substance of the paper. The standard ratio-form CSFs of contest theory lead to well-behaved competition, but they cannot guarantee an efficient outcome: part of what a firm gains from its own spending is taken from a rival, so a firm's private return to spending exceeds the social return. We show that this gap does not have to vanish for the equilibrium to be optimal, and we identify a CSF, together with two simple instruments, under which equilibrium and optimum coincide on every instance.

\subsection{Our Contributions}
\label{sec:contri}

We extend the awareness model of \citet{bimpikis2016competitive} to $m$ firms in a two-timescale setting, where consumers update their awareness quickly and firms revise their budgets slowly, and we bound the number of fast steps that awareness needs to settle (\Cref{thm:T-bound}). On the slow timescale, we show that best-response dynamics (BRD) converge at a geometric rate to a Nash equilibrium (NE) whenever the induced utilities are strongly concave and smooth (\Cref{thm:BRD-conv}). We then turn to the planner's problem. We first identify a subclass of the standard ratio-form CSFs, which we call Regular RCSFs, that satisfies these conditions. Their equilibrium need not maximize welfare, because a firm's spending lowers its rivals' attention shares, and we show that a constant per-unit charge does not fix this (\Cref{thm:RCSF_impossibility}, \Cref{rem:alignment-reading}). We next give a sufficient condition for the equilibrium to coincide with the welfare optimum (\Cref{prop:alignment}). To satisfy it, we scale each firm's ratio share by a function of the total spending at the consumer; we call this class \emph{scaled ratio CSFs} (SRCSFs), and we show that within it the condition holds exactly when the scaling function is a power of the total spending, which gives the CSF we call the \emph{Power RCSF} (\Cref{lem:isoelastic}, \Cref{prop:power-align}). The Power RCSF still needs two additions. Its equilibrium is optimal only when the firms exhaust their budgets, which the designer cannot control, so we add a constant per-unit charge whose revenue is returned to the firms (\Cref{prop:budget_bind_cond}, \Cref{prop:tax-alignment}). Its curvature is also unbounded near zero spending, which breaks the convergence guarantee, so we require every firm to spend a small minimum amount on each consumer, and we show that this floor leaves the equilibrium unchanged (\Cref{prop:unbounded}, \Cref{lem:explicit-floor}, \Cref{lem:power-SB}). With both additions, BRD converge to a unique, symmetric equilibrium that maximizes $p$-mean social welfare for every $p\leqslant 1$ at the same time, so the Price of Anarchy (PoA) equals one (\Cref{thm:welfare-max}, \Cref{cor:poa}). \Cref{tab:results} summarizes how each stage of the design adds one guarantee.

\begin{table}[t]
\centering
\small
\renewcommand{\arraystretch}{1.25}
\begin{tabular}{@{}>{\raggedright\arraybackslash}p{3.3cm}>{\raggedright\arraybackslash}p{2.4cm}>{\raggedright\arraybackslash}p{1.9cm}>{\raggedright\arraybackslash}p{3.5cm}>{\raggedright\arraybackslash}p{3.1cm}@{}}
\toprule
\textbf{Design} & \textbf{Sufficient for BRD convergence} & \textbf{Unique, symmetric NE} & \textbf{NE is welfare optimal} & \textbf{Results} \\
\midrule
Regular RCSF
  & \tikzcmark
  & \tikzcmark
  & \tikzxmark \ \ not guaranteed;\newline a constant charge does not fix it
  & \Cref{lem:rcsf-sb,lem:rcsf-unique-symmetric}, \Cref{thm:RCSF_impossibility}, \Cref{rem:alignment-reading} \\ \hline
Power RCSF
  & \tikzxmark \ \ curvature unbounded near zero spending
  & \tikzcmark
  & only if budgets are exhausted at the optimum; never when $n\leqslant mC$
  & \Cref{prop:alignment,prop:power-align,prop:budget_bind_cond,prop:unbounded}, \Cref{lem:power-unique-sym} \\ \hline
Power RCSF\newline + charge $\tau$ and BB payment
  & \tikzxmark
  & \tikzcmark
  & \tikzcmark
  & \Cref{prop:tax-alignment} \\ \hline
Power RCSF\newline + charge $\tau$ and BB payment\newline + participation floor $\varepsilon<\bar\varepsilon$
  & \tikzcmark
  & \tikzcmark
  & \tikzcmark \ \ for every $p\leqslant 1$,\newline so $\mathrm{PoA}=1$
  & \Cref{lem:explicit-floor,lem:power-SB}, \Cref{thm:welfare-max}, \Cref{cor:poa} \\
\bottomrule
\end{tabular}
\caption{Guarantees obtained at each stage of the CSF design. The fast-timescale dynamics converge for every CSF (\Cref{thm:T-bound}), and best-response dynamics converge whenever the induced utilities have the SB property (\Cref{thm:BRD-conv}). Each row adds one instrument to remove the failure in the row above it. BB denotes the budget-balanced redistribution of the charge in \Cref{eq:bb}.}
\label{tab:results}
\end{table}

\section{Related Work}
\label{sec:related}

Our work draws on several literatures: competition for influence on networks, the theory of contests, models of competitive marketing, the efficiency of proportional allocation rules, and learning dynamics in games. We discuss them in turn.

\paragraph{Competitive influence on networks.}
Influence maximization was introduced by \citet{kempe2003maximizing}, and its competitive version studies firms that choose seed consumers in a shared network \citep{bharathi2007competitive,alon2010note,goyal2012competitive}. \citet{tzoumas2012game} analyze a two-player diffusion game under a linear threshold model, and \citet{zuo2022online} study an online version with bandit feedback. \citet{goyal2012competitive} also bound the inefficiency of equilibria in their contagion model. These works model competition through the choice of seeds; we instead study the allocation of continuous budgets.

\paragraph{Budget allocation and advertising on networks.}
A parallel line of work studies firms that spread budgets over consumers. \citet{fazeli2012game,fazeli2012targeted,fazeli2016competitive} study the trade-off between seeding and product quality, and \citet{masucci2014strategic,masucci2017advertising}, \citet{varma2018marketing,varma2019allocating}, and \citet{ansari2019competitive} study related allocation games. \citet{maehara2015} show that a multi-advertiser budget allocation game is a potential game with a Price of Anarchy of two. Closest to us, \citet{bimpikis2016competitive} model targeted advertising by two firms through the awareness dynamics that we extend to $m$ firms, and they find that in equilibrium firms invest inefficiently high, with an inefficiency that grows with the centrality of the consumers they target. Our impossibility result for the Regular RCSF is consistent with this finding; our contribution is to ask how the allocation rule can be designed to remove the inefficiency.

\paragraph{Welfare and fairness in networked marketing.}
Several works add welfare or fairness objectives to influence maximization. \citet{lu2013bang} and \citet{wang2022fairness} study fair allocation of seeds or budgets among advertisers, \citet{rahmattalabi2021fair} and \citet{becker2022fairness} design fair influence maximization through welfare objectives and randomization, and \citet{banerjee2019maximizing,banerjee2020maximizing} maximize consumer welfare under diffusion models. These works optimize allocations from the point of view of a planner or a host, and do not study how firms that choose their own spending can be led to an efficient equilibrium.

\paragraph{Contest success functions and welfare.}
Ratio-form contests go back to \citet{tullock1980efficient}, and \citet{skaperdas1996contest} characterizes them axiomatically; \citet{konrad2009strategy} surveys the theory. \citet{corchon2010foundations} derive CSFs from the decision problem of a contest administrator, and \citet{szidarovszky1997existence} establish existence and uniqueness of equilibrium in rent-seeking games. A recurring conclusion of this literature is that competition for a fixed prize dissipates resources. The closest work to our design question is \citet{chung1996}, who studies contests in which the prize grows with aggregate effort, as the market-expansion factor $f(B_i)$ of our SRCSFs does; he shows that the resulting positive externality does not offset the negative one, and that aggregate effort remains socially excessive. We show that a specific functional form, the iso-elastic one, makes the two forces exactly offsettable by a constant charge. In industrial organization, \citet{dixit1978advertising} show that competitive advertising is socially excessive, and the classical remedy for such externalities is a corrective charge \citep{pigou1920economics,baumol1972taxation}. Our charge is of this type, with a budget-balanced redistribution similar to that of \citet{murhekar2024incentives}.

\paragraph{Market-share attraction models.}
In marketing, a brand's market share is often modeled as its attraction divided by the total attraction of all brands \citep{bell1975market}, which is the ratio form of contest theory. Game-theoretic analyses of these models characterize equilibrium marketing efforts \citep{karnani1985,monahan1987}, and \citet{federgruen2009competition} study generalized attraction models in which total demand also responds to aggregate attraction. Our SRCSF has the same factorized structure, with $f$ playing the role of total demand. To our knowledge, the observation that a constant elasticity of $f$ is what aligns the competitive equilibrium with the welfare optimum does not appear in this literature.

\paragraph{Efficiency of proportional allocation.}
\citet{johari2004efficiency} show that when a divisible resource is allocated in proportion to bids, the equilibrium attains at least three quarters of the optimal aggregate utility, and that this bound is tight. Our setting differs in that the total amount of attention is itself endogenous, which is what allows a suitable design to reach full efficiency.

\paragraph{Dynamics and uniqueness in games and contests.}
Uniqueness of equilibrium in concave games under diagonal strict concavity goes back to \citet{rosen1965existence}, and projection methods for the resulting variational inequalities converge linearly under strong monotonicity and Lipschitz continuity \citep{facchinei2003finite}. \citet{murhekar2024incentives} use gradient dynamics of this kind to reach welfare-maximizing equilibria in federated learning. In contests, \citet{ghosh2023best} and \citet{elkind2024continuous} analyze best-response dynamics in Tullock contests with convex costs, and \citet{ewerhart2020unique} prove uniqueness of equilibrium in contests with incomplete information. We contribute convergence guarantees for $m$ firms in a two-timescale model, and identify the regularity failure that makes a participation floor necessary for the iso-elastic CSF.

\section{Preliminaries}
\label{sec:model}

We consider a market with $m$ competing firms (players) denoted by $\mathscr{M} = \{1, \ldots, m\}$, each having an equal and fixed marketing budget $C>0$\footnote{The equal-budgets assumption ensures that the equilibrium can be the welfare maximizer. Our convergence results do not use it. See \Cref{rem:unequal-budgets} for more details.} that they strategically allocate across $n$ vertices (consumers) denoted by $\mathscr{N} = \{1, 2, \ldots, n\}$, in a directed weighted social network $\mathscr{G} = (\mathscr{N}, \mathscr{E})$. The adjacency matrix $\mathscr{E} = (e_{ij})_{i,j \in \mathscr{N}}$ is row-stochastic, with $e_{ii} = 0$ for all $i$, and $e_{ij}$ representing the influence of vertex $j$ on vertex $i$. We consider \textit{awareness} as a cardinal metric of a consumer's attention towards a particular firm. For each firm $s \in \mathscr{M}$ and each consumer $i \in \mathscr{N}$, we define an \emph{awareness} $x_{s,i} \in [0,1]$ that captures how prominently firm $s$ occupies consumer $i$'s attention relative to its competitors. We allow $\sum_s x_{s,i} \leqslant 1$, with the residual mass $1 - \sum_s x_{s,i}$ corresponding to the consumer being unaware of any of the $m$ firms. Each firm aims to maximize the overall awareness of its brand across the network, subject to its budget constraints.

\subsection{A Two-Timescale Communication Model}
\label{sec:timescales}

Real advertising markets evolve at two distinct rates. Consumers interact with each other (via word-of-mouth, social-media reshares, and direct content consumption) on a fast timescale, often hourly or daily. Firms revise their advertising budgets and targeting strategies on a slower timescale, often quarterly or annually. Hence, we model the interaction between firms' strategic decisions and the evolution of customer awareness using a two-timescale framework.

\textbf{Slow timescale (strategy updates).}
The slow timescale is indexed by $k = 0, 1, 2, \ldots$. At the start of each period $k$, firm $s$ commits to a budget vector $\mathbf{b}_s(k) = (b_{s,1}(k), \ldots, b_{s,n}(k))^\top \in \mathbb{R}^n_{\geqslant 0}$ representing the allocation of its marketing budget across the $n$ vertices in the network, and this allocation remains fixed for the duration of period $k$. The set of feasible budget vectors for firm $s$ is $\mathcal{B}_s = \{\mathbf{b}_s \in \mathbb{R}^n_{\geqslant 0}: \sum_i b_{s,i} \leqslant C\}$. The joint feasible set $\mathcal{B} = \prod_s \mathcal{B}_s$ is a convex and compact and we denote the joint strategy profile of all firms by the matrix $\mathbf{b} \coloneqq (\mathbf{b}_s, \mathbf{b}_{-s})$ where $\mathbf{b}_{-s}$ denotes the allocation vectors of all competing firms other than $s$.

\textbf{Fast timescale (awareness dynamics).}
Within each period $k$, awareness evolves on a fast timescale indexed by $t = 0, 1, \ldots, T$. At every fast-timescale step, each consumer receives an influence that mixes two sources: the awareness of its neighbours, with weight $\alpha \in (0,1)$, and the direct effect of the firms' advertising, with weight $1-\alpha$. We call $\alpha$ the \emph{peer-influence weight}. Our model is inspired by the awareness framework of \citet{bimpikis2016competitive}, but unlike theirs it is deterministic. The awareness of vertex $i$ toward firm $s$ at fast-timescale step $t$ of period $k$ is updated as
\begin{equation}
\label{eq:awareness_update_eqn}
x_{s,i}(t) = \frac{1}{t}\Bigl[\alpha \sum_{j \in \mathscr{N}_i} e_{ij}\, x_{s,j}(t-1) + (1-\alpha)\, h_{s,i}(\mathbf{b}(k))\Bigr] + \Bigl(1 - \frac{1}{t}\Bigr) x_{s,i}(t-1), \qquad t \geqslant 1,
\end{equation}
where $\mathscr{N}_i$ is the set of neighbours of $i$, and $x_{s,i}(0)\in[0,1]$ is the awareness at the start of the period. The \textit{contest success function} (CSF) $h_{s,i}(\mathbf{b}(k)) \coloneqq h(b_{s,i}(k), \mathbf{b}_{-s,i}(k))$ specifies the share of consumer $i$'s attention that firm $s$ captures through direct advertising, given the budgets all firms have invested in $i$; in contest-theoretic terms, it is firm $s$'s success probability at $i$. We require $\sum_{s \in \mathscr{M}} h_{s,i}(\mathbf b(k)) \leqslant 1$, and we use the shorthand $h_{s,i}(\mathbf{b})$ when the dependence on $k$ is clear.

The update \eqref{eq:awareness_update_eqn} gives weight $1-1/t$ to the awareness accumulated so far and weight $1/t$ to the newest influence. This has a simple interpretation. Writing $y_{s,i}(t')$ for the bracketed term in \eqref{eq:awareness_update_eqn} at step $t'$ and multiplying \eqref{eq:awareness_update_eqn} by $t$ gives $t\,x_{s,i}(t) = y_{s,i}(t) + (t-1)\,x_{s,i}(t-1)$, and unrolling this identity down to $t=1$ yields
\[
x_{s,i}(t) = \frac{1}{t}\sum_{t'=1}^{t} y_{s,i}(t').
\]
Thus the awareness at step $t$ is the average of all the influences received so far, each counting equally. We use this to model awareness as cumulative exposure: an influence received early in the period is not forgotten when newer ones arrive. The dynamics also respect the interpretation of awareness as a share of attention. Each $y_{s,i}(t')$ is a convex combination of the numbers $x_{s,j}(t'-1)$ and $h_{s,i}$, because the weights $\alpha e_{ij}$ and $1-\alpha$ are non-negative and sum to one by row-stochasticity of $\mathscr{E}$. Summing over firms, $\sum_s y_{s,i}(t') \leqslant \alpha \max_j \sum_s x_{s,j}(t'-1) + (1-\alpha)\sum_s h_{s,i} \leqslant 1$ whenever $\sum_s x_{s,j}(t'-1)\leqslant1$ for all $j$. By induction on $t$, if every $x_{s,i}(0)$ lies in $[0,1]$ and $\sum_s x_{s,i}(0) \leqslant 1$ for every $i$, then the same holds at every step $t$.

\paragraph{Steady-state awareness.}
\label{sec:awareness}

Fix a slow-timescale period $k$ and a firm $s$, and write $\mathbf h_s := \mathbf h_s(\mathbf b(k))$ for the vector $(h_{s,i}(\mathbf b(k)))_{i\in\mathscr N}$, which is constant throughout the period. In vector form, \eqref{eq:awareness_update_eqn} reads $\mathbf x_s(t) = \frac1t\bigl[\alpha\mathscr E\mathbf x_s(t-1)+(1-\alpha)\mathbf h_s\bigr] + \bigl(1-\frac1t\bigr)\mathbf x_s(t-1)$. A vector $\mathbf x$ is left unchanged by this update if and only if $\mathbf x = \alpha\mathscr E\mathbf x + (1-\alpha)\mathbf h_s$, since the terms involving $1/t$ cancel. Because $\mathscr{E}$ is row-stochastic, the spectral radius of $\alpha\mathscr E$ is at most $\alpha < 1$, so $I-\alpha\mathscr E$ is invertible and this equation has exactly one solution,
\begin{equation}
\label{eq:steady-state}
\mathbf{x}_s^*(k) = (1-\alpha)(I - \alpha \mathscr{E})^{-1} \mathbf{h}_s\bigl(\mathbf{b}_s(k), \mathbf{b}_{-s}(k)\bigr),
\end{equation}
We call $\mathbf x_s^*(k)$ the \emph{steady-state awareness} of firm $s$ in period $k$. It has the same form as the steady state derived by \citet[Theorem 1]{bimpikis2016competitive} for two firms. The next theorem shows that the fast dynamics converge to it from any starting point, and it quantifies how many fast-timescale steps this takes.

\begin{theorem}[Convergence of the fast-timescale dynamics]
\label{thm:T-bound}
Fix a period $k$ and a firm $s$, and let $\mathbf{x}_s^*$ be the steady state in \Cref{eq:steady-state}. For every initial awareness $\mathbf x_s(0)\in[0,1]^n$, the iterates of \Cref{eq:awareness_update_eqn} satisfy the following:
\begin{enumerate}
\item[(a)] \emph{Convergence with an explicit rate.} For every $T\geqslant1$,
\[
\|\mathbf{x}_s(T) - \mathbf{x}_s^*\|_\infty \;\leqslant\; P_T\,\|\mathbf{x}_s(0) - \mathbf{x}_s^*\|_\infty,
\qquad\text{where}\quad P_T := \prod_{t=1}^{T}\Bigl(1-\frac{1-\alpha}{t}\Bigr) \;\leqslant\; (T+1)^{-(1-\alpha)} .
\]
In particular, $\mathbf x_s(T)\to\mathbf x_s^*$ as $T\to\infty$.
\item[(b)] \emph{Number of steps.} For every accuracy $\epsilon_0\in(0,1]$, running $T \geqslant \epsilon_0^{-1/(1-\alpha)}$ fast-timescale steps guarantees $\|\mathbf{x}_s(T) - \mathbf{x}_s^*\|_\infty \leqslant \epsilon_0$. Thus $T = O\bigl(\epsilon_0^{-1/(1-\alpha)}\bigr)$ steps suffice.
\item[(c)] \emph{Tightness.} The factor $P_T$ cannot be improved: it equals $\Gamma(T+\alpha)/\bigl(\Gamma(\alpha)\Gamma(T+1)\bigr)$, which behaves like $T^{-(1-\alpha)}/\Gamma(\alpha)$ as $T\to\infty$, and the bound in (a) holds with equality whenever $\mathbf x_s(0)-\mathbf x_s^*$ has equal coordinates.
\end{enumerate}
\end{theorem}

\begin{psketch}
Subtracting the fixed-point equation from \eqref{eq:awareness_update_eqn} shows that the error $\mathbf e(t) := \mathbf x_s(t)-\mathbf x_s^*$ obeys the linear recursion $\mathbf e(t) = A(t)\,\mathbf e(t-1)$ with $A(t) := (1-\tfrac1t)I + \tfrac{\alpha}{t}\mathscr E$. Because $\mathscr E$ is non-negative and row-stochastic, every row of $A(t)$ is non-negative and sums to $1-(1-\alpha)/t$, which is therefore its $\ell_\infty$ operator norm. Multiplying these norms over $t=1,\dots,T$ gives the factor $P_T$. Using $\log(1-x)\leqslant -x$ and the bound $\sum_{t=1}^T 1/t \geqslant \log(T+1)$ on the harmonic sum then gives $P_T\leqslant(T+1)^{-(1-\alpha)}$, and (b) follows because both $\mathbf x_s(0)$ and $\mathbf x_s^*$ lie in $[0,1]^n$. For (c), a vector with equal coordinates is an eigenvector of every $A(t)$, and a Gamma-function product identity evaluates $P_T$ exactly. The complete proof is in \Cref{app:T-bound}.
\end{psketch}

In the rest of the paper, we take the awareness of firm $s$ at the end of period $k$ to be the steady state $\mathbf x_s^*(k)$, and \Cref{thm:T-bound} quantifies the error of this idealization after $T$ fast steps. The bound also gives explicit guidance for spacing the slow-timescale epochs. When peer influence is weak (small $\alpha$), the exponent $1/(1-\alpha)$ is close to $1$, and about $1/\epsilon_0$ fast steps suffice. As $\alpha\to1$, the exponent grows, so more fast steps are needed for the same accuracy: stronger peer influence means that each new influence moves awareness less, and the dynamics settle more slowly. For every $\alpha$, the required number of steps grows as the desired accuracy $\epsilon_0$ decreases.

\begin{remark}[On the equal weighting of past influences]
\label{rem:equal-weights}
The polynomial rate in \Cref{thm:T-bound} is a consequence of the equal-weight memory of \Cref{eq:awareness_update_eqn}. If recent influences were instead weighted more heavily, for instance by replacing $1/t$ with a constant weight $\gamma\in(0,1]$, the same argument would give the error factor $\bigl(1-\gamma(1-\alpha)\bigr)^T$, that is, geometric convergence in $O(\log(1/\epsilon_0))$ steps. The equal-weight model is therefore the harder case for the separation of timescales, and the guarantee of \Cref{thm:T-bound} carries over, with a better rate, to discounted variants.
\end{remark}

\subsection{Firms' Utilities and the Awareness-Competition Game (ACG)}
\label{sec:utilities}

We define the utility (payoff) of a firm as the difference between its consolidated awareness over the network and the cost incurred through advertising \citep{ghosh2023best,elkind2024continuous}. This formulation captures the trade-off between maximizing awareness and minimizing expenditure. The utility of firm $s$ at the end of period $k$ is

\begin{equation}
\label{eq:utility}
u_s(\mathbf{b}(k)) = \mathbf{1}^\top\mathbf x_s^*(k) - \mathbf{1}^\top \mathbf{b}_s(k)
= \sum_{i \in \mathscr{N}} M_i\, h_{s,i}(\mathbf{b}(k)) - \sum_{i \in \mathscr{N}} b_{s,i}(k),
\end{equation}
where we drop the period index $k$ when it is clear, and
\begin{equation}
\label{eq:Mi}
M_i := (1-\alpha)\bigl[\mathbf{1}^\top (I - \alpha \mathscr{E})^{-1}\bigr]_i .
\end{equation} The weight $M_i$ is a scaled version of the \emph{absorption centrality} of vertex $i$ used by \citet{bimpikis2016competitive}: it measures how much of the attention captured at $i$ eventually reaches the whole network.
A firm $s$ strategically allocates its budget across vertices to maximize this utility by balancing awareness gains against advertising costs in a competitive environment. We call the induced game between firms the \emph{awareness competition game} (ACG).

\subsection{Design Desiderata}

We now define the equilibrium concept, the planner's objective, and the measure of inefficiency that connects them.

\begin{definition}[Pure-strategy Nash equilibrium, PSNE]
\label{def:nash-eq}
A budget profile $\mathbf{b}^* = (\mathbf{b}_s^*, \mathbf{b}_{-s}^*) \in \mathcal{B}$ is a \emph{PSNE} if $u_s(\mathbf{b}_s^*, \mathbf{b}_{-s}^*) \geqslant u_s(\mathbf{b}_s, \mathbf{b}_{-s}^*)$ for every $s \in \mathscr{M}$ and every $\mathbf{b}_s \in \mathcal{B}_s$.
\end{definition}

We shorten this to Nash equilibrium (NE) where it is clear from context. While the competing firms reach an NE, the goal of a planner (network policymaker or regulator) is to ensure some \emph{social welfare}. We consider the generalized $p$-mean welfare function, which provides a unifying framework for different welfare criteria.

\begin{definition}[$p$-mean social welfare, \citealp{barman2020tight}]
\label{def:p-mean}
Let $\mathcal B_{\geqslant0}:=\{\mathbf b\in\mathcal B: u_s(\mathbf b)\geqslant0\ \forall s\in\mathscr M\}$. For $\mathbf b\in\mathcal B_{\geqslant0}$, the $p$-mean social welfare is
\begin{equation}
    W_p(\mathbf{b}) = \left( \frac{1}{m} \sum_{s \in \mathscr{M}} u_s(\mathbf{b})^p \right)^{\frac{1}{p}},
    \label{eq:Wp}
\end{equation}
where $p\in(-\infty,1]$ controls the trade-off between equity and efficiency.
For $p<0$ we set $W_p(\mathbf b):=0$ whenever some $u_s(\mathbf b)=0$, the limiting value. 
\end{definition}

Special cases include (a) utilitarian welfare ($p=1$): $W_{1}(\mathbf{b}) = \frac{1}{m} \sum_{s} u_s(\mathbf{b})$, (b) egalitarian welfare ($p \to -\infty$): $W_\text{egal}(\mathbf{b}) = \min_{s} u_s(\mathbf{b})$, and (c) Nash social welfare ($p \to 0$): $W_\text{NSW}(\mathbf{b}) = (\prod_{s} u_s(\mathbf{b}))^{1/m}$. The planner optimizes over $\mathcal{B}_{\geqslant 0}$ because $u_s^p$ is not real for $u_s < 0$ and non-integer $p$ and a firm can guarantee $u_s \geqslant 0$ by spending nothing, so every NE of the ACG lies in $\mathcal{B}_{\geqslant 0}$. The set $\mathcal{B}_{\geqslant 0}$ is compact and $W_p$ is continuous on it, so the maxima exist. Note that $W_1$ is defined on all of $\mathcal{B}$, and we use it there.

\begin{definition}[Price of Anarchy]
\label{def:poa}
Let $\mathcal{B}^\star \subseteq \mathcal{B}_{\geqslant 0}$ be the set of Nash equilibria of the ACG. The \emph{Price of Anarchy} for $W_p$ is
\[
\mathrm{PoA} = \frac{\max_{\mathbf{b} \in \mathcal{B}_{\geqslant 0}} W_p(\mathbf{b})}{\min_{\mathbf{b} \in \mathcal{B}^\star} W_p(\mathbf{b})},
\]
the ratio of the optimal welfare to the welfare at the worst equilibrium. A larger value means more inefficiency from competition, and $\mathrm{PoA}=1$ means that every equilibrium is optimal.
\end{definition}

A Price of Anarchy of one is only meaningful if the firms can find the equilibrium. Before asking which CSFs make the equilibrium efficient, we therefore study how firms reach an equilibrium in the first place.

\section{Best Response Dynamics of ACGs and Convergence to Nash}
\label{sec:brd}

We now ask how the firms reach an equilibrium. A firm does not know its rivals' future budgets, and in practice it observes only how its own utility responds to small changes in its spending. A natural model of play is therefore an adjustment process. In each slow-timescale period, every firm observes the current budget profile, and the awareness it induces, and moves its own budget in a direction that increases its utility, taking its rivals' budgets as given. This is the idea behind \emph{best-response dynamics} (BRD) and we use a gradient-based form of it, in which every firm simultaneously takes a small step along the gradient of its own utility.

In terms of the two timescales of \Cref{sec:model}, at the start of period $k$ each firm commits to $\mathbf{b}_s(k)$, awareness evolves on the fast timescale to the steady state in \Cref{eq:steady-state}, which takes the number of fast-timescale steps given by \Cref{thm:T-bound}, and at the start of period $k+1$ each firm updates its budget using the utilities realized in period $k$. We show in this section that under certain conditions, the ACG converges to an NE when firms play BRD. This section makes no reference to a planner and focuses solely on the behavior of firms competing under a CSF whose induced utilities satisfy these conditions. The CSFs which make that equilibrium efficient are addressed in  \Cref{sec:welfare,sec:regular_rcsf,sec:scaled}. %

\paragraph{Feasible sets.}
For a lower boundary $\ell\geqslant 0$ with $n\ell\leqslant C$, define
\begin{equation}\label{eq:Ds}
\mathcal D_s^\ell:=\Bigl\{\mathbf b_s\in\mathbb R^n:\ b_{s,i}\geqslant \ell\ \ \forall i\in\mathscr N,\ \ \textstyle\sum_{i\in\mathscr N}b_{s,i}\leqslant C\Bigr\},
\qquad
\mathcal D^\ell:=\prod_{s\in\mathscr M}\mathcal D_s^\ell .
\end{equation}
Throughout this section $\ell=0$, so that $\mathcal D_s^\ell=\mathcal B_s$ and $\mathcal D^\ell=\mathcal B$. 
Each $\mathcal D_s^\ell$ is non-empty, closed and convex, and hence so is $\mathcal D^\ell$.

Algorithm~\ref{alg:BRD} implements BRD for the ACG. Every firm moves its budget in the direction of the gradient of its own utility, corrected by a term $\boldsymbol\mu_s(k)$ that keeps the new budget feasible. The corrected gradient of firm $s$ is $\mathbf g_s(\mathbf b(k),\boldsymbol\mu_s(k))\in\mathbb R^{n}$ with $i$th component
\begin{equation}\label{eq:gradient}
g_{s,i}(\mathbf b(k),\boldsymbol\mu_s(k))=\frac{\partial u_s(\mathbf b(k))}{\partial b_{s,i}}+\mu_{s,i}(k),
\end{equation}
where $\boldsymbol\mu_s(k) \in \mathbb{R}^n$ is the smallest correction, in Euclidean norm, for which the update in line~\ref{line:brd_update_rule} stays feasible:
\begin{equation}\label{eq:mu}
\boldsymbol\mu_s(k)\in\arg\min\Bigl\{\|\boldsymbol\mu\|_2:\ \mathbf b_s(k)+\gamma\bigl(\nabla_{\mathbf b_s}u_s(\mathbf b(k))+\boldsymbol\mu\bigr)\in\mathcal D_s^\ell\Bigr\}.
\end{equation}

\begin{algorithm}[t!]
\caption{Best Response Dynamics}
\label{alg:BRD}
\begin{algorithmic}[1]
\Require Initial budgets $\mathbf{b}_s(0) \in \mathcal{D}_s^\ell$ $\forall s \in \mathscr{M}$; step size $\gamma > 0$; number of iterations $K_\epsilon$.
\For{each iteration $k = 0, 1, \ldots, K_\epsilon-1$}
  \State Compute $\mathbf{g}_s(\mathbf{b}(k), \boldsymbol{\mu}_s(k)) = \nabla_{\mathbf{b}_s} u_s(\mathbf{b}(k)) + \boldsymbol{\mu}_s(k)$ for all $s \in \mathscr{M}$ \Comment{see \Cref{eq:gradient}}
  \label{line:grad}
  \State Update: $\mathbf{b}_s(k+1) \gets \mathbf{b}_s(k) + \gamma\,\mathbf{g}_s(\mathbf{b}(k), \boldsymbol{\mu}_s(k))$, for all $s \in \mathscr{M}$
  \label{line:brd_update_rule}
\EndFor
\Ensure $\mathbf{b}^* \gets \mathbf{b}(K_\epsilon)$ \Comment{Nash equilibrium}
\end{algorithmic}
\end{algorithm}

Lemma~\ref{lem:mu-unique} (proof in \cref{app:mu-unique}) below shows that the correction in \eqref{eq:mu} is well defined, and it explains what the correction does. 
In particular, if the uncorrected gradient step is feasible, then $\boldsymbol\mu_s(k)=\mathbf 0$; otherwise, $\boldsymbol\mu_s(k)$ provides only the correction needed to restore feasibility.

\begin{lemma}
\label{lem:mu-unique}
For every $\mathbf b\in\mathcal D^\ell$ and every $s\in\mathscr M$, the minimization in \eqref{eq:mu} admits a unique solution, denoted by $\boldsymbol\mu_s(\mathbf b)$. The correction used in
Algorithm~\ref{alg:BRD} is therefore given by $\boldsymbol\mu_s(k)=\boldsymbol\mu_s(\mathbf b(k)).$
\end{lemma}

\begin{psketch}
Since the new budget equals the uncorrected step plus $\gamma\boldsymbol\mu$, minimizing $\|\boldsymbol\mu\|_2$ over feasible updates is the same as finding the feasible budget nearest to the uncorrected step, that is, the Euclidean projection onto the non-empty, closed, convex set $\mathcal D_s^\ell$, which exists and is unique. See \Cref{app:mu-unique}.
\end{psketch}

For analytical convenience, we vectorize the joint budget matrix $\mathbf b$ into a column vector $\mathbf b\in\mathbb R^{mn}$ and stack the corrections and corrected gradients as $\boldsymbol\mu(k):=[\boldsymbol\mu_s(k)]_{s\in\mathscr M}$ and $\mathbf g(\mathbf b(k),\boldsymbol\mu(k)):=[\mathbf g_s(\mathbf b(k),\boldsymbol\mu_s(k))]_{s\in\mathscr M}\in\mathbb R^{mn}$. 
The following \cref{lem:grad-NE} (proof in Appendix~\ref{app:grad-NE}) relates the stationary points of BRD to Nash equilibria. 

\begin{lemma}\label{lem:grad-NE}
Suppose that each $u_s$ is concave and continuously differentiable in $\mathbf b_s$. Then a profile $\mathbf b^\star\in\mathcal D^\ell$ satisfies $\mathbf g_s(\mathbf b^\star,\boldsymbol\mu_s(\mathbf b^\star))=\mathbf 0$ for all $s\in\mathscr M$ if and only if $\mathbf b^\star$ is a pure-strategy Nash equilibrium of the game played on $\mathcal D^\ell$.
\end{lemma}

\begin{psketch}
The corrected gradient vanishes exactly when the projected step returns to $\mathbf b_s^\star$, which, by the characterization of projections onto convex sets, is the first-order optimality condition for firm $s$'s best-response problem. Concavity makes that condition sufficient as well as necessary. See \Cref{app:grad-NE}.
\end{psketch}

The above lemma clarifies that the proposed best-response dynamics cannot become stationary at a non-equilibrium profile, where some firm still has a profitable deviation, including a reallocation of its budget across consumers, as the dynamics do not converge to a trivial zero-gradient state.

\subsection{Convergence of Best Response Dynamics}
\label{sec:BRDconvergence}

In this section, we present the conditions on the utility functions and analyze the convergence of the BRD to an NE.
We define the block Hessian $G(\mathbf{b}) = [G_{s_1, s_2}]_{s_1,s_2 \in \mathscr{M}} \in \mathbb{R}^{mn \times mn}$ by $[G_{s_1, s_2}]_{i,j} = \partial^2 u_{s_1}(\mathbf{b}) / (\partial b_{s_2, j}\, \partial b_{s_1, i})$.

\begin{definition}[SB property]
\label{def:SB}
The utilities $\{u_s(\mathbf{b})\}_{s \in \mathscr{M}}$ satisfy the \emph{SB property} on a convex and compact set $\mathcal{D}^\ell$ if there exist constants $\lambda, L > 0$ such that, for every $\mathbf b \in \mathcal{D}^\ell$,
\begin{enumerate}
\item \emph{Strong concavity:} The matrix $G(\mathbf{b})$ is $\lambda$-negative semi-definite, i.e., the symmetric part of $G(\mathbf b)$ satisfies $\tfrac12\bigl(G(\mathbf b)+G(\mathbf b)^\top\bigr)+\lambda I_{mn}\preceq0$.
\item \emph{Bounded operator norm:} $\| G(\mathbf{b}) \|_{\text{op}} \leqslant L$, where $\|G\|_{\text{op}} := \inf\{c>0 : \|G\mathbf{v}\| \leqslant c\|\mathbf{v}\|\ \forall \mathbf{v}\}$.
\end{enumerate}
\end{definition}

These are the canonical strong-concavity and smoothness conditions used in the analysis of best-response dynamics and gradient-based algorithms \citep{murhekar2024incentives}. \emph{Strong concavity} ensures each firm's utility is strongly concave in its own strategy, yielding a stable equilibrium, while a \emph{bounded norm} guarantees that utility gradients do not change abruptly, so that fixed-step gradient updates converge.
Since the SB property is a property of the induced utilities, it must be verified separately for each CSF considered.

With agents having SB property, the BRD converges to an NE, as formalized in the following theorem.
Note that the convergence of BRD to a point where $\|\mathbf g(\mathbf b,\boldsymbol{\mu})\|_2$ is arbitrarily small implies convergence to an NE in the limit~(\Cref{lem:grad-NE}).

\begin{theorem}[Convergence to NE]\label{thm:BRD-conv}
Suppose the agent utilities satisfy the SB property (\Cref{def:SB}) on a convex feasible set $\mathcal{D}^\ell$ as in \eqref{eq:Ds} containing all iterates. Then \Cref{alg:BRD} converges to a Nash equilibrium $\mathbf{b}^\star$. More precisely, for any $\epsilon > 0$ and any initial point $\mathbf{b}(0) \in \mathcal D^\ell$, after $K_\epsilon > \frac{2L^2}{\lambda^2} \log \bigl(\|\mathbf{g}(\mathbf{b}(0), \boldsymbol{\mu}(0))\|_2/\epsilon \bigr)$ iterations the gradients satisfy $\| \mathbf g(\mathbf{b}(K_\epsilon), \boldsymbol{\mu}(K_\epsilon) ) \|_2 < \epsilon$, when the step size is $\gamma = \lambda/L^2$.
\end{theorem}

\begin{proof}[Proof sketch]
All iterates lie in $\mathcal D^\ell$ by the construction of \eqref{eq:mu}. We track the $\ell_2$ norm of the update direction $\mathbf g(\mathbf b(k),\boldsymbol\mu(k))$. Keeping the correction $\boldsymbol\mu(k)$ fixed, a first-order expansion in integral form gives $\mathbf g(\mathbf b(k+1),\boldsymbol\mu(k))=(I+\gamma\bar G(k))\,\mathbf g(\mathbf b(k),\boldsymbol\mu(k))$, where $\bar G(k)$ is the average of $G$ along the segment from $\mathbf b(k)$ to $\mathbf b(k+1)$, which lies in $\mathcal D^\ell$ by convexity and so inherits both SB bounds (\cref{def:SB}). Substituting these bounds and choosing $\gamma=\lambda/L^2$ gives the contraction inequality $\|\mathbf{g}(\mathbf{b}(k), \boldsymbol{\mu}(k))\|_2^2 \leqslant \left(1 - \lambda^2/{L^2} \right)^k \|\mathbf{g}(\mathbf{b}(0), \boldsymbol{\mu}(0))\|_2^2$, implying exponential convergence.  Thus, we can conclude that \Cref{alg:BRD} reaches a Nash equilibrium $\mathbf b^\star$ with $\|\mathbf g(\mathbf b(K_\epsilon),\boldsymbol{\mu}(K_\epsilon))\|_2<\epsilon$ after $K_\epsilon=\mathcal O\!\left(\log\frac{1}{\epsilon}\right)$ iterations.
The complete proof is deferred to \Cref{app:BRD-proof}.
\end{proof}

Since each step of \Cref{alg:BRD} is $\gamma$ times the corrected gradient, \Cref{thm:BRD-conv} implies that the step lengths shrink geometrically, so the iterates form a Cauchy sequence and converge to some $\mathbf b^\star$ in the closed set $\mathcal D^\ell$. The corrected gradient is continuous in $\mathbf b$, so it vanishes at $\mathbf b^\star$, and by \Cref{lem:grad-NE}, $\mathbf b^\star$ is a Nash equilibrium. In particular, the SB property guarantees that a Nash equilibrium exists.

\Cref{thm:BRD-conv} alone does not produce a welfare guarantee. An equilibrium may exist and be reached by BRD while remaining inefficient relative to the social optimum. The following sections examine this welfare gap and develops the conditions under which it can be eliminated.

\section{Equilibrium Need Not Be Efficient}
\label{sec:welfare}

We now introduce the planner, whose goal is to choose a CSF that shapes firms' incentives so that their strategic behavior leads to a welfare-maximizing outcome. We first show by example that competition under a natural CSF can settle at a very inefficient equilibrium. \Cref{sec:regular_rcsf} then shows that the standard ratio-form CSFs, although they make BRD converge to a unique symmetric equilibrium, cannot guarantee efficiency, and identifies the reason. From that reason we derive a sufficient condition for efficiency (\Cref{sec:alignment}), find the CSF that satisfies it (\Cref{sec:scaled}), and add two instruments that make it work on every instance: a budget-balanced charge (\Cref{sec:pigou}) and a participation floor (\Cref{sec:floor}). \Cref{sec:main} then brings these pieces together to establish that the resulting CSF admits a unique, symmetric, welfare-maximizing Nash equilibrium.

\begin{example}[An inefficient equilibrium]
\label{ex:counter}
Consider two firms $A$ and $B$ competing over a fully connected network of five agents, with the influence matrix $\mathscr{E}$ in \Cref{tab:E}. The peer-influence weight $\alpha = 0.5$, which gives the centralities $M=(0.810,\,1.210,\,1.131,\,0.958,\,0.891)$. Each firm has budget $C=2$, and the platform uses the ratio-form CSF \citep{skaperdas1996contest}:
\[
h_{s,i}(\mathbf{b}) = \frac{e^{2b_{s,i}}-1}{\sum_{r \in \mathscr{M}} \bigl(e^{2b_{r,i}}-1\bigr) + 0.5}.
\]

\begin{table}[h]
\centering
\small
\caption{Influence matrix $\mathscr{E}$ for \Cref{ex:counter}.}
\label{tab:E}
\begin{tabular}{c|ccccc}
\toprule
& 1 & 2 & 3 & 4 & 5 \\
\midrule
1 & 0 & 0.3901 & 0.3003 & 0.2456 & 0.0640 \\
2 & 0.0669 & 0 & 0.3715 & 0.2578 & 0.3037 \\
3 & 0.0149 & 0.7005 & 0 & 0.1534 & 0.1313 \\
4 & 0.1407 & 0.2334 & 0.4025 & 0 & 0.2234 \\
5 & 0.4340 & 0.0989 & 0.2072 & 0.2599 & 0 \\
\bottomrule
\end{tabular}
\end{table}

Let $\left(\mathbf{b}_A^\text{opt}, \mathbf{b}_ B^\text{opt}\right)= \argmax_{\mathbf{b}} W_p(\mathbf{b})$ denote the optimal budgets. 
The NE is the symmetric profile $\mathbf b^\star_A=\mathbf b^\star_B=(0.230,\,0.448,\,0.399,\,0.302,\,0.268)$, with utilities $u_A(\mathbf{b}^\star) = u_B(\mathbf{b}^\star) = 0.320$, so that $W_p(\mathbf b^\star)=0.320$ for every $p\in (-\infty, 1]$. The utilitarian welfare-maximizing profiles are $\mathbf{b}_A^\text{opt} =(0,0.368,0.349,0,0.279)$ and $ \mathbf{b}_B^\text{opt} = (0.253,0,0,0.300,0)$, which give  $W_1 = 0.813$. At the egalitarian optimum $\mathbf{b}_A^\text{opt} =(0.162, 0, 0.348, 0, 0.279)$ and $ \mathbf{b}_B^\text{opt} = (0.067, 0.369, 0, 0.300, 0)$, the utilities $u_A\approx u_B\approx0.803$, so $W_p\approx0.803$ for every $p\in(-\infty,1]$. The Nash social welfare optimum $\mathbf{b}_A^\text{opt} =(0.253, 0, 0, 0.300, 0.279)$ and $ \mathbf{b}_B^\text{opt} = (0, 0.369, 0.348, 0, 0)$ similarly gives $W_\text{NSW} = 0.811$. 

\end{example}

In this example, both firms engage in strictly positive spending at equilibrium and contest every consumer, while the planner avoids this duplicated spending by assigning each consumer to one firm. The resulting duplication leads to a substantial welfare loss, with $\mathrm{PoA} \gtrsim 2.5$. This inefficiency stems from the structure of the existing CSF, which encourages firms to spend resources that do not contribute to higher total welfare. To overcome this, we next explore how an appropriately designed CSF can align equilibrium behavior with collective social welfare, starting with the most widely used class. 

\section{First Attempt: the Regular RCSF}
\label{sec:regular_rcsf}

The natural place to look for a good CSF is the class most commonly studied in contest theory. We focus on the \emph{ratio-form} CSF, in which a firm's success at a vertex is its own score relative to the total score of all firms competing for that vertex. This is the canonical axiomatic CSF of \citet{skaperdas1996contest}, characterized by monotonicity, anonymity, and consistency, and is the standard choice in the contest literature. So our first attempt is to find whether a CSF from this class ensures both BRD convergence to NE and welfare maximization.

Following \citet{bimpikis2016competitive}, we additively regularize the denominator by a constant $\delta > 0$, which models an outside option that absorbs unallocated attention and keeps the CSF well defined even when all budgets vanish. This gives the \emph{Ratio CSF} (RCSF),
\begin{equation}
\label{eq:csf_form}
h_{s,i}(\mathbf{b}) = \frac{f(b_{s,i})}{\sum_{r \in \mathscr{M}} f(b_{r,i}) + \delta},
\end{equation}
where $f$ is a scalar \emph{generator} applied to each firm's budget allocation $b_{s,i}$. A platform that scores ads by $f(\text{budget})$ and allocates impressions in proportion to scores implements an RCSF.

Not every RCSF is well behaved. We single out a subclass, defined by three conditions on $f$ over the interval $[0,C]$ of possible budgets, which suffices for BRD to converge to a unique symmetric equilibrium. The CSF of \mbox{\Cref{ex:counter}} is itself an RCSF, but its generator $e^{kb_{s,i}} - 1$ is convex, so it does not belong to this subclass.

\begin{definition}[Regular RCSF]
\label{def:regular_rcsf}
An RCSF is \emph{Regular} if its generator $f$ satisfies the following conditions, where all suprema are over budget $b\in[0,C]$.
\begin{enumerate}[label=(R\arabic*)]
\item \label{cond1}\emph{Concave increasing generator:} $f(0) = 0$, $f'(b) > 0$, and $f''(b) < 0$.
\item \label{cond2}\emph{Bounded slope and curvature:} $\sup f' < \infty$ and $\sup |f''| < \infty$.
\item \label{cond3}\emph{Curvature dominates coupling:}
\begin{equation}
\label{eq:strongconc}
\sup \frac{f'}{-f''} \cdot \sup f' \;<\; \frac{\delta}{m-1}.
\end{equation}
\end{enumerate}
\end{definition}

The three conditions are easy to check. \Cref{cond1} asks for a concave, increasing, normalized generator, \Cref{cond2} keeps the second derivatives of the utilities bounded, and \Cref{cond3} asks the generator's own curvature to be strong enough, relative to its slope and the number of rivals, to dominate the coupling between firms. For the saturating exponential $f(b) = 1 - e^{-\beta b}$ with $\beta>0$, we have $f'/(-f'') \equiv 1/\beta$ and $\sup f' = \beta$, so the left side of \Cref{eq:strongconc} equals $1$ and the CSF is Regular whenever $\delta > m-1$. The logarithmic generator $f(b)=\log(1+\beta b)$ is Regular when $\delta > (m-1)(1+\beta C)$. The power generator $f(b)=b^q$ \citep{tullock1980efficient} satisfies Condition~\ref{cond1}, but for $q<1$ its slope and curvature are unbounded near zero, and for $q=1$, since $f''=0$, it has no curvature at all.

The next two lemmas establish the equilibrium properties of the Regular RCSF. \Cref{lem:rcsf-sb} (proof in \Cref{app:rcsf-sb}) shows that BRD converges, and \Cref{lem:rcsf-unique-symmetric} (proof in \Cref{app:RRCSF_sym_unique}) characterizes the equilibrium it reaches.

\begin{lemma}[Regular RCSF induces the SB property]
\label{lem:rcsf-sb}
If the CSF is a Regular RCSF (\Cref{def:regular_rcsf}), then the induced utilities satisfy the SB property (\Cref{def:SB}) on $\mathcal D^\ell = \mathcal{B}$. Consequently, by \cref{thm:BRD-conv} the best response dynamics (\Cref{alg:BRD}) converges to a Nash equilibrium $\mathbf b^\star$.
\end{lemma}

\begin{psketch}
Each firm's success at consumer $i$ depends only on the budgets placed on $i$, so after reordering coordinates the block Hessian is block diagonal, with one $m\times m$ block per consumer. \cref{cond3} makes every block strictly diagonally dominant with a negative diagonal, which by Gershgorin's circle theorem gives strong concavity, and \Cref{cond2} bounds every entry, which gives the norm bound. See \Cref{app:rcsf-sb}.
\end{psketch}

\begin{lemma}[Unique and symmetric NE for Regular RCSF]
\label{lem:rcsf-unique-symmetric}
For an RCSF whose generator $f$ satisfies \Cref{cond1}, the ACG has a unique Nash equilibrium, and it is symmetric: $\mathbf{b}_1^\star = \cdots = \mathbf{b}_m^\star$.
\end{lemma}

\begin{psketch}
Strict concavity of each firm’s utility, together with symmetry, compactness, and convexity of the strategy spaces, gives existence of a symmetric NE by \cite{Cheng2004}. To rule out asymmetric NE, the Regular RCSF satisfies the marginal dominance inequality: if $b_{p,i}>b_{q,i}$, then $\partial h_{p,i}/\partial b_{p,i}<\partial h_{q,i}/\partial b_{q,i}$. Combined with the first-order conditions of the firm $M_i\partial h_{s,i}/\partial b_{s,i}=1+\lambda_s-\mu_{s,i}$, this implies that a firm spending more at some vertex has a strictly smaller budget multiplier. There are two cases for an asymmetric profile, and in both cases the first-order conditions together with the marginal dominance inequality give a contradiction. Hence every NE is symmetric. At a symmetric profile, the marginal return $\partial h_{s,i}/\partial b_{s,i}$ is strictly decreasing in the common budget, so the symmetric first-order conditions have exactly one solution, giving a unique symmetric NE. 
See \Cref{app:RRCSF_sym_unique}.
\end{psketch}

The Regular RSCF therefore secures a well-behaved equilibrium. Nevertheless, this class is fundamentally incapable of implementing socially optimal equilibria. Before establishing this impossibility result in the following \Cref{thm:RCSF_impossibility}, we introduce the notation and auxiliary quantities used in its proof and in the analysis that follows.

\begin{definition}[Private and social marginal returns, and the wedge]
\label{def:wedge}
At a symmetric profile with common value $b_{s,i} = t_i$ for every firm $s \in \mathscr M$, define
\[
a(t_i) := \left.\frac{\partial h_{s,i}}{\partial b_{s,i}}\right|_{\mathrm{sym}}, \qquad d(t_i) := \left.\frac{\partial h_{r,i}}{\partial b_{s,i}}\right|_{\mathrm{sym}}\ (r\neq s), \qquad \psi(t_i) := \sum_{r\in\mathscr{M}}\left.\frac{\partial h_{r,i}}{\partial b_{s,i}}\right|_{\mathrm{sym}} = a(t_i) + (m-1)d(t_i).
\]
Here $a(t_i)$ is the \emph{private marginal return} to firm $s$ from its own spending at vertex $i$, $d(t_i)$ is the marginal effect of that spending on a single rival, and $\psi(t_i)$ is the \emph{social marginal return}, the marginal effect on the sum of all $m$ firms' success probabilities at $i$. The \emph{externality wedge} is
\begin{equation}
   w(t_i) := a(t_i) - \psi(t_i) = -(m-1)\, d(t_i).
   \label{eq:wedge}
\end{equation}
\end{definition}

By symmetry, these quantities do not depend on which firm is considered. The private and social marginal returns determine the respective first-order conditions of the firm and the planner, and the wedge \(w(t_i)\) measures the difference between these marginal incentives. \Cref{thm:RCSF_impossibility} shows that this wedge can prevent the Nash equilibrium from being welfare optimal.

\begin{theorem}[Regular RCSFs cannot ensure that equilibrium is welfare optimal]
\label{thm:RCSF_impossibility}
Let $h$ be a Regular RCSF. Then there exist instances of the awareness-competition game under this CSF for which the unique symmetric Nash equilibrium $\mathbf{b}^\star$ of \Cref{lem:rcsf-unique-symmetric} fails to maximize the utilitarian welfare $W_1$ over the feasible set $\mathcal{B}$.
\end{theorem}

\begin{psketch}
Utilitarian welfare $W_1$ depends on the budgets only through the per-consumer totals of budgets $b_{s,i}$ and scores $f(b_{s,i})$, and by Jensen’s inequality and concavity of $f$, symmetrizing any feasible profile weakly increases $W_1$ while preserving aggregate spending. Hence it suffices to compare the symmetric NE with the welfare optimum among symmetric profiles. At a symmetric profile, the firm and planner’s first-order conditions involve the private marginal return $a(t_i)$ and social marginal return $\psi(t_i)$, and for $t_i>0$ are $M_i a(t_i)=1+\lambda_s$ and $M_i\psi(t_i)=1+\hat\lambda$, respectively. For the Regular RCSF, the wedge $w(t_i)=a(t_i)-\psi(t_i)>0$, and the ratio $a(t_i)/\psi(t_i)=1+(m-1)f(t_i)/\delta$ is strictly increasing in $t_i$. If the equilibrium funds vertices with different centralities, the two first-order conditions therefore cannot hold simultaneously. If all centralities are equal, we construct an instance whose welfare optimum is positive and leaves the budget unexhausted. Then both budget multipliers are zero, which would require $a(t)=\psi(t)$, contradicting the strictly positive wedge. Thus, in either case, the unique symmetric NE is not welfare-optimal. The complete proof is in \Cref{app:RCSF_impossibility}.
\end{psketch}

\begin{remark}[The failure is structural]
\label{rem:wedge-structural}
The wedge is the familiar \emph{business-stealing externality}: part of what a firm gains from its spending is taken from a rival, so it counts in the firm's own utility but cancels in total welfare. \Cref{thm:RCSF_impossibility} shows that it is strictly positive for every Regular RCSF and every symmetric profile with positive budgets, so the failure does not depend on a particular generator $f$ or regularizer $\delta$. The wedge would vanish only for $\delta = 0$, but then $\sum_s h_{s,i} \equiv 1$ and, whenever budgets are exhausted, total welfare no longer depends on the allocation, which makes the design problem vacuous.
\end{remark}

\cref{thm:RCSF_impossibility} shows that the canonical Regular RCSF cannot guarantee that the Nash equilibrium reached by BRD is welfare-optimal. It already fails for the utilitarian welfare criterion $p=1$. Since our design goal is stronger, namely an equilibrium maximizing $W_p$ for every $p\in(-\infty,1]$, ruling out the utilitarian case is sufficient to rule out the full-family requirement. We therefore characterize the conditions under which the firms' and planner's incentives can align. 

\subsection{The alignment criterion of social optimality and NE}
\label{sec:alignment}

The externality wedge \eqref{eq:wedge} need not vanish for the planner's and firms' incentives to be aligned as requiring $w\equiv0$ would imply $d\equiv0$, meaning that a firm's spending has no effect on any rival, which is no longer a contest. 
For the equilibrium to coincide with the welfare optimum, the stationarity conditions in the KKT conditions of each firm's best-response problem must be satisfied by the profile that satisfies the stationarity conditions in the KKT conditions of the planner's problem. This requires an alignment condition between the private and social marginal returns.
The following proposition gives the exact condition for this alignment in terms of the social and private marginal returns $\psi(t_i)$ and $a(t_i)$ at symmetric profiles,\footnote{The proof of \Cref{thm:RCSF_impossibility} shows that, for the Regular RCSF, replacing a feasible profile with its corresponding symmetric profile weakly increases welfare. This motivates considering symmetric profiles in the alignment analysis. The symmetry restriction is subsequently verified for the CSF identified by this analysis.} as defined in \Cref{def:wedge}.

\begin{proposition}[Alignment criterion for social optimality and NE]
\label{prop:alignment}
Consider a CSF for which the utility of a firm $s$, i.e. $u_s$ is concave in $\mathbf{b}_s$, and whose private and social marginal returns (\Cref{def:wedge}) at symmetric profiles $b_{s,i} = t_i$ satisfy the proportional-marginals condition
\begin{equation}
\label{eq:proportional}
    \psi(t_i) = c\, a(t_i),
\end{equation}
for a constant $c \in (0,1)$. Let $\mathbf{b}^{\mathrm{opt}}$ be a symmetric profile that maximizes $W_1$,
$b^{\mathrm{opt}}_{s,i} = t^{\mathrm{opt}}_i$ for every firm $s \in \mathscr{M}$. Then:
\begin{enumerate}
  \item[(i)] if $\sum_i t^{\mathrm{opt}}_{i} = C$, that is, every firm exhausts its budget, then        $\mathbf{b}^{\mathrm{opt}}$ is a Nash equilibrium;
  \item[(ii)] conversely, if $\mathbf{b}^{\mathrm{opt}} \neq \mathbf{0}$ and
        $\mathbf{b}^{\mathrm{opt}}$ is a Nash equilibrium, then
        $\sum_i t^{\mathrm{opt}}_{i} = C$.
\end{enumerate}
\end{proposition}

\begin{psketch}
Substituting $\psi=ca$ into the planner's condition $M_i\psi(t_i)-1=\hat\lambda-\hat\mu_i$ and dividing by $c$ yields the firm's condition $M_ia(t_i)-1=\lambda_s-\mu_{s,i}$ with the relabeled multipliers $\lambda_s=(\hat\lambda+1-c)/c$ and $\mu_{s,i}=\hat\mu_i/c$. All requirements on these multipliers are inherited from the planner's, except complementary slackness for the budget: since $c<1$, the multiplier $\lambda_s$ is strictly positive, so the budget must be exhausted. Concavity makes the firm's conditions sufficient. For (ii), an unexhausted budget would force both budget multipliers to zero, and then $\psi=a$ at any funded consumer, which contradicts $c<1$. The complete proof is in \Cref{app:alignment}.
\end{psketch}

\begin{remark}
\label{rem:alignment-reading}
Since $\psi=a+(m-1)d$, the condition $\psi=ca$ is equivalent to $d(t_i)/a(t_i) = (c-1)/(m-1)$ for all $t_i$. For $c<1$ the cross effect $d$ stays negative, so the contest retains genuine rivalry. The condition therefore requires the ratio of the cross effect to the own effect to remain constant across spending levels. For the Regular RCSF, this ratio is 
$-f(t_i)/[(m-1)f(t_i)+\delta]$, which depends on $t_i$ for every $\delta>0$, with the only constant-ratio case $\delta=0$ rendering the design problem vacuous as noted in \Cref{rem:wedge-structural}. This gives a second view of why the Regular RCSF fails.
\end{remark}

\Cref{prop:alignment} separates the design of a CSF that aligns NE and social optimality, into two tasks. The first is to construct a CSF with proportional private and social marginal returns, which is determined by the functional form of the CSF and is addressed in the next section. The second is to ensure that the firms exhaust their full budget at the welfare optimum, which, as we will see, depends on the underlying instance rather than on the CSF and cannot be controlled by the designer.

\section{Second Attempt: Scaled ratio CSFs}
\label{sec:scaled}

We now construct a CSF satisfying the alignment criterion in \Cref{prop:alignment}. Since a constant ratio $d/a$ is unattainable inside the Ratio CSF form \eqref{eq:csf_form}, we enlarge the class while preserving its ratio structure as it makes the rule a contest. Specifically, we scale each firm's contest share by a function of the total budget spent on the consumer.

\begin{definition}[Scaled Ratio CSF]
\label{def:srcsf}
A \emph{Scaled Ratio CSF }(SRCSF) has the form
\begin{equation}
\label{eq:SRCSF}
h_{s,i}(\mathbf{b}) = f(B_i)\cdot\frac{b_{s,i}}{B_i}, \qquad B_i:=\sum_{r\in\mathscr M}b_{r,i},
\end{equation}
for an increasing, concave function $f:[0,mC]\to[0,1]$ with $f(0)=0$. When $B_i=0$ we set $h_{s,i}(\mathbf b):=0$, which is the continuous extension. 
\end{definition}

Summing \eqref{eq:SRCSF} over firms gives $\sum_sh_{s,i}=f(B_i)$, so $f(B_i)$ is the total attention the platform releases at vertex $i$ when firms spend $B_i$ there. The second factor is the classical Tullock share \citep{tullock1980efficient}. The two factors separate the two economic forces cleanly: $f$ governs how much attention competition creates in total, and the share governs how it is divided. 
At a symmetric profile, $b_{s,i}=t_i=B_i/m$, we derive $\psi,a$ (\Cref{def:wedge}) as follows:
\begin{equation}
\label{eq:srcsf-a-psi}
  \psi(t_i) = f'(B_i), \qquad
  a(t_i) = \frac{f'(B_i)}{m} + \frac{(m-1)f(B_i)}{mB_i},
\end{equation}
and their ratio is
\begin{equation}
\label{eq:c_PRCSF}
\frac{\psi(t_i)}{a(t_i)} = \frac{m\,f'(B_i)\,B_i}{f'(B_i)\,B_i+(m-1)f(B_i)},
\end{equation}
which can be expressed in terms of the elasticity of $f$, which we define below.

\begin{definition}[Elasticity and iso-elasticity]
\label{def:elasticity}
The \emph{elasticity} of a differentiable positive function $f$ at $B_i$ is the ratio of the relative change in $f(B_i)$ to the relative change in $B_i$ \citep{simon1994mathematics}:
\[
e(B_i) := \frac{d\log f(B_i)}{d\log B_i} = \frac{f'(B_i)\,B_i}{f(B_i)} 
\]
We call $f$ \emph{iso-elastic} if $e(B_i)$ does not depend on $B_i$.
\end{definition}

In words, $e(B_i)$ measures the percentage change in total attention $f(B_i)$ released at a vertex $i$ resulting from a one-percent change in aggregate budget spending $B_i$. Iso-elasticity means that this percentage is constant across budget levels.

\begin{lemma}[Iso-elasticity is necessary and sufficient]
\label{lem:isoelastic}
Within the SRCSF class \eqref{eq:SRCSF}, the proportional-marginals condition \eqref{eq:proportional} of \Cref{prop:alignment} holds if and only if $f$ is iso-elastic, that is, $f(B_i)=\kappa B_i^\eta$ for constants $\kappa>0$ and $\eta\in(0,1]$.
\end{lemma}
\begin{psketch}
Dividing \Cref{eq:c_PRCSF} through by $f(B_i)$ gives $\psi/a = m\,e(B_i)/(e(B_i)+m-1)$, a strictly increasing function of the elasticity, so the ratio is constant exactly when the elasticity is. Solving $f'/f=\eta/B_i$ gives $f=\kappa B_i^\eta$, and the requirements on $f$ force $\eta\in(0,1]$. The complete proof is in \Cref{app:isoelastic}.
\end{psketch}

\Cref{lem:isoelastic} identifies the functional form required for alignment within the SRCSF class. In particular, the total attention released by the platform at a vertex must have constant elasticity with respect to aggregate budget spending at that vertex. We therefore restrict attention to this iso-elastic subclass and we name the resulting CSF after its power-function form.

\begin{definition}[Power RCSF]
\label{def:power_rcsf}
For $\eta\in(0,1)$ and $0<\kappa\leqslant(mC)^{-\eta}$, the \emph{Power-scaled Ratio CSF }(Power RCSF) is the SRCSF \eqref{eq:SRCSF} with $f(B_i)=\kappa B_i^\eta$, that is
\begin{equation}
\label{eq:power_rcsf}
h_{s,i}(\mathbf{b}) = \kappa\,b_{s,i}\,B_i^{\eta-1}.
\end{equation}
\end{definition}

The cap on $\kappa$ ensures $\sum_sh_{s,i}=\kappa B_i^\eta\leqslant\kappa(mC)^\eta\leqslant1$ since $B_i\leqslant mC$ and $\eta>0$,. The restriction $\eta<1$ keeps the rule a contest, because then a rival's spending strictly reduces a firm's success:
\begin{equation}
\label{eq:power-derivs}
\frac{\partial h_{s,i}}{\partial b_{s,i}} = \kappa B_i^{\eta-2}\bigl[B_i+(\eta-1)b_{s,i}\bigr] > 0,
\qquad
\frac{\partial h_{s,i}}{\partial b_{r,i}} = \kappa(\eta-1)\, b_{s,i}\, B_i^{\eta-2} < 0
\quad (r \neq s),
\end{equation}
The parameter $\eta$ interpolates between two degenerate extremes. As $\eta\to0$, the total attention $\kappa B_i^\eta$ becomes constant and the rule reduces to the classical Tullock contest, where firms' spending only redistributes attention among competitors. As $\eta\to1$, we get $h_{s,i}=\kappa b_{s,i}$, so a firm's success depends only on its own spending, eliminating rivalry among firms.

\begin{proposition}[Power RCSF has proportional marginals]
\label{prop:power-align}
For the Power RCSF (\Cref{def:power_rcsf}), the induced utility of each firm is concave in its own strategy, and the proportional marginals  condition \eqref{eq:proportional} of \Cref{prop:alignment} holds with
\begin{equation}
\label{eq:c-value}
c = \frac{m\eta}{m+\eta-1}\in(0,1).
\end{equation}
\end{proposition}

\begin{psketch}
Concavity follows from the sign of the second derivative of $h_{s,i}$ in $b_{s,i}$. Since the elasticity of $\kappa B_i^\eta$ is identically $\eta$, the formula in the proof of \Cref{lem:isoelastic} gives $c=m\eta/(\eta+m-1)$, which lies in $(0,1)$ because $\eta<1$. The complete proof is in \Cref{app:power-align}.
\end{psketch}

\begin{remark}
For the Power RCSF, the total attention $\sum_s h_{s,i}=f(B_i)=\kappa B_i^\eta$ depends on aggregate spending when $\eta\in(0,1)$, ensuring total welfare remains allocation-dependent. Unlike Regular RCSF, the Power RCSF therefore attains the constant ratio $c$ in \Cref{eq:c-value} and hence the constant ratio $d/a = (c-1)/(m-1)$ without falling into the degenerate case of $\delta=0$ flagged in \Cref{rem:wedge-structural,rem:alignment-reading}.
\label{rem:const-charge}
\end{remark}

\Cref{prop:power-align} establishes the proportional marginals condition identified in \Cref{prop:alignment}. The remaining condition to be verified is full budget exhaustion at the optimum.
The next proposition characterizes exactly when full budgets are exhausted for the Power RCSF, and shows that the condition depends on the primitives of the game instance such as the number of vertices $n$, the total budget $C$ and the scaled network centralities $M_i$, none of which the designer controls.

\begin{proposition}[When are budgets exhausted?]
\label{prop:budget_bind_cond}
Fix the Power RCSF with parameters $\eta \in (0,1)$ and $\kappa \leqslant (mC)^{-\eta}$, and write $\theta := 1/(1-\eta) > 1$ and $\|M\|_\theta := (\sum_i M_i^\theta)^{1/\theta}$. Then:
\begin{enumerate}[label=\textup{(\roman*)}]
\item \label{bb:i} The symmetric welfare-maximizing profile exhausts every firm's budget if and only if
\begin{equation}
\label{eq:budget_bind_cond}
  \sum_{i\in\mathscr{N}}\bigl(\kappa\eta M_i\bigr)^{\theta} \;\geqslant\; mC .
\end{equation}
\item \label{bb:ii} If \eqref{eq:budget_bind_cond} holds for some admissible $\kappa$, then $\eta\,\|M\|_{\theta} \geqslant mC$.
\item \label{bb:iii} Consequently, if $n \leqslant mC$, then \emph{no} admissible pair $(\kappa,\eta)$ satisfies \eqref{eq:budget_bind_cond}: on such instances no Power RCSF can align equilibrium with the welfare optimum.
\end{enumerate}
\end{proposition}

\begin{psketch}
Utilitarian welfare depends only on the totals $B_i$ and equals $\frac1m[\kappa\sum_iM_iB_i^\eta-\sum_iB_i]$, whose unconstrained maximizer is $B_i=(\kappa\eta M_i)^\theta$; the budget is exhausted exactly when these totals add up to at least $mC$, which is (i). Part (ii) substitutes the largest admissible $\kappa$. For (iii), $\|M\|_\theta\leqslant\sum_iM_i=n$ by \Cref{lem:sumM}, so (ii) fails when $n\leqslant mC$; the optimum is then unexhausted and non-zero, and part (ii) of \Cref{prop:alignment} rules out alignment. The complete proof is in \Cref{app:budget_bind_cond}.
\end{psketch}

The above proposition shows that full budget exhaustion cannot be treated as a general consequence of choosing the Power RCSF as condition \eqref{eq:budget_bind_cond} depends on the underlying instance, rather than the CS design parameters. Thus, even after identifying the appropriate CSF within the SRCSF class, alignment of welfare and NE cannot be guaranteed for all instances, as shown in \cref{prop:budget_bind_cond}\ref{bb:iii}. This motivates a different approach to restore alignment.

\subsection{Adding a Corrective Charge}
\label{sec:pigou}

We have seen that the Power RCSF can align private and social marginal returns (\Cref{prop:power-align}), but the full budget exhaustion condition \eqref{eq:budget_bind_cond} needed for alignment depends on the ACG instance and cannot be guaranteed by the CSF design alone (\Cref{prop:budget_bind_cond}). We therefore take a different route and modify the firms' incentives directly.
In this paper, we therefore restore alignment via a corrective charge on advertising expenditure, following the standard \emph{Pigouvian} logic of pricing that is prevalent in the literature~\citep{pigou1920economics,baumol1972taxation}. The firm's marginal incentive can be aligned with the planner's social marginal return without requiring the firms to exhaust their budgets.

Concretely, let the platform levy a per-unit charge $\tau\geqslant0$ on advertising spend and redistribute the resulting charge revenue. Let $\rho_s$ be the \emph{rebate} received by firm $s$ and require the scheme to be \emph{budget balanced},
\begin{equation}
\label{eq:bb}
\sum_{s\in\mathscr{M}}\rho_s(\mathbf{b}) \;=\; \tau\sum_{s\in\mathscr{M}}\sum_i b_{s,i},
\end{equation}
so that the platform retains no charge revenue. Budget balance is an aggregate condition and does not require each firm to receive its own charge revenue. A concrete scheme satisfying this condition is
\begin{equation}
\label{eq:rebate}
\rho_s(\mathbf{b}_{-s}) := \frac{\tau}{m-1}\sum_{r\neq s}\sum_i b_{r,i},
\end{equation}
which credits firm $s$ with an equal share of of the charge payments made by its rivals. Thus each firm's payment is fully redistributed across the other $m-1$ firms. Schemes of this kind are used for the same purpose by \citet{murhekar2024incentives}. Importantly, $\rho_s$ depends only on rivals' expenditures and is therefore a lump sum from firm $s$'s perspective. Firm $s$ consequently maximizes the \textit{taxed utility}
\begin{equation}
    \label{eq:pigouvian_utility}
    \widetilde{u}_s(\mathbf{b}) = \sum_iM_ih_{s,i}(\mathbf{b}) - (1+\tau)\sum_ib_{s,i} + \rho_s(\mathbf{b}_{-s}),
\end{equation}
From here on, whenever a charge is in place, the utilities in the welfare function of \Cref{def:p-mean}, and in the set $\mathcal B_{\geqslant0}$, are the taxed utilities $\widetilde u_s$. By budget balance, the fees and rebates cancel in the sum, $\sum_{s}\widetilde{u}_s(\mathbf{b}) = \sum_{s} u_s(\mathbf{b}),$ 
so utilitarian welfare, and with it the planner's first-order conditions, is unchanged, while each firm now faces the unit cost $1+\tau$ instead of $1$. The charge therefore moves the firms' first-order conditions toward the planner's without requiring budgets to be exhausted.
The next proposition, which is the analogue of \Cref{prop:alignment}, establishes the corresponding alignment result. 

\begin{proposition}[Alignment of welfare and NE with a corrective charge]
\label{prop:tax-alignment}
Consider a CSF whose induced taxed utility $\widetilde u_s$ \eqref{eq:pigouvian_utility} is concave in $\mathbf{b}_s$, and which satisfies the proportional-marginals condition \eqref{eq:proportional} for a constant $c\in(0,1)$. Let the firms face a budget-balanced charge at the rate
\begin{equation}
\label{eq:tau}
\tau \;=\; \frac{1}{c}-1 .
\end{equation}
Then every symmetric profile that maximizes $W_1$  is a Nash equilibrium, whether or not budgets are exhausted.
\end{proposition}
\begin{psketch}
With the charge, the firm's first order condition is $M_ia(t_i)-(1+\tau)=\lambda_s-\mu_{s,i}$. Substituting $\psi=ca$ into the planner's condition and dividing by $c$ gives exactly this, with $1+\tau=1/c$ and the multipliers $\lambda_s=\hat\lambda/c$ and $\mu_{s,i}=\hat\mu_i/c$. Unlike in \Cref{prop:alignment}, the budget multiplier now carries no additive offset, so every complementary slackness condition is inherited from the planner's. The complete proof is in \Cref{app:tax-alignment}.
\end{psketch}

Note that the Regular RCSF fails the proportional marginals condition \eqref{eq:proportional} (\Cref{rem:alignment-reading}), and the argument of \Cref{thm:RCSF_impossibility} goes through with $1+\tau$ in place of $1$, so no constant charge restores alignment for that class.

Two features of the charge \eqref{eq:tau} are useful. 
It is a constant per-unit rate, which follows from the constant ratio $c$ in \Cref{eq:proportional}, and for the Power RCSF it depends only on $m$ and $\eta$, not on the network or the budgets, so the platform can compute it without knowing the instance.
Before turning to the convergence dynamics, we first characterize the equilibrium under the Power RCSF under a common charge in \cref{lem:power-unique-sym} below. 

\begin{lemma}[Unique and symmetric Nash equilibrium for the Power RCSF]
\label{lem:power-unique-sym}
Consider the Power RCSF (\Cref{def:power_rcsf}) with every firm facing the same unit advertising charge $1+\tau\geqslant1$. Then the induced game admits a unique and symmetric Nash equilibrium. Writing $\theta := 1/(1-\eta)$ and $K := \kappa\, m^{\eta-2}(m+\eta-1)$, the equilibrium $\mathbf b^\star$ is
\begin{equation}
\label{eq:tstar}
b_{s,i}^\star = t_i^\star = \beta_i(\Gamma^\star), \quad \forall s \in \mathscr M; \qquad \beta_i(\Gamma) := \Bigl(\frac{M_iK}{\Gamma}\Bigr)^{\theta},
\end{equation}
where $\Gamma^\star\geqslant1+\tau$ is the unique value with $S(\Gamma^\star)=\min\{S(1+\tau),\,C\}$ and $S(\Gamma):=\sum_i\beta_i(\Gamma)$.
\end{lemma}
\begin{psketch}
The argument has the same two parts as \Cref{lem:rcsf-unique-symmetric}, with the unit cost $1$ replaced by $1+\tau$, and three details change. First, the marginal dominance inequality is re-derived directly from \eqref{eq:power-derivs}: because swapping two firms' budgets at a vertex leaves $B_i$ unchanged, the firm bidding strictly more there has the strictly smaller marginal return, and the same case analysis on multipliers then rules out every asymmetric profile. Second, at a symmetric profile $a(t)=Kt^{\eta-1}$, so $a'(t)=K(\eta-1)t^{\eta-2}<0$ and $a$ is strictly decreasing, hence invertible, which gives \eqref{eq:tstar}. Third, $a(t)\to+\infty$ as $t\to0^+$ because $\eta<1$, so no coordinate of the symmetric candidate can sit at the lower boundary and $S$ is strictly decreasing on all of $(0,\infty)$, whence $\Gamma^\star$ is unique. The complete proof is in \Cref{app:power-unique}.
\end{psketch}

The above lemma establishes the uniqueness and symmetry of the equilibrium under the Power RCSF, and gives an explicit expression for the equilibrium budgets used in the results that follow. We now turn to the question of whether
this equilibrium can be reached through best-response dynamics (\cref{alg:BRD}).

\subsection{BRD to NE: Challenges and the need for a participation floor}
\label{sec:floor}

\Cref{thm:BRD-conv} guarantees convergence of BRD only when the utilities have the SB property, and in particular when their second derivatives are bounded. This subsection shows that the Power RCSF fails the SB property on the unrestricted feasible set $\mathcal{B}_s$. The failure arises near $B_i=0$ from the same constant-elasticity feature of the Power RCSF, $f(B_i)=\kappa B_i^\eta$, that enables alignment of welfare and NE.
Since $\eta<1$, both $f'(B_i)=\kappa\eta B_i^{\eta-1}$ and the curvature of $f$ become unbounded as $B_i\to0$. The proposition below formalizes this failure and motivates the participation floor introduced thereafter.

\begin{proposition}[Unbounded curvature without a floor]
\label{prop:unbounded}
Let $n\geqslant2$, $m\geqslant2$, $\eta\in(0,1)$, $\kappa>0$, and let $\mathcal{B}=\prod_s\{\mathbf{b}_s\geqslant\mathbf{0}:\sum_ib_{s,i}\leqslant C\}$ be the unrestricted feasible set. Let $H_i\in\mathbb{R}^{m\times m}$ denote the vertex-$i$ Hessian block, $(H_i)_{sr}=M_i\,\partial^2h_{s,i}/(\partial b_{r,i}\partial b_{s,i})$. Then for every vertex $i$,
\[
\sup_{\mathbf{b}\in\mathcal{B}}\|H_i(\mathbf{b})\|_{\mathrm{op}} = +\infty.
\]
Consequently the induced utilities do not satisfy the SB property (\Cref{def:SB}) on $\mathcal{B}$: no finite bound $L$ can satisfy the bounded-operator-norm condition, and the step size $\gamma=\lambda/L^2$ of \Cref{thm:BRD-conv} is not defined.
\end{proposition}
\begin{psketch}
Let every firm place $B_i/m$ on consumer $i$ and spend the rest elsewhere. A diagonal entry of $H_i$ then has magnitude proportional to $B_i^{\eta-2}$, which diverges as $B_i\to0$ because $\eta-2<0$. The complete proof is in \Cref{app:unbounded}.
\end{psketch}

The singularity arises as $B_i=\sum_s b_{s,i}\to0$, when all firms simultaneously do not spend any budget at vertex $i$, a standard technical difficulty in Tullock-style (ratio-form) contests \citep{tullock1980efficient}. By contrast, a zero budget spent by one firm does not create a singularity when at least one rival maintains positive spending. Previous work handles this difficulty either by modifying the best-response correspondence at zero effort \citep{ghosh2023best,elkind2024continuous} or by introducing a temporary positive lower bound and subsequently letting it vanish \citep{ewerhart2020unique}. In our setting, the singularity is more consequential because the Power RCSF has unbounded curvature as aggregate spending $B_i$ approaches zero. We therefore introduce a participation constraint to keep the process away from this singular region, following the spirit of minimal-effort participation constraints used as a substantive design feature in \citet{megidish2013allocation}.

\begin{definition}[Floored game]
\label{def:floor}
Every firm must invest at least $\varepsilon>0$ on every consumer, which can be read as an upfront participation fee of $n\varepsilon$ charged by the platform, and allocate the rest of its budget freely. 
Each firm's feasible set becomes the \emph{floored set}
$\mathcal{B}_s^\varepsilon := \mathcal D_s^\varepsilon$ of \Cref{eq:Ds}, with $\mathcal{B}^\varepsilon:=\prod_{s}\mathcal{B}_s^\varepsilon$, and \Cref{alg:BRD} is run with $\mathcal D_s^\ell=\mathcal B_s^\varepsilon$, that is, with the lower boundary $\ell=\varepsilon$ in \eqref{eq:Ds} in place of $\ell=0$. On $\mathcal{B}^\varepsilon$, every consumer receives total spending $B_i\in[m\varepsilon,mC]$.
\end{definition}

Restricting the strategy space can alter both the equilibrium and the optimum, and therefore may affect the alignment between them as established in \Cref{prop:tax-alignment}. The next lemma shows that, for a sufficiently small floor,
the equilibrium remains unaffected, and provides the corresponding threshold. We say the floor is \emph{inactive} at a profile if every coordinate satisfies $b_{s,i}>\varepsilon$, so that the constraint $b_{s,i}\geqslant\varepsilon$ does not bind and its multiplier is zero.

\begin{lemma}[Threshold for an inactive floor]
\label{lem:explicit-floor}
Let $\theta=1/(1-\eta)$, $K=\kappa m^{\eta-2}(m+\eta-1)$, and let $\mathbf b^\star$ be the equilibrium of \Cref{lem:power-unique-sym} on the unrestricted set. Since the equilibrium is symmetric, write $b_{s,i}^\star = t_i^\star, \; \forall s \in \mathscr M$ and define
\begin{equation}
\label{eq:epsbar}
\bar\varepsilon \;=\; \min\left\{\ \Bigl(\frac{K\,\min_iM_i}{1+\tau}\Bigr)^{\theta},\ \ C\,\frac{\min_iM_i^{\theta}}{\sum_jM_j^{\theta}}\ \right\} \;>\;0.
\end{equation}
Then for every $\varepsilon\in(0,\bar\varepsilon)$, we have $t^\star_i>\varepsilon$ for all $i$, so the floor is inactive at the equilibrium. Moreover $\mathcal{B}^\varepsilon$ is non-empty, since $n\varepsilon<C$.
\end{lemma}

\begin{psketch}
By \Cref{lem:power-unique-sym}, equilibrium spending is proportional to $M_i^\theta$ across consumers, so $\min_it_i^\star$ equals the total equilibrium spending $S(\Gamma^\star)$ times $\min_iM_i^\theta/\sum_jM_j^\theta$. If the budget is not exhausted, the total is $S(1+\tau)$ and this gives the first term of \Cref{eq:epsbar}; if it is exhausted, the total is $C$ and this gives the second. In each case, the relevant term is also the smaller one. The complete proof is in \Cref{app:explicit-floor}.
\end{psketch}

\paragraph{Why a floor.} The floor is one of several possible remedies for the unbounded curvature of \Cref{prop:unbounded}. The role of the participation floor, including the alternatives considered, is discussed in detail in \Cref{app:floor-why}.
The next lemma shows that the Power RCSF satisfies the SB Property (\Cref{def:SB}) for the floored game (\Cref{def:floor}). 

\begin{lemma}[The Power RCSF satisfies SB on the floored set]
\label{lem:power-SB}
On $\mathcal{B}^\varepsilon$ and with any common unit advertising charge $1+\tau\geqslant1$, the utilities induced by the Power RCSF satisfy the SB property (\Cref{def:SB}, with $\mathcal{D}^\ell = \mathcal{B}^\varepsilon$) with the constants
\[
\widetilde{\lambda} := \kappa(1-\eta)\,\chi(m,\eta)\,(mC)^{\eta-2}\min_iM_i \;>\; 0, \qquad \widetilde{L} := 2m\,\kappa(1-\eta)\,(m\varepsilon)^{\eta-2}\max_iM_i \;<\; \infty,
\]
where $\chi(m,\eta) := \dfrac{1}{2}\Big[m+\eta-\sqrt{m(m-2\eta+\eta^2)}\Big] > 0$ for $\eta\in(0,1)$, $m\geqslant2$.
\end{lemma}

\begin{psketch}
As for the Regular RCSF, it suffices to bound each block $H_i$. For the Power RCSF, every row of $H_i$ has equal off-diagonal entries, which gives $H_i = c_0(\mathbf{v}\mathbf{1}^\top + B_i I_m)$ with $c_0:=M_i\kappa(\eta-1)B_i^{\eta-3}<0$, $v_s:=B_i-(2-\eta)b_{s,i}$. 
The symmetric part of $\mathbf{v}\mathbf{1}^\top$ has rank at most two, with eigenvalues $\tfrac12(\mathbf1^\top\mathbf v\pm\sqrt m\|\mathbf v\|_2)$ and $0$; bounding $\|\mathbf v\|_2$ using $\sum_sb_{s,i}^2\leqslant B_i^2$ yields $\lambda_{\min}\bigl(\tfrac12(H_i+H_i^\top)/c_0\bigr)\geqslant\chi(m,\eta)B_i$, and multiplying by $c_0<0$ with $B_i\leqslant mC$ gives $\widetilde\lambda$. For the norm, every entry of $H_i$ is at most $2M_i\kappa(1-\eta)B_i^{\eta-2}$ in magnitude, and $B_i\geqslant m\varepsilon$ gives $\widetilde L$. The complete proof is in \Cref{app:power-SB}.
\end{psketch}

The SB property is therefore restored by imposing a sufficiently small participation floor, without changing the equilibrium characterization. Its only cost is a deterioration in the convergence rate as the resulting smoothness bound $\widetilde L$ grows like $\varepsilon^{\eta-2}$ as the floor shrinks, so in practice one takes $\varepsilon$ close to, but below, $\bar\varepsilon$.
We now bring together the above results to establish the main convergence and welfare result for the Power RCSF in the next section.
 
\section{BRD convergence and welfare-optimality: Best of both worlds}
\label{sec:main}

We now combine the pieces. The Power RCSF has proportional private and social marginal returns with the constant $c=m\eta/(m+\eta-1)$ (\Cref{prop:power-align}), and within the SRCSF class it is the only CSF with this property (\Cref{lem:isoelastic}). The budget-balanced charge at rate $\tau=1/c-1$ makes the welfare optimum an equilibrium without requiring budgets to be exhausted (\Cref{prop:tax-alignment}), which removes the obstacle of \Cref{prop:budget_bind_cond}. The participation floor restores the SB property (\Cref{lem:power-SB}), and below the threshold $\bar\varepsilon$ it is inactive at the equilibrium (\Cref{lem:explicit-floor}), which is established to be unique and symmetric (\Cref{lem:power-unique-sym}).  The theorem below combines these results to provide the three guarantees for the Power RCSF.

\begin{theorem}[Welfare-optimal NE under Power RCSF]
\label{thm:welfare-max}
Let $h$ be the Power RCSF (\Cref{def:power_rcsf}) played on the floored set $\mathcal{B}^\varepsilon$ (\Cref{def:floor}) with a budget-balanced charge (\Cref{eq:bb}), and suppose
\begin{itemize}
\item[\textbf{(F)}] \emph{Participation floor.} $0<\varepsilon<\bar\varepsilon$, with $\bar\varepsilon$ as in \cref{eq:epsbar}; and
\item[\textbf{(T)}] \emph{Charge.} $\displaystyle \tau=\frac1c-1=\frac{(m-1)(1-\eta)}{m\eta}$, with $c$ as in \cref{eq:c-value}.
\end{itemize}
Then the following hold:
\begin{enumerate}
\item[\textup{(a)}] \emph{Convergence.} The best-response dynamics (\Cref{alg:BRD}, run with $\mathcal D^\ell=\mathcal B^\varepsilon$) with step size $\gamma=\widetilde\lambda/\widetilde L^2$ converges to a Nash equilibrium $\mathbf b^\star$ of the floored game, reaching $\|\mathbf g(\mathbf b(K_\epsilon),\boldsymbol\mu(K_\epsilon))\|_2<\epsilon$ after $K_\epsilon=O\bigl((\widetilde L/\widetilde\lambda)^2\log(1/\epsilon)\bigr)$ iterations, with $\widetilde\lambda,\widetilde L$ as in \Cref{lem:power-SB}. 
\item[\textup{(b)}] \emph{Uniqueness and symmetry.}
The floored game has a unique Nash equilibrium $\mathbf b^\star$, which is symmetric and coincides with the unique Nash equilibrium of the unrestricted game. Moreover, $\widetilde u_s(\mathbf b^\star)>0$ for every
$s$, so $\mathbf b^\star\in\mathcal B_{\geqslant0}$.
\item[\textup{(c)}] \emph{Welfare optimality.} For every $p\in(-\infty,1]$, $\mathbf b^\star$ maximizes the $p$-mean welfare $W_p$ over $\mathcal{B}_{\geqslant0}$, and hence also over $\mathcal{B}_{\geqslant0}\cap\mathcal B^\varepsilon$.
\end{enumerate}
\end{theorem}

\begin{proof}

\textbf{(a)} 
By ~\cref{lem:power-SB}, the induced utilities satisfy the SB property on $\mathcal B^{\varepsilon}$ with the constants $\widetilde\lambda$ and $\widetilde L$, so~\cref{thm:BRD-conv} applies with $\mathcal D^\ell=\mathcal B^{\varepsilon}$. $\mathcal B^\varepsilon$ is convex and compact and is non-empty, because $n\varepsilon<C$ by \Cref{lem:explicit-floor}. Every iterate of \Cref{alg:BRD} lies in $\mathcal B^\varepsilon$ by construction of the correction $\boldsymbol \mu_s$ in \cref{eq:mu}, and the iterates converge to a Nash equilibrium $\mathbf b^\star$ of the floored game at the stated rate $\gamma =\widetilde\lambda/\widetilde L^2$; the iteration count follows from the bound in \Cref{thm:BRD-conv}, since $2(\widetilde L/\widetilde\lambda)^2\log(\|\mathbf g(\mathbf b(0),\boldsymbol\mu(0))\|_2/\epsilon)=O\bigl((\widetilde L/\widetilde\lambda)^2\log(1/\epsilon)\bigr)$

\medskip
\noindent\textbf{(b)} \Cref{lem:power-unique-sym} gives uniqueness and symmetry on the unrestricted set $\mathcal{B}$. 
The same argument applies verbatim on $\mathcal{B}^\varepsilon$ with the lower boundary $0$ replaced by $\varepsilon$: in \emph{Part I}, the marginal dominance inequality in \cref{eq:MD} does not involve the boundary value, the multiplier $\mu_{s,i}$ now enforces $b_{s,i}\geqslant\varepsilon$ with slackness $\mu_{s,i}(b_{s,i}-\varepsilon)=0$, and $b^\star_{p,i_0} > b^\star_{q,i_0} \geqslant \varepsilon$ gives $b^\star_{p,i_0} > \varepsilon$, so $\mu^\star_{p,i_0} = 0$ and the case analysis is unchanged; and in \emph{Part II}, $\beta_i(\Gamma)$ is replaced by $\max\{(M_iK/\Gamma)^{\theta},\varepsilon\}$, so that the total budget $S_\varepsilon(\Gamma)=\sum_i\max\{\beta_i(\Gamma),\varepsilon\}$ is continuous and
non-increasing in $\Gamma$, and is strictly decreasing whenever at least one coordinate lies above the floor. 
Since $n\varepsilon<C$ by \Cref{lem:explicit-floor}, if $S_\varepsilon(1+\tau)>C$, then $\lim_{\Gamma\to\infty}S_\varepsilon(\Gamma)=n\varepsilon<C$, hence the level $C$ is crossed at a finite $\Gamma^\star>1+\tau$ in the strictly decreasing region, which gives a unique $\Gamma^\star$; if
$S_\varepsilon(1+\tau)\leqslant C$, then $\Gamma^\star=1+\tau$. Thus in either case the floored game has a unique symmetric equilibrium.

By assumption \textbf{(F)} and \Cref{lem:explicit-floor}, the unrestricted unique and symmetric solution $\mathbf b^{\star}_s$ already lies strictly above $\varepsilon$  at every vertex, i.e. $b_{s,i}^\star = t_i^\star >\varepsilon$, and hence $\mathbf b^\star\in\mathcal B^\varepsilon$ and no floor constraint is active at it. Each $\mathbf b^{\star}_s$ is a best response to $\mathbf b^{\star}_{-s}$ over $\mathcal B_s$, and since it lies in the smaller set $\mathcal B^\varepsilon_s\subseteq\mathcal B_s$, it is also a best response over $\mathcal B^\varepsilon_s$. So $\mathbf b^\star$ is also a NE of the floored game. By the uniqueness established above, it is the unique NE of the floored game. Hence the floor does not affect the equilibrium.

For positivity of the taxed utility \eqref{eq:pigouvian_utility}, note that at the symmetric profile the rebate \eqref{eq:rebate} returns to each firm exactly what it paid, $\rho_s=\tau\sum_it^\star_i$, so
$\widetilde u_s(\mathbf b^\star)=\kappa m^{\eta-1}\sum_iM_i(t^\star_i)^{\eta}-\sum_it^\star_i$.
Stationarity gives $M_iK(t^\star_i)^{\eta-1}=\Gamma^\star$, i.e.\ $M_i(t^\star_i)^{\eta}=\Gamma^\star t^\star_i/K$, and substituting with $K=\kappa m^{\eta-2}(m+\eta-1)$ gives
\begin{equation}
\label{eq:ustar}
\widetilde u_s(\mathbf b^\star) = \widetilde u(\mathbf b^\star) = \Bigl(\sum_i t^\star_i\Bigr)\Bigl[\frac{m\,\Gamma^\star}{m+\eta-1}-1\Bigr], \quad \forall s \in \mathscr M
\end{equation}
By \Cref{lem:power-unique-sym}, $\Gamma^\star\geqslant1+\tau=1/c=(m+\eta-1)/(m\eta)$, and since $\eta<1$ we have $(m+\eta-1)/(m\eta)>(m+\eta-1)/m$, so the bracket in \eqref{eq:ustar} is strictly positive. Hence $\widetilde u(\mathbf b^\star)>0$ for every $s$ and by \Cref{def:p-mean}, we can conclude that $\mathbf{b}^\star$ lies in $\mathcal B_{\geqslant0}$.

\medskip
\noindent\textbf{(c)} \emph{Step 1: the utilitarian optimum without the floor.} For the taxed utilities $\widetilde u_s$, the charge and rebate cancel in the sum of the utilties, so the utilitarian welfare ($W_p$ for $p=1$) is 
\[
W_1(\mathbf{b}) = \frac{1}{m}\sum_s \widetilde u_s(\mathbf{b}) = \frac1m\Bigl[\kappa\sum_iM_iB_i^{\eta} - \sum_iB_i\Bigr],
\]
which depends on $\mathbf{b}$ only through the total budget $B_i=\sum_sb_{s,i}$. Over $\mathcal{B}$ the attainable totals are exactly the vectors $\{\mathbf B: B_i\geqslant 0,\ \sum_iB_i\leqslant mC\}$: any feasible $\mathbf b$ produces such totals and any such totals are realized by the symmetric split $b_{s,i}=B_i/m$, which is feasible for each firm. This objective is separable and strictly concave in $\mathbf B$, and we call its maximizer $\mathbf B^{\mathrm{opt}}$: let $\mathbf b^{\mathrm{opt}}$ be its symmetric split, with $t^{\mathrm{opt}}_i=B^{\mathrm{opt}}_i/m$. Then $W_1(\mathbf b)\leqslant W_1(\mathbf b^{\mathrm{opt}})$ for every $\mathbf b\in\mathcal{B}$, since $W_1$ doesn't depend on how the total budget $B_i^{\mathrm{opt}}$ is split among firms.

\emph{Step 2: $\mathbf b^{\mathrm{opt}}$ is the equilibrium.} By \Cref{prop:power-align} the Power RCSF has proportional marginals with constant $c$, and each $u_s$, hence each $\widetilde u_s$, is concave in $\mathbf b_s$. 
Every total $B^{\mathrm{opt}}_i$ is strictly positive, because in the planner's first-order condition $M_i\psi(t_i) - 1 = \hat\lambda - \hat\mu_{i}$, the marginal welfare 
$M_i\psi(t_i) = M_i\kappa\eta B_i^{\eta-1}$ grows without bound as $B_i\to0$ and $M_i\geqslant1-\alpha>0$ by \cref{lem:sumM}; so the planner's first-order conditions are well defined at $\mathbf b^{\mathrm{opt}}$.
By assumption \textbf{(T)} the charge satisfies $1+\tau=1/c$. \Cref{prop:tax-alignment} therefore applies and $\mathbf b^{\mathrm{opt}}$ is a Nash equilibrium of the game without the floor. By part (b) the Nash equilibrium is unique, so $\mathbf b^\star=\mathbf b^{\mathrm{opt}}$, maximizes $W_1$ over $\mathcal B$, and also over $\mathcal B^\varepsilon\subseteq\mathcal B$, to which it belongs.

\emph{Step 3: extension to all $p\in(-\infty,1]$.} For $\mathbf b\in\mathcal B_{\geqslant0}$, $W_p(\mathbf b)$ is the power mean of order $p$ of the non-negative numbers $\widetilde u_1(\mathbf b),\dots,\widetilde u_m(\mathbf b)$. By the power mean inequality, a power mean of order $p\leqslant 1$ never exceeds the arithmetic mean, with the limiting conventions of \Cref{def:p-mean} for $p\leqslant0$, so $W_p(\mathbf b)\leqslant W_1(\mathbf b)$.
At the equilibrium this inequality is an equality. Since $\mathbf{b}^\star$ is symmetric, every firm holds an identical allocation and receives a rebate equal to its own payment ($\rho_s=\tau\sum_it^\star_i$), so all utilities coincide at the common value $\tilde u(\mathbf{b}^\star)>0$ of \cref{eq:ustar} in part (b); and any power mean of $m$ equal numbers equals that number, so $W_p(\mathbf b^\star)=W_1(\mathbf b^\star)$. Steps~1 and~2 give $W_1(\mathbf b)\leqslant W_1(\mathbf b^{\mathrm{opt}})=W_1(\mathbf b^\star)$ for every $\mathbf b\in\mathcal B$. Combining all three inequalities together, we obtain
\[
W_p(\mathbf{b})\;\leqslant\; W_1(\mathbf{b})\;\leqslant\; W_1(\mathbf{b}^\star)\;=\;W_p(\mathbf{b}^\star)
\qquad \forall \; \mathbf b\in\mathcal{B}_{\geqslant0}, \; p\in(-\infty,1].
\]
Since $\mathbf b^\star\in\mathcal B_{\geqslant0}\cap\mathcal B^\varepsilon$ by part (b), it maximizes $W_p$ over $\mathcal B_{\geqslant0}$ and over $\mathcal{B}_{\geqslant0}\cap\mathcal B^\varepsilon$.
By \Cref{def:p-mean}, the planner's problem is restricted to $\mathcal{B}_{\geqslant0}$, so profiles outside this set need not be considered.
\end{proof}

\begin{corollary}
\label{cor:poa}
Under the conditions of \Cref{thm:welfare-max}, $\mathrm{PoA} = 1$ for every $p \in (-\infty, 1]$ simultaneously.
\end{corollary}

\begin{remark}[Unequal budgets]\label{rem:unequal-budgets}
Suppose firm $s$ has its own budget $C_s$, and replace $mC$ by $\sum_sC_s$ in the cap on $\kappa$ and in the constant $\tilde\lambda$ of Lemma~\ref{lem:power-SB}, and $C$ by $\min_sC_s$ in the requirement $n\varepsilon<C$ that keeps the floored set non-empty. Theorem~\ref{thm:T-bound} does not involve budgets, Theorem~\ref{thm:BRD-conv} only needs $\mathcal D^\ell$ to be convex and compact, and Lemma~\ref{lem:power-SB} holds with this replacement; hence best-response dynamics still converge to a Nash equilibrium of the floored game. What fails is symmetry: the case analysis in the proofs of \Cref{lem:rcsf-unique-symmetric,lem:power-unique-sym} uses equal budgets, so the equilibrium is in general asymmetric. Characterizing equilibrium and welfare in this setting is left as future work.
\end{remark}

\section{Conclusion}
\label{sec:conclusion}

We developed a two-timescale framework to study how multiple competing firms allocate advertising budgets over social networks to maximize brand awareness, with convergence-rate guarantees on the fast timescale that formally justify the separation of dynamics. Under standard conditions, firms’ best-response dynamics (BRD) converge to a pure-strategy Nash equilibrium (NE), which need not coincide with the social optimum. We therefore study whether a platform can design a contest success function (CSF), which determines how attention is allocated among firms, so that competition leads to a socially optimal outcome.

For standard ratio-form CSFs, BRD converges to a unique symmetric equilibrium, but equilibrium behaviour cannot in general be aligned with social welfare. We show that this failure is structural: firms’ private marginal returns to spending are strictly larger than the corresponding social marginal returns. We then show that this gap can be accounted for through a broader class of scaled ratio CSFs. For NE and welfare to align, it suffices that private and social marginal returns are proportional, and within the class of scaled ratio CSFs, this condition singles out the Power RCSF.
The Power RCSF alone is not sufficient: a budget-balanced charge removes the need for firms to exhaust their budgets, while a small participation floor restores the regularity needed for convergence of BRD. Together, these ingredients yield dynamics that converge to a unique, symmetric equilibrium maximizing every $p$-mean welfare with $p\leqslant 1$, so the Price of Anarchy is one. Our results show that the allocation rule itself can be a useful design instrument: rather than eliminating competition, a platform can shape how competition translates into socially desirable outcomes.

A natural direction for future work is to allow firms to have heterogeneous budgets. The BRD and its convergence guarantee continue to apply, but the equilibrium need not be symmetric in general. Consequently, our welfare characterization does not extend directly to heterogeneous budgets. Characterizing equilibrium and welfare in this setting is an interesting direction for future work.

\bibliographystyle{plainnat}
\bibliography{refs}

\appendix
\section{Appendix to \texorpdfstring{\Cref{sec:awareness}}{Section 3}}

\subsection{Proof of Theorem~\ref{thm:T-bound}: Convergence of the Fast-Timescale Dynamics}
\label{app:T-bound}

Throughout the proof we fix the slow-timescale period $k$ and the firm $s$, and we suppress both from the notation where no confusion arises. In particular, we write $\mathbf h := \mathbf h_s(\mathbf b(k))$ for the vector of CSF values, which stays constant during the period, $\mathbf x(t)$ for $\mathbf x_s(t)$, and $\mathbf x^*$ for the steady state \eqref{eq:steady-state}. In vector form, the dynamics \eqref{eq:awareness_update_eqn} read
\begin{equation}
\label{eq:app-fast-vector}
\mathbf x(t) = \frac1t\Bigl[\alpha\mathscr E\,\mathbf x(t-1) + (1-\alpha)\,\mathbf h\Bigr] + \Bigl(1-\frac1t\Bigr)\mathbf x(t-1), \qquad t\geqslant1 .
\end{equation}
We work with the $\ell_\infty$ norm $\|\mathbf v\|_\infty=\max_i|v_i|$ on $\mathbb R^n$. Its induced operator norm on matrices is the maximum absolute row sum, $\|A\|_\infty=\max_i\sum_j|a_{ij}|$, and it is sub-multiplicative: $\|A\mathbf v\|_\infty\leqslant\|A\|_\infty\|\mathbf v\|_\infty$ for every matrix $A$ and vector $\mathbf v$.

\emph{Step 1: the steady state lies in $[0,1]^n$.}
We first record a property of $\mathbf x^*$ that is used in part (b). Since $\mathscr E$ is row-stochastic, the spectral radius of $\alpha\mathscr E$ is at most $\alpha\|\mathscr E\|_\infty=\alpha<1$, so the Neumann series $\sum_{\ell\geqslant0}(\alpha\mathscr E)^\ell$ converges to $(I-\alpha\mathscr E)^{-1}$. Every term of this series has non-negative entries, so the matrix $M:=(1-\alpha)(I-\alpha\mathscr E)^{-1}$ is non-negative. Its row sums equal one. Indeed, $\mathscr E\mathbf 1=\mathbf 1$ gives $(\alpha\mathscr E)^\ell\mathbf 1=\alpha^\ell\mathbf 1$ for every $\ell$, and summing the geometric series yields $M\mathbf 1=(1-\alpha)\sum_{\ell\geqslant0}\alpha^\ell\,\mathbf 1=\mathbf 1$. Hence every coordinate of $\mathbf x^*=M\mathbf h$ is a weighted average of the coordinates of $\mathbf h$, and since $\mathbf h\in[0,1]^n$, we conclude that $\mathbf x^*\in[0,1]^n$.

\emph{Step 2: the error obeys a linear recursion.}
Define the error $\mathbf e(t):=\mathbf x(t)-\mathbf x^*$. By construction, $\mathbf x^*$ satisfies the fixed-point equation
\begin{equation}
\label{eq:app-fixed-point}
\mathbf x^* = \alpha\mathscr E\,\mathbf x^* + (1-\alpha)\,\mathbf h .
\end{equation}
We subtract $\mathbf x^*$ from both sides of \eqref{eq:app-fast-vector}. On the right-hand side, we write $\mathbf x^*=\frac1t\mathbf x^*+(1-\frac1t)\mathbf x^*$ and replace the first copy by the right-hand side of \eqref{eq:app-fixed-point}. Grouping the terms that multiply $1/t$ and $1-1/t$, the constant term $(1-\alpha)\mathbf h$ cancels, and we obtain
\[
\mathbf e(t)=\frac1t\Bigl[\alpha\mathscr E\bigl(\mathbf x(t-1)-\mathbf x^*\bigr)\Bigr]+\Bigl(1-\frac1t\Bigr)\bigl(\mathbf x(t-1)-\mathbf x^*\bigr).
\]
In other words, the error evolves by the linear recursion
\begin{equation}
\label{eq:app-error-recursion}
\mathbf e(t)=A(t)\,\mathbf e(t-1),\qquad\text{where}\quad A(t):=\Bigl(1-\frac1t\Bigr)I+\frac{\alpha}{t}\,\mathscr E .
\end{equation}

\emph{Step 3: the norm of each step.}
For $t\geqslant1$ the coefficients $1-1/t$ and $\alpha/t$ are non-negative, and $\mathscr E$ has non-negative entries, so every entry of $A(t)$ is non-negative. Its row sums can therefore be computed without absolute values. Using the row-stochasticity of $\mathscr E$, that is, $\sum_j e_{ij}=1$ for every $i$, the $i$th row sum is
\[
\sum_j\Bigl[\Bigl(1-\frac1t\Bigr)\delta_{ij}+\frac{\alpha}{t}\,e_{ij}\Bigr]=1-\frac1t+\frac{\alpha}{t}=1-\frac{1-\alpha}{t},
\]
where $\delta_{ij}$ is the Kronecker delta. All rows have the same sum, so
\begin{equation}
\label{eq:app-norm-A}
\|A(t)\|_\infty=1-\frac{1-\alpha}{t}\qquad\text{and, in fact,}\qquad A(t)\,\mathbf 1=\Bigl(1-\frac{1-\alpha}{t}\Bigr)\mathbf 1 .
\end{equation}
Since $\alpha\in(0,1)$, this number lies in $(0,1)$ for every $t\geqslant1$, so each step strictly contracts the error.

\emph{Step 4: proof of part (a).}
Applying the recursion \eqref{eq:app-error-recursion} successively for $t=T,T-1,\dots,1$ gives $\mathbf e(T)=A(T)A(T-1)\cdots A(1)\,\mathbf e(0)$. By sub-multiplicativity of the operator norm and \eqref{eq:app-norm-A},
\begin{equation}
\label{eq:app-PT}
\|\mathbf e(T)\|_\infty\leqslant\prod_{t=1}^T\|A(t)\|_\infty\,\|\mathbf e(0)\|_\infty=P_T\,\|\mathbf e(0)\|_\infty,\qquad P_T:=\prod_{t=1}^T\Bigl(1-\frac{1-\alpha}{t}\Bigr).
\end{equation}
It remains to bound $P_T$. Every factor of $P_T$ is positive, so we may take logarithms. By the elementary inequality $\log(1-x)\leqslant-x$, valid for every $x<1$, applied with $x=(1-\alpha)/t$, we get
\[
\log P_T=\sum_{t=1}^T\log\Bigl(1-\frac{1-\alpha}{t}\Bigr)\leqslant-(1-\alpha)\sum_{t=1}^T\frac1t .
\]
We now bound the harmonic sum from below by an integral. On each interval $[t,t+1]$ we have $1/t\geqslant1/x$, so $\sum_{t=1}^T\frac1t\geqslant\int_1^{T+1}\frac{dx}{x}=\log(T+1)$. Substituting this, and using $1-\alpha>0$, we obtain $\log P_T\leqslant-(1-\alpha)\log(T+1)$, that is,
\begin{equation}
\label{eq:app-PT-bound}
P_T\leqslant(T+1)^{-(1-\alpha)} .
\end{equation}
Together with \eqref{eq:app-PT}, this proves the inequality in part (a). Since $(T+1)^{-(1-\alpha)}\to0$ as $T\to\infty$, it also shows that $\mathbf x(T)\to\mathbf x^*$.

\emph{Step 5: proof of part (b).}
By hypothesis $\mathbf x(0)\in[0,1]^n$, and by Step~1 also $\mathbf x^*\in[0,1]^n$. The difference of two vectors in $[0,1]^n$ has every coordinate in $[-1,1]$, so $\|\mathbf e(0)\|_\infty\leqslant1$. Combining this with part (a), we have $\|\mathbf e(T)\|_\infty\leqslant(T+1)^{-(1-\alpha)}$. The right-hand side is at most $\epsilon_0$ exactly when $(T+1)^{1-\alpha}\geqslant1/\epsilon_0$, that is, when $T+1\geqslant\epsilon_0^{-1/(1-\alpha)}$. This holds in particular whenever $T\geqslant\epsilon_0^{-1/(1-\alpha)}$, which proves part (b).

\emph{Step 6: proof of part (c).}
We first show that the bound \eqref{eq:app-PT} can hold with equality. Suppose $\mathbf e(0)=c\,\mathbf 1$ for some constant $c$. By the second identity in \eqref{eq:app-norm-A}, the vector $\mathbf 1$ is an eigenvector of every $A(t)$, with eigenvalue $1-(1-\alpha)/t$. Applying the recursion \eqref{eq:app-error-recursion} $T$ times therefore gives $\mathbf e(T)=P_T\,c\,\mathbf 1$ exactly, so $\|\mathbf e(T)\|_\infty=P_T\|\mathbf e(0)\|_\infty$. Hence no factor smaller than $P_T$ can serve in part (a) for all initial conditions.

We next evaluate $P_T$ exactly, to identify its rate of decay. Writing each factor as $(t-(1-\alpha))/t$ and setting $c_0:=1-\alpha\in(0,1)$,
\[
P_T=\frac{\prod_{t=1}^T(t-c_0)}{\prod_{t=1}^T t}=\frac{\prod_{t=1}^T(t-c_0)}{T!} .
\]
The numerator is evaluated by the following identity.

\begin{lemma}[Gamma product identity]
\label{lem:gamma-product}
For every $c_0\in(0,1)$ and every positive integer $T$, $\displaystyle\prod_{t=1}^T(t-c_0)=\frac{\Gamma(T+1-c_0)}{\Gamma(1-c_0)}$.
\end{lemma}
\begin{proof}
We use the functional equation of the Gamma function, $\Gamma(z+1)=z\,\Gamma(z)$ for every $z>0$. Applying it first with $z=T-c_0$, then with $z=T-1-c_0$, and so on down to $z=1-c_0$, we obtain
\[
\Gamma(T+1-c_0)=(T-c_0)\,\Gamma(T-c_0)=(T-c_0)(T-1-c_0)\,\Gamma(T-1-c_0)=\cdots=\Gamma(1-c_0)\prod_{t=1}^T(t-c_0).
\]
Every argument used is positive because $c_0<1$. Dividing both sides by $\Gamma(1-c_0)>0$ gives the identity.
\end{proof}

Applying \Cref{lem:gamma-product}, and using $T!=\Gamma(T+1)$ and $1-c_0=\alpha$, we obtain
\[
P_T=\frac{\Gamma(T+\alpha)}{\Gamma(\alpha)\,\Gamma(T+1)} .
\]
Finally, the standard asymptotic ratio of Gamma functions, $\Gamma(T+a)/\Gamma(T+b)\sim T^{a-b}$ as $T\to\infty$ \citep[Eq.~6.1.46]{abramowitz1948handbook}, applied with $a=\alpha$ and $b=1$, gives
\[
P_T\sim\frac{T^{-(1-\alpha)}}{\Gamma(\alpha)}\qquad\text{as }T\to\infty .
\]
So the error factor decays exactly at the polynomial rate $T^{-(1-\alpha)}$, and the upper bound \eqref{eq:app-PT-bound} used in parts (a) and (b) is sharp up to the constant factor $1/\Gamma(\alpha)$, which lies in $(0,1)$ for $\alpha\in(0,1)$. \hfill$\square$

\section{Appendix to \texorpdfstring{\Cref{sec:brd}}{Section 4}}

\subsection{Proof of \Cref{lem:mu-unique}.}
\label{app:mu-unique}
For $\mathbf b\in\mathcal D^\ell$ and a correction $\boldsymbol\mu\in\mathbb R^n$, from \cref{alg:BRD} line \ref{line:brd_update_rule}, we denote by $\mathbf b_s^{\boldsymbol\mu}:=\mathbf b_s+\gamma\bigl(\nabla_{\mathbf b_s}u_s(\mathbf b)+\boldsymbol\mu\bigr)$, the updated budget that $\boldsymbol\mu$ produces for firm $s \in \mathscr M$. In particular, $\mathbf b_s^{\mathbf 0}=\mathbf b_s+\gamma\nabla_{\mathbf b_s}u_s(\mathbf b)$ is the \emph{uncorrected budget update}, and $\mathbf b_s^{\boldsymbol\mu_s(k)}=\mathbf b_s(k+1)$ when $\mathbf b=\mathbf b(k)$. By definition, the minimisation in \eqref{eq:mu} is the minimisation of $\|\boldsymbol\mu\|_2$ over the set of admissible corrections
\[
\mathcal K_s:=\bigl\{\boldsymbol\mu\in\mathbb R^n:\ \mathbf b_s^{\boldsymbol\mu}\in\mathcal D_s^\ell\bigr\},
\]
that is, over the corrections whose updated budget is feasible. 

\emph{Step 1: the correction in \eqref{eq:mu} is a projection.}
Since $\mathbf b_s^{\boldsymbol\mu}=\mathbf b_s^{\mathbf 0}+\gamma\boldsymbol\mu$, consider the map $T:\mathcal K_s\to\mathcal D_s^\ell,\qquad T(\boldsymbol\mu):=\mathbf b_s^{\mathbf 0}+\gamma\boldsymbol\mu=\mathbf b_s^{\boldsymbol\mu}$,
which maps each admissible correction to the budget it produces which lies in $\mathcal D_s^\ell$ by the definition of $\mathcal K_s$. 
The map $T$ is a bijection, with inverse $T^{-1}(\mathbf b_s')=\boldsymbol{\mu}=\frac{\mathbf b_s'-\mathbf b_s^{\mathbf 0}}{\gamma}=\frac{\mathbf b_s'-\mathbf b_s}{\gamma}-\nabla_{\mathbf b_s}u_s(\mathbf b),$ for all $\mathbf b_s'\in\mathcal D_s^\ell$.
Thus the minimum-norm correction problem in \eqref{eq:mu}, $\min_{\boldsymbol\mu\in\mathcal K_s}\|\boldsymbol\mu\|_2$ is equivalent to $\min_{\mathbf b_s'\in\mathcal D_s^\ell}\bigl\|\mathbf b_s'-\mathbf b_s^{\mathbf 0}\bigr\|_2,$
i.e. the Euclidean projection of $\mathbf b_s^{\mathbf 0}$ onto the convex set $\mathcal D_s^\ell$: $\Pi_{\mathcal D_s^\ell}\bigl(\mathbf b_s^{\mathbf 0}\bigr):=\arg\min_{\mathbf b_s'\in\mathcal D_s^\ell}\bigl\|\mathbf b_s'-\mathbf b_s^{\mathbf 0}\bigr\|_2$.

\emph{Step 2: existence and uniqueness.}
The set $\mathcal D_s^\ell$ is non-empty, closed and convex. By the projection theorem \citep[Prop.~2.1.3]{bertsekas1999nonlinear}, the projection of any point onto a non-empty, closed and convex set exists and is unique. Hence $\Pi_{\mathcal D_s^\ell}(\mathbf b_s^{\mathbf 0})$ exists and is unique. Since $T$ is a bijection, exactly one admissible correction produces this budget, namely $\boldsymbol\mu_s(\mathbf b)=T^{-1}\bigl(\Pi_{\mathcal D_s^\ell}(\mathbf b_s^{\mathbf 0})\bigr)$ which by Step~1, is the unique solution of \eqref{eq:mu}.

\emph{Step 3: a characterisation of the minimiser.}
The projection theorem also characterises the projection: 
\[
\mathbf b_s^{\boldsymbol\mu}=\Pi_{\mathcal D_s^\ell}\bigl(\mathbf b_s^{\mathbf 0}\bigr)
\quad\Longleftrightarrow\quad
\bigl(\mathbf b_s^{\mathbf 0}-\mathbf b_s^{\boldsymbol\mu}\bigr)^\top\bigl(\mathbf b_s'-\mathbf b_s^{\boldsymbol\mu}\bigr)\leqslant0\qquad \; \forall \; \mathbf b_s'\in\mathcal D_s^\ell.
\]
From the update $\mathbf b_s^{\boldsymbol\mu}=\mathbf b_s^{\mathbf 0}+\gamma\boldsymbol\mu$ we have $\mathbf b_s^{\mathbf 0}-\mathbf b_s^{\boldsymbol\mu}=-\gamma\boldsymbol\mu$. Substituting this into the inequality and dividing by $-\gamma<0$, we conclude, using Step~2, that an admissible correction $\boldsymbol\mu$ is the minimiser $\boldsymbol\mu_s(\mathbf b)$ if and only if
\begin{equation}
\label{eq:mu-VI}
\boldsymbol\mu^\top\bigl(\mathbf b_s'-\mathbf b_s^{\boldsymbol\mu}\bigr)\geqslant0\qquad\; \forall \;\mathbf b_s'\in\mathcal D_s^\ell .
\end{equation}
In words, the minimal correction, which points from the uncorrected step to its nearest feasible budget, makes a non-obtuse angle with every direction pointing from that budget into the feasible set; otherwise a small move towards some feasible budget would bring the updated budget closer to the uncorrected step, and the correction would not be minimal. \hfill$\square$

\subsection{Proof of \cref{lem:grad-NE}.}
\label{app:grad-NE}

($\Leftarrow$) \emph{Every Nash equilibrium has zero corrected gradient.} Let $\mathbf b^\star$ be a Nash equilibrium of the game on $\mathcal D^\ell$, fix a firm $s$ and any $\mathbf b_s'\in\mathcal D_s^\ell$. For $t\in(0,1]$, the budget $\mathbf b^\star_s+t(\mathbf b_s'-\mathbf b^\star_s)$ lies in $\mathcal D_s^\ell$ by convexity. Since $\mathbf b^\star_s$ is a best response to $\mathbf b^\star_{-s}$, this budget cannot increase firm $s$'s utility:
\[
u_s\bigl(\mathbf b^\star_s+t(\mathbf b_s'-\mathbf b^\star_s),\mathbf b^\star_{-s}\bigr)\leqslant u_s(\mathbf b^\star) 
\]
Dividing by $t$,
\[
\frac{u_s\bigl(\mathbf b^\star_s+t(\mathbf b_s'-\mathbf b^\star_s),\mathbf b^\star_{-s}\bigr)-u_s(\mathbf b^\star)}{t}\leqslant0\qquad\text{for every }t\in(0,1].
\]
As $t\downarrow0$, the left-hand side converges to the directional derivative of $u_s(\cdot,\mathbf b^\star_{-s})$ at $\mathbf b^\star_s$ in the direction $\mathbf b_s'-\mathbf b^\star_s$, which equals $\nabla_{\mathbf b_s}u_s(\mathbf b^\star)^\top(\mathbf b_s'-\mathbf b^\star_s)$ because $u_s$ is continuously differentiable. This gives the first-order condition
\begin{equation}\label{eq:foc-NE}
\nabla_{\mathbf b_s}u_s(\mathbf b^\star)^\top\bigl(\mathbf b_s'-\mathbf b^\star_s\bigr)\leqslant0\qquad\text{for all }\mathbf b_s'\in\mathcal D_s^\ell .
\end{equation}
Now consider the candidate correction $\hat{\boldsymbol\mu}:=-\nabla_{\mathbf b_s}u_s(\mathbf b^\star)$. Its updated budget is $\mathbf b_s^{\hat{\boldsymbol\mu}}=\mathbf b^\star_s+\gamma\bigl(\nabla_{\mathbf b_s}u_s(\mathbf b^\star)-\nabla_{\mathbf b_s}u_s(\mathbf b^\star)\bigr)=\mathbf b^\star_s\in\mathcal D_s^\ell$, so $\hat{\boldsymbol\mu}$ is admissible. Moreover, for every $\mathbf b_s'\in\mathcal D_s^\ell$,
\[
\hat{\boldsymbol\mu}^\top\bigl(\mathbf b_s'-\mathbf b_s^{\hat{\boldsymbol\mu}}\bigr)=-\nabla_{\mathbf b_s}u_s(\mathbf b^\star)^\top\bigl(\mathbf b_s'-\mathbf b^\star_s\bigr)\geqslant0
\]
by \eqref{eq:foc-NE}; that is, $\hat{\boldsymbol\mu}$ satisfies \cref{app:mu-unique}'s condition \eqref{eq:mu-VI} (Proof of \cref{lem:mu-unique}). By \cref{lem:mu-unique}, the minimizer is unique and characterised by \eqref{eq:mu-VI}, so $\boldsymbol\mu_s(\mathbf b^\star)=\hat{\boldsymbol\mu}$. Hence
\[
\mathbf g_s\bigl(\mathbf b^\star,\boldsymbol\mu_s(\mathbf b^\star)\bigr)=\nabla_{\mathbf b_s}u_s(\mathbf b^\star)+\hat{\boldsymbol\mu}=\mathbf 0 .
\]

($\Rightarrow$) \emph{Zero corrected gradient implies a Nash equilibrium.} Suppose $\mathbf g_s(\mathbf b^\star,\boldsymbol\mu_s(\mathbf b^\star))=\mathbf 0$ for every $s\in\mathscr M$, and fix $s$. By the definition \eqref{eq:gradient}, $\nabla_{\mathbf b_s}u_s(\mathbf b^\star)+\boldsymbol\mu_s(\mathbf b^\star)=\mathbf 0$, i.e.\ $\boldsymbol\mu_s(\mathbf b^\star)=-\nabla_{\mathbf b_s}u_s(\mathbf b^\star)$, and the corresponding updated budget is $\mathbf b_s^{\boldsymbol\mu_s(\mathbf b^\star)}=\mathbf b^\star_s+\gamma\cdot\mathbf 0=\mathbf b^\star_s$. Since $\boldsymbol\mu_s(\mathbf b^\star)$ is the minimiser, it satisfies \eqref{eq:mu-VI} from the proof of \cref{lem:mu-unique} (\cref{app:mu-unique}):
\[
-\nabla_{\mathbf b_s}u_s(\mathbf b^\star)^\top\bigl(\mathbf b_s'-\mathbf b^\star_s\bigr)\geqslant0,\quad\text{i.e.}\quad\nabla_{\mathbf b_s}u_s(\mathbf b^\star)^\top\bigl(\mathbf b_s'-\mathbf b^\star_s\bigr)\leqslant0\qquad\text{for all }\mathbf b_s'\in\mathcal D_s^\ell .
\]
Since $u_s$ is concave and differentiable in $\mathbf b_s$, it lies below its tangent plane at $\mathbf b^\star_s$: for every $\mathbf b_s'\in\mathcal D_s^\ell$,
\[
u_s(\mathbf b_s',\mathbf b^\star_{-s})\leqslant u_s(\mathbf b^\star)+\nabla_{\mathbf b_s}u_s(\mathbf b^\star)^\top\bigl(\mathbf b_s'-\mathbf b^\star_s\bigr)\leqslant u_s(\mathbf b^\star).
\]
Hence $\mathbf b^\star_s$ is a best response to $\mathbf b^\star_{-s}$ over $\mathcal D_s^\ell$, i.e. $\mathbf b^\star$ is a Nash equilibrium. \hfill$\square$

\subsection{Proof of \texorpdfstring{\Cref{thm:BRD-conv}}{Theorem 2}}
\label{app:BRD-proof}

We stack the own-gradients as $\mathbf g(\mathbf b):=\bigl[\nabla_{\mathbf b_s}u_s(\mathbf b)\bigr]_{s\in\mathscr M}\in\mathbb R^{mn}$, so that $\mathbf g(\mathbf b,\boldsymbol\mu)=\mathbf g(\mathbf b)+\boldsymbol\mu$. By definition of the block Hessian, $[G_{s_1,s_2}(\mathbf b)]_{i,j}=\partial^2u_{s_1}/(\partial b_{s_2,j}\partial b_{s_1,i})$, which is the derivative of the $(s_1,i)$ entry of $\mathbf g(\mathbf b)$ with respect to $b_{s_2,j}$; that is, $G(\mathbf b)$ is the Jacobian matrix of $\mathbf g(\mathbf b)$.

\emph{Step 1: Each utility is concave in its own budget.} Strong concavity in Definition~\ref{def:SB} gives $\mathbf v^\top G(\mathbf b)\,\mathbf v\leqslant-\lambda\|\mathbf v\|_2^2$ for every $\mathbf v\in\mathbb R^{mn}$. Consider a $\mathbf v$ that is zero outside the block of firm $s$, with block $\mathbf v_s$. Then $\mathbf v^\top G(\mathbf b)\mathbf v=\mathbf v_s^\top G_{s,s}(\mathbf b)\mathbf v_s$, and $G_{s,s}(\mathbf b)$ is the Hessian of $u_s$ with respect to $\mathbf b_s$. Thus, $\mathbf v_s^\top G_{s,s}(\mathbf b)\mathbf v_s\leqslant-\lambda\|\mathbf v_s\|_2^2<0$ for all $\mathbf v_s\neq\mathbf 0$ and all $\mathbf b\in\mathcal D^\ell$. Along any segment in the convex set $\mathcal D_s^\ell$, the second derivative of $u_s(\cdot,\mathbf b_{-s})$ is therefore negative, so $u_s$ is concave in $\mathbf b_s$ on $\mathcal D_s^\ell$. Hence Lemma~\ref{lem:grad-NE} applies.

\emph{Step 2: All iterates are feasible.}
By assumption, $\mathbf b(0)\in\mathcal D^\ell$. Suppose $\mathbf b(k)\in\mathcal D^\ell$. By \cref{lem:mu-unique}, each correction $\boldsymbol\mu_s(k)$ exists, and it satisfies the constraint in \eqref{eq:mu}, which says precisely that $\mathbf b_s(k+1)=\mathbf b_s(k)+\gamma\,\mathbf g_s(k)\in\mathcal D_s^\ell$. Hence $\mathbf b(k+1)\in\mathcal D^\ell$, and by induction $\mathbf b(k)\in\mathcal D^\ell$ for all $k\geqslant0$. Since $\mathcal D^\ell$ is convex, the whole segment between two consecutive iterates also lies in $\mathcal D^\ell$.

\emph{Step 3: Geometric decay of the corrected gradient.}
The convergence of the budget update rule in line~\ref{line:brd_update_rule} of Algorithm~\ref{alg:BRD} is analyzed through the $L^2$ norm of the update direction. 
Keep the correction $\boldsymbol\mu(k)$ fixed and move from $\mathbf b(k)$ to $\mathbf b(k+1)$. Since $\boldsymbol\mu(k)$ does not depend on the point at which $\mathbf g(\mathbf b)$ is evaluated,
\begin{equation}
    \mathbf g(\mathbf b(k+1),\boldsymbol\mu(k))-\mathbf g(\mathbf b(k),\boldsymbol\mu(k))=\mathbf g(\mathbf b(k+1))-\mathbf g(\mathbf b(k)).
    \label{eq:expansion}
\end{equation}

Since $\mathbf g(\mathbf b)$ is a vector-valued map, we use its integral form. Let $\phi(\sigma):=\mathbf g(\mathbf b(k)+\sigma (\mathbf b(k+1)-\mathbf b(k)))$ for $\sigma\in[0,1]$. The points $\mathbf b(k)+\sigma(\mathbf b(k+1)-\mathbf b(k))$ lie in $\mathcal D^\ell$ (Step~2), and $\mathbf g(\mathbf b)$ is continuously differentiable there with Jacobian $G$, so $\phi'(\sigma)=G(\mathbf b(k)+\sigma(\mathbf b(k+1)-\mathbf b(k)))\,(\mathbf b(k+1)-\mathbf b(k))$. The fundamental theorem of calculus gives
\[
\mathbf g(\mathbf b(k+1))-\mathbf g(\mathbf b(k))=\int_0^1\phi'(\sigma)\,d\sigma=\bar G(k)\,(\mathbf b(k+1)-\mathbf b(k)),
\]
where
\begin{equation}\label{eq:Gbar}
\bar G(k):=\int_0^1G\bigl(\mathbf b(k)+\sigma(\mathbf b(k+1)-\mathbf b(k))\bigr)\,d\sigma
\end{equation}

is the average Hessian-like matrix along the path from $\mathbf b(k)$ to $\mathbf b(k+1)$. Since every point of the segment lies in $\mathcal D^\ell$, the averaged matrix $\bar G(k)$ inherits both SB bounds of \cref{def:SB}. For every $\mathbf v\in\mathbb R^{mn}$,
\begin{equation}\label{eq:Gbar-bounds}
\mathbf v^\top\bar G(k)\mathbf v=\int_0^1\mathbf v^\top G(\cdot)\,\mathbf v\,d\sigma\leqslant-\lambda\|\mathbf v\|_2^2,
\qquad
\|\bar G(k)\mathbf v\|_2\leqslant\int_0^1\|G(\cdot)\,\mathbf v\|_2\,d\sigma\leqslant L\|\mathbf v\|_2 .
\end{equation}
Thus, \eqref{eq:expansion} becomes

\begin{equation}
    \mathbf{g}(\mathbf{b}(k+1), \boldsymbol{\mu}(k)) = \mathbf{g}(\mathbf{b}(k), \boldsymbol{\mu}(k)) + \bar G(k) (\mathbf{b}(k+1) - \mathbf{b}(k))
\end{equation}

Substituting $\mathbf b(k + 1) - \mathbf b(k)$ from line~\ref{line:brd_update_rule} of Algorithm~\ref{alg:BRD}, we get:
\[\mathbf{g}(\mathbf{b}(k+1), \boldsymbol{\mu}(k)) = \mathbf{g}(\mathbf{b}(k), \boldsymbol{\mu}(k)) + \gamma \bar G(k) \mathbf{g}(\mathbf{b}(k), \boldsymbol{\mu}(k))\]
Taking the squared norm of both sides and expanding, we have
\begin{align}
    &\|\mathbf{g}(\mathbf{b}(k+1), \boldsymbol{\mu}(k))\|_2^2 \; = \; \|\mathbf{g}(\mathbf{b}(k), \boldsymbol{\mu}(k)) + \gamma \bar G(k) \mathbf{g}(\mathbf{b}(k), \boldsymbol{\mu}(k)) \|_2^2 \notag \\
    &= \|\mathbf{g}(\mathbf{b}(k), \boldsymbol{\mu}(k))\|_2^2 \; + \; 2\gamma \mathbf{g}(\mathbf{b}(k), \boldsymbol{\mu}(k))^T \bar G(k) \mathbf{g}(\mathbf{b}(k), \boldsymbol{\mu}(k)) + \gamma^2 \|\bar G(k) \mathbf{g}(\mathbf{b}(k), \boldsymbol{\mu}(k))\|_2^2
    \label{sq_eq}
\end{align}
From the assumption of strong concavity of utility applied with $\mathbf v = \mathbf{g}(\mathbf{b}(k), \boldsymbol{\mu}(k))$,
\begin{equation}
    \mathbf{g}(\mathbf{b}(k), \boldsymbol{\mu}(k))^T \bar G(k) \mathbf{g}(\mathbf{b}(k), \boldsymbol{\mu}(k)) \leqslant -\lambda \|\mathbf{g}(\mathbf{b}(k), \boldsymbol{\mu}(k))\|_2^2 
    \label{concave_assump}
\end{equation}
From the assumption that the operator norm of the averaged Hessian-like matrix is bounded, $\|\bar G(k)\|_{\text{op}} \leqslant L$ and using the standard property of operator norms $\|A \mathbf{v}\|_2 \leqslant \|A\|_{\text{op}} \|\mathbf{v}\|_2$, we obtain the following inequalities
\begin{align}
    \|\bar G(k) \mathbf{g}(\mathbf{b}(k), \boldsymbol{\mu}(k))\|_2 \leqslant L \|\mathbf{g}(\mathbf{b}(k), \boldsymbol{\mu}(k))\|_2 \notag \\
    \|\bar G(k) \mathbf{g}(\mathbf{b}(k), \boldsymbol{\mu}(k))\|_2^2 \leqslant L^2 \|\mathbf{g}(\mathbf{b}(k), \boldsymbol{\mu}(k))\|_2^2
    \label{op_norm}
\end{align}
Substituting the bounds from (\ref{concave_assump}) and (\ref{op_norm}) in (\ref{sq_eq}):
\begin{equation}
    \|\mathbf{g}(\mathbf{b}(k+1), \boldsymbol{\mu}(k))\|_2^2 \leqslant \|\mathbf{g}(\mathbf{b}(k), \boldsymbol{\mu}(k))\|_2^2 (1 - 2\gamma \lambda + \gamma^2 L^2)
    \label{new_sq}
\end{equation}
For convergence, the coefficient of the right-hand side must be less than one, leading to the step-size condition
\[1 - 2\gamma \lambda + \gamma^2 L^2 < 1 \quad \Rightarrow  \gamma < \frac{2\lambda}{L^2}\]
Hence, choose the step size as $\gamma = \frac{\lambda}{L^2}$.

\emph{Updating the correction cannot increase the norm: }
From the proof of \cref{lem:mu-unique} (\cref{app:mu-unique}), $\boldsymbol\mu_s(k)$ satisfies \eqref{eq:mu-VI} with updated budget $\mathbf b_s(k+1)$, and $\boldsymbol\mu_s(k+1)$ satisfies it with updated budget $\mathbf b_s(k+2)$. Take $\mathbf b_s'=\mathbf b_s(k+2)$ in the first and $\mathbf b_s'=\mathbf b_s(k+1)$ in the second; both are feasible by Step~2. Since $\mathbf b_s(k+2)-\mathbf b_s(k+1)=\gamma\,\mathbf g_s(\mathbf b(k+1),\boldsymbol\mu_s(k+1))$ by line~\ref{line:brd_update_rule} (\cref{alg:BRD}), summing over $s$ and dividing by $\gamma>0$ gives
\[
\boldsymbol\mu(k+1)^\top\mathbf g(\mathbf b(k+1),\boldsymbol\mu(k+1))\;\leqslant\;0\;\leqslant\;\boldsymbol\mu(k)^\top\mathbf g(\mathbf b(k+1),\boldsymbol\mu(k+1)).
\]
Writing $\mathbf g(\mathbf b(k+1),\boldsymbol\mu(k+1))=\mathbf g(\mathbf b(k+1))+\boldsymbol\mu(k+1)$, applying these two inequalities and then the Cauchy--Schwarz inequality,
\begin{align*}
    \|\mathbf g(\mathbf b(k+1),\boldsymbol\mu(k+1))\|_2^2\;
    &\leqslant\;\mathbf g(\mathbf b(k+1),\boldsymbol\mu(k+1))^\top\bigl(\mathbf g(\mathbf b(k+1))+\boldsymbol\mu(k)\bigr)\; \\
    &\leqslant\;\|\mathbf g(\mathbf b(k+1),\boldsymbol\mu(k+1))\|_2\,\|\mathbf g(\mathbf b(k+1),\boldsymbol\mu(k))\|_2 ,
\end{align*}
and dividing by $\|\mathbf g(\mathbf b(k+1),\boldsymbol\mu(k+1))\|_2$ (the claim is trivial if it is zero) gives:

\begin{equation}\label{eq:mu-update}
\|\mathbf g(\mathbf b(k+1),\boldsymbol\mu(k+1))\|_2\leqslant\|\mathbf g(\mathbf b(k+1),\boldsymbol\mu(k))\|_2 .
\end{equation}
Thus, Eq. (\ref{new_sq}) simplifies to
\[\|\mathbf{g}(\mathbf{b}(k+1), \boldsymbol{\mu}(k+1))\|_2^2 \leqslant \|\mathbf{g}(\mathbf{b}(k), \boldsymbol{\mu}(k))\|_2^2 \left(1 - \frac{\lambda^2}{L^2} \right)\]
where $\left(1 - \frac{\lambda^2}{L^2} \right) < 1$.
Applying this recursively over $k$ iterations yields
\begin{equation}
    \|\mathbf{g}(\mathbf{b}(k), \boldsymbol{\mu}(k))\|_2^2 \leqslant \left(1 - \frac{\lambda^2}{L^2} \right)^k \|\mathbf{g}(\mathbf{b}(0), \boldsymbol{\mu}(0)\|_2^2
    \label{eq:contraction}
\end{equation}

Under the inequality $(1-x)^k \leqslant e^{-xk}$ and taking the square roots, we obtain an exponential rate of convergence
\begin{equation}
    \|\mathbf{g}(\mathbf{b}(k), \boldsymbol{\mu}(k))\|_2 \leqslant e^{-\frac{\lambda^2}{2L^2} k} \|\mathbf{g}(\mathbf{b}(0),\boldsymbol{\mu}(0))\|_2
\end{equation}

\emph{Step 4: Iteration count.}
The number of iterations $k = K_\epsilon$ required to achieve a desired error tolerance $\epsilon > 0$ is
\[    e^{-\frac{\lambda^2}{2L^2} K_\epsilon} \|\mathbf{g}(\mathbf{b}(0),\boldsymbol{\mu}(0))\|_2 = \epsilon\]
\begin{equation}
    \Rightarrow K_\epsilon = \frac{2L^2}{\lambda^2} \log \left(\frac{\|\mathbf{g}(\mathbf{b}(0), \boldsymbol{\mu}(0)\|_2}{\epsilon} \right)
\end{equation}
\emph{Step 5: Convergence of the iterates to a Nash equilibrium.}
By line~\ref{line:brd_update_rule} (\cref{alg:BRD}) and \eqref{eq:contraction}, $$\|\mathbf b(k+1)-\mathbf b(k)\|_2=\gamma\,\|\mathbf g(\mathbf b(k),\boldsymbol\mu(k))\|_2\leqslant\gamma\,q^k\,\|\mathbf g(\mathbf b(0),\boldsymbol\mu(0))\|_2$$ 
with $q:=(1-\lambda^2/L^2)^{1/2}<1$. 

Hence, for a step $l>k$, the triangle inequality gives 
\[
\|\mathbf b(l)-\mathbf b(k)\|_2\leqslant\sum_{t=k}^{l-1}\|\mathbf b(t+1)-\mathbf b(t)\|_2 \leqslant 
\sum_{t=k}^{l-1}\gamma\,q^t\,\|\mathbf g(\mathbf b(0),\boldsymbol\mu(0))\|_2
\]
and the geometric series sum ($\sum_{t=k}^{l-1} q^t = q^k (1-q^{(l-k)})/(1-q) \leqslant q^k/(1-q)$) gives
\[
\|\mathbf b(l)-\mathbf b(k)\|_2\leqslant\frac{\gamma q^k}{1-q}\|\mathbf g(\mathbf b(0),\boldsymbol\mu(0))\|_2\xrightarrow[k\to\infty]{}0 .
\]
Hence, for every $\epsilon>0$, we can choose $K$ sufficiently large such that 
\[
\frac{\gamma q^K}{1-q}\|\mathbf g(\mathbf b(0),\boldsymbol\mu(0))\|_2 \leqslant \epsilon
\]
Then for all $l>k\geqslant K$, $\|\mathbf b(l)-\mathbf b(k)\|_2 \leqslant \epsilon$.
So $(\mathbf b(k))_k$ is a Cauchy sequence, and by the completeness of $\mathbb R^{mn}$ it converges to some $\mathbf b^\star\in\mathbb R^{mn}$. Since every $\mathbf b(k)$ lies in the closed set $\mathcal D^\ell$ (Step 2), $\mathbf b^\star\in\mathcal D^\ell$.

To see that $\mathbf b^\star$ is a Nash equilibrium, for a firm $s$, let $\mathbf b_s'\in\mathcal D_s^\ell$. By the characterization \eqref{eq:mu-VI} in the proof of \cref{lem:mu-unique}, the correction $\boldsymbol\mu_s(k)=\mathbf g_s(\mathbf b(k),\boldsymbol\mu_s(k))-\mathbf g_s(\mathbf b(k))$ satisfies \eqref{eq:mu-VI} with updated budget $\mathbf b_s(k+1)$:
\[
\bigl(\mathbf g_s(\mathbf b(k),\boldsymbol\mu_s(k))-\mathbf g_s(\mathbf b(k))\bigr)^\top\bigl(\mathbf b_s'-\mathbf b_s(k+1)\bigr)\geqslant0\qquad \forall \; k.
\]
As $k\to\infty$, the corrected gradient tends to $\mathbf 0$ (Step~3), $\mathbf g_s(\mathbf b(k))\to\mathbf g_s(\mathbf b^\star)$ by continuity of the gradient, and $\mathbf b_s(k+1)\to\mathbf b^\star_s$; hence
\[
\mathbf g_s(\mathbf b^\star)^\top(\mathbf b_s'-\mathbf b^\star_s)\leqslant0 \qquad \forall \; \mathbf b_s' \in \mathcal{D}_s^\ell
\] 
Since $u_s$ is concave in $\mathbf b_s$ (Step~1), $u_s(\mathbf b_s',\mathbf b^\star_{-s})\leqslant u_s(\mathbf b^\star)+\mathbf g_s(\mathbf b^\star)^\top(\mathbf b_s'-\mathbf b^\star_s)\leqslant u_s(\mathbf b^\star)$. So every firm is best-responding at $\mathbf b^\star$, which is therefore a Nash equilibrium. Equivalently, by \cref{lem:grad-NE}, $\mathbf g(\mathbf b^\star,\boldsymbol\mu(\mathbf b^\star))=\mathbf 0$. This also follows directly from the continuity of the corrected gradient: by the proof of \cref{lem:mu-unique}, $\mathbf g_s(\mathbf b,\boldsymbol\mu_s(\mathbf b))=\bigl[\Pi_{\mathcal D_s^\ell}\bigl(\mathbf b_s+\gamma\nabla_{\mathbf b_s}u_s(\mathbf b)\bigr)-\mathbf b_s\bigr]/\gamma$, which is continuous in $\mathbf b$ because the projection onto a non-empty, closed, convex set is non-expansive and $\nabla_{\mathbf b_s}u_s$ is continuous; hence $\mathbf g(\mathbf b^\star,\boldsymbol\mu(\mathbf b^\star))=\lim_k\mathbf g(\mathbf b(k),\boldsymbol\mu(k))=\mathbf 0$ by Step~3. 
Thus, the best response dynamics converges to a Nash Equilibrium $\mathbf b^\star$ in $K_\epsilon = \mathcal{O} \left(\log \frac{1}{\epsilon} \right)$ iterations. \hfill$\square$

\section{Appendix to \texorpdfstring{\Cref{sec:welfare}}{Section 5}}

\subsection{Reduction of the SB property to vertex-wise blocks}
\label{app:reduction}

Both \Cref{lem:rcsf-sb} and \Cref{lem:power-SB} rest on the same reduction, which we record once. The block Hessian of \Cref{def:SB} has entries $[G_{s_1,s_2}]_{i,j} = \partial^2u_{s_1}/(\partial b_{s_2,j}\partial b_{s_1,i}) = M_i\,\partial^2h_{s_1,i}/(\partial b_{s_2,j}\partial b_{s_1,i})$, since the cost term is linear. Because $h_{s_1,i}$ depends only on the budgets allocated at vertex $i$, this vanishes whenever $j\neq i$. Reordering coordinates from firm-major order $\mathbf{b} = (b_{1,1}, \ldots, b_{1,n}, \ldots. b_{m,1}, \ldots, b_{m,n})^\top$ to vertex-major order $\widetilde{\mathbf{b}} = (b_{1,1}, \ldots, b_{m,1}, \ldots. b_{1,n}, \ldots, b_{m,n})^\top$ therefore turns $G$ into the block-diagonal matrix
\[
\widetilde{G} = \mathrm{block\text{-}diag}(H_1,\ldots,H_n),
\qquad
(H_i)_{s_1,s_2} = M_i\,\frac{\partial^2h_{s_1,i}}{\partial b_{s_2,i}\,\partial b_{s_1,i}}, \quad s_1,s_2\in\mathscr{M} .
\]
This reordering is an orthogonal permutation, so $G$ and $\widetilde{G}$ have identical spectra and identical operator norms. Hence the SB property (\Cref{def:SB}) for some constants $\widetilde{\lambda}, \widetilde{L} > 0$, can be rewritten as:
\begin{equation}
\label{eq:modified-SB}
    \tfrac12\bigl(\widetilde{G}(\mathbf b)+\widetilde{G}(\mathbf b)^\top\bigr)+\widetilde{\lambda} I_{mn}\preceq0
  \;\; \text{(Strong Concavity)} \quad \text{and} \quad \|\widetilde{G}\|_{\mathrm{op}} \leqslant \widetilde{L} \;\; \text{(Bounded Norm)}.
\end{equation}
Since $\widetilde{G}$ is block-diagonal, the spectrum of its symmetric part is the union of the spectra of the symmetric parts of the blocks $H_i$ and $\|\tilde{G}\|_{\mathrm{op}}=\max_{i\in\mathscr N}\|H_i\|_{\mathrm{op}}$.
Thus, \eqref{eq:modified-SB} can be verified block by block: for each $i\in\mathscr  N$, suppose: $\frac{1}{2}(H_i+H_i^\top)\preceq -\widetilde\lambda_i I_m$ and $\|H_i\|_{\mathrm{op}}\leqslant \widetilde L_i$. The common SB constants can then be chosen as $\widetilde\lambda:=\min_{i}\widetilde\lambda_i$ and $\widetilde L:=\max_{i}\widetilde L_i$ so that the SB property can be stated uniformly as:
\begin{equation}
\label{eq:blockwise-SB}
\tfrac{1}{2}(H_i + H_i^\top) \preceq -\widetilde{\lambda}\, I_m \quad \text{and} \quad \|H_i\|_{\mathrm{op}} \leqslant \widetilde{L}, \qquad \forall\, i \in \mathscr{N}.
\end{equation}

\subsection{Proof of \texorpdfstring{\Cref{lem:rcsf-sb}}{Lemma 4}: the Regular RCSF induces the SB property}
\label{app:rcsf-sb}

By \Cref{app:reduction} it suffices to establish the two conditions in \eqref{eq:blockwise-SB} for each vertex block $H_i$.
A convenient \emph{sufficient} test for the first condition of \eqref{eq:blockwise-SB} (strong concavity) is strict diagonal dominance as defined below, which by \textit{Gershgorin's circle theorem}, guarantees the required spectral bound. The converse does not hold in general.

\begin{definition}[Strict diagonal dominance]
A matrix $H_i\in\mathbb{R}^{m\times m}$ is said to satisfy the $(\tilde{\lambda},m)$-strict diagonal dominance condition if for some $\tilde{\lambda}>0$,
\[
|(H_i)_{ss}|-\sum_{r\neq s}|(H_i)_{sr}|
\geqslant \tilde{\lambda},
\qquad \forall s\in\mathscr{M},
\]
where $|(H_i)_{ss}|$ is the diagonal magnitude and $|(H_i)_{sk}|$ is the magnitude of the off diagonal entries.
\label{def:diag_dom}
\end{definition}

For a vertex $i$ and firm $r \in \mathscr{M}$, we will denote $f_r = f(b_{r,i})$, $f_r'=f'(b_{r,i})$, $f_r''=f''(b_{r,i})$ and $S=\sum_{r\in\mathscr{M}}f_r+\delta$. The required second derivatives for the Regular RCSF are
\begin{equation}
    \frac{\partial^2 h_{s,i}}{\partial b_{s,i}^2} = \frac{(S-f_s)}{S^2}\Bigl(f_s'' - \frac{2(f_s')^2}{S}\Bigr),
    \qquad
    \frac{\partial^2 h_{s,i}}{\partial b_{r,i}\,\partial b_{s,i}} = \frac{f_s'\,f_r'\,(2f_s-S)}{S^3}, \quad r\neq s.
    \label{eq:curv}
\end{equation}
Since $f$ is increasing ($f_s' >0$) and concave ($f_s'' <0$) and $S-f_s=\sum_{r\neq s}f_r+\delta\geqslant\delta>0$, $\frac{\partial^2 h_{s,i}}{\partial b_{s,i}^2} < 0$. We use the elementary bounds $S\geqslant\delta$ and $S-f_s\geqslant\delta$, both from $f\geqslant0$.

\begin{claim}[Strong concavity via diagonal dominance]
\label{lem:dd}
Under conditions~\ref{cond1} and~\ref{cond3} of \Cref{def:regular_rcsf}, each block $H_i$ is strictly diagonally dominant with negative diagonal, and there is a constant $\widetilde{\lambda}>0$ with $\tfrac12(H_i+H_i^\top)\preceq-\widetilde{\lambda}I_m$ for every $i$. Hence $\tfrac12\bigl(\widetilde{G}(\mathbf b)+\widetilde{G}(\mathbf b)^\top\bigr)+\widetilde{\lambda} I_{mn}\preceq0$.
\end{claim}
\begin{proof}
Fix a row $s$. By \eqref{eq:curv} and $S-f_s\geqslant\delta$, the diagonal magnitude satisfies
\[
\Bigl|\frac{\partial^2h_{s,i}}{\partial b_{s,i}^2}\Bigr| = \frac{(S-f_s)}{S^2}\Bigl(\frac{2(f_s')^2}{S}-f_s''\Bigr) \geqslant \frac{(S-f_s)(-f_s'')}{S^2} \geqslant \frac{\delta(-f_s'')}{S^2}.
\]
For the off-diagonal entries of row $s$, using $|2f_s-S|\leqslant S$ (as $0\leqslant f_s\leqslant S$) and $\sum_{k\neq s}f_k'\leqslant(m-1)\sup_{\mathcal{B}}f'$,
\[
\sum_{r\neq s}\Bigl|\frac{\partial^2h_{s,i}}{\partial b_{r,i}\partial b_{s,i}}\Bigr| = \frac{f_s'|2f_s-S|}{S^3}\sum_{r\neq s}f_r' \leqslant \frac{f_s'(m-1)\sup_{\mathcal{B}}f'}{S^2}.
\]
Subtracting,
\begin{equation}
    \Bigl|\frac{\partial^2h_{s,i}}{\partial b_{s,i}^2}\Bigr| - \sum_{r\neq s}\Bigl|\frac{\partial^2h_{s,i}}{\partial b_{r,i}\partial b_{s,i}}\Bigr| \geqslant \frac{1}{S^2}\Bigl[\delta(-f_s'') - (m-1)f_s'\sup_{\mathcal{B}}f'\Bigr].
    \label{eq:diag_diff}
\end{equation}
Condition~\ref{cond3} of \Cref{def:regular_rcsf} gives 
\[
    (m-1)\, \frac{f_s'}{-f_s''}\, \sup f' \leqslant (m-1) \sup \frac{f'}{-f''}\, \sup f' < \delta
\]
\[
    \implies (m-1) f_s'\, \sup f' < \delta\,(-f_s'')
\]
Hence in \eqref{eq:diag_diff}, the RHS term
\[
    \delta\, (-f_s'') - (m-1) f_s'\, \sup f' > 0
\] 
On the compact set $\mathcal{B}$, we have $S\leqslant mf(C)+\delta$ (since $b_{s,i} \leqslant C$) so $\frac{1}{S^2}\geqslant \frac{1}{(mf(C)+\delta)^2}$. Define the smallest possible uniform diagonal-dominance gap:
\[
\Delta :=
\min_{\substack{\mathbf b\in\mathcal B\\
s\in\mathscr M,\;i\in\mathcal N}}
\left\{
\delta\bigl(-f_s''\bigr)
-(m-1)f_s'\sup f'
\right\}.
\]
Substituting in \Cref{def:diag_dom},
\[
|(H_i)_{ss}|-\sum_{k\neq s}|(H_i)_{sk}| = M_i \Big[ \left| \frac{\partial^2 h_{s,i}}{\partial b_{s,i}^2} \right| - \sum_{k \neq s} \left| \frac{\partial^2 h_{s,i}}{\partial b_{k,i}\, \partial b_{s,i}} \right|\Big] \geqslant  \frac{M_i\Delta}{(mf(C)+\delta)^2}
\]
Since $M_i\geqslant 1- \alpha >0$ (\cref{lem:sumM}) and $\Delta > 0$,
\[
|(H_i)_{ss}|-\sum_{k\neq s}|(H_i)_{sk}|\geqslant  \frac{(1-\alpha)\Delta}{(mf(C)+\delta)^2} =: \widetilde{\lambda} > 0
\]

The same bounds apply to the column sums, because the off-diagonal entries of column $s$ are bounded by the same quantity with the roles of $f_s'$ and $f_r'$ exchanged, and $f_r'\leqslant\sup f'$. So the symmetric part $\tfrac12(H_i + H_i^\top)$ inherits strict diagonal dominance with negative diagonal. Consequently, all eigenvalues of $\tfrac12(H_i + H_i^\top)$ are bounded above by $-\widetilde{\lambda}$ implying $\tfrac{1}{2}(H_i + H_i^\top) \preceq -\widetilde{\lambda}\, I_m$ (\eqref{eq:blockwise-SB}). By the reordering above, $\tfrac12\bigl(\widetilde{G}(\mathbf b)+\widetilde{G}(\mathbf b)^\top\bigr)+\widetilde{\lambda} I_{mn}\preceq0$.
\end{proof}

\begin{claim}[Bounded operator norm]
\label{lem:bounded}
Under conditions~\ref{cond1} and~\ref{cond2} of \Cref{def:regular_rcsf}, every entry of $H_i$ is bounded in magnitude by $M_i\varkappa$, where $\varkappa = 2(\sup f')^2/\delta^2 + \sup|f''|/\delta$. Hence $\|H_i\|_{\mathrm{op}}\leqslant\|H_i\|_F\leqslant mM_i\varkappa$, and $\|\widetilde{G}\|_{\mathrm{op}} \leqslant \widetilde{L}$ with $\widetilde{L}:=m \varkappa \max_i M_i<\infty$.
\end{claim}
\begin{proof}
Using $S-f_s\leqslant S$, $1/S\leqslant1/\delta$ and $1/S^2\leqslant1/\delta^2$ in \eqref{eq:curv},
\[
\Bigl|\frac{\partial^2h_{s,i}}{\partial b_{s,i}^2}\Bigr| \leqslant \frac{1}{S}\Bigl(\frac{2(f_s')^2}{S}+|f_s''|\Bigr) \leqslant \frac{2(\sup f')^2}{\delta^2}+\frac{\sup|f''|}{\delta} = \varkappa,
\]
while the off-diagonal entries satisfy $|\partial^2h_{s,i}/(\partial b_{k,i}\partial b_{s,i})|\leqslant(\sup f')^2/S^2 \leqslant (\sup f')^2/\delta^2\leqslant\varkappa$. Thus every entry of $H_i$ is at most $M_i \varkappa$ in magnitude, so $\|H_i\|_F\leqslant mM_i\varkappa$ and $\|H_i\|_{\mathrm{op}}\leqslant\|H_i\|_F \leqslant m \varkappa \max_i M_i := \widetilde{L}$. Thus in \eqref{eq:modified-SB}, $\|\widetilde G\|_{\mathrm{op}}\leqslant \widetilde{L} < \infty$.
\end{proof}

\Cref{lem:dd,lem:bounded} establish \Cref{eq:blockwise-SB}, so by \Cref{app:reduction} the utilities have the SB property (Definition~\ref{def:SB}) on $\mathcal D^\ell = \mathcal B$ with the constants $\lambda=\widetilde{\lambda}$ and $L=\widetilde{L}$. Hence, by \cref{thm:BRD-conv}, the best-response dynamics (\Cref{alg:BRD}) converges to a Nash equilibrium $\mathbf b^\star$ with step size $\gamma=\widetilde\lambda/\widetilde L^2$, reaching $\|\mathbf g(\mathbf b(K_\epsilon),\boldsymbol\mu(K_\epsilon))\|_2<\epsilon$ after $K_\epsilon=O\bigl((\widetilde L/\widetilde\lambda)^2\log(1/\epsilon)\bigr)$ iterations.
\hfill$\square$

\subsection{Proof of \texorpdfstring{\Cref{lem:rcsf-unique-symmetric}}{Lemma 5}: uniqueness and symmetry for the RCSF}
\label{app:RRCSF_sym_unique}

\emph{Part (I): every NE is symmetric.}

Following \cite{Cheng2004},the game admits a pure strategy symmetric equilibrium because it is a symmetric game with compact and convex strategy spaces and continuous and strictly-concave utility function (from condition 1 of \cref{def:regular_rcsf}, the Regular RCSF $h_{s,i}(\mathbf b)$ is strictly concave which implies that each firm's utility $u_s(\mathbf b)$ is also strictly concave in its own decision vector $\mathbf b_s$ since it combines a concave total awareness term $\sum_i M_i h_{s,i}(\mathbf b)$ and the negative of a convex (linear) cost term $\sum_i b_{s,i}$. 
Now it remains to show that the game doesn't have any asymmetric equilibria. First, we derive an inequality which we call the \emph{marginal dominance (MD) inequality} which shows that at any vertex, the firm bidding strictly more has the strictly smaller marginal return there. This inequality will be used crucial later in ruling out asymmetric equilibria from the game.

\emph{Step 1: marginal dominance inequality.} Fix a vertex $i$ and two distinct firms $p, q \in \mathscr{M}$. Write $b_{p,i} = x$, $b_{q,i} = y$ with $0 \leqslant y < x \leqslant C$, and let $S = \sum_{r \neq p,q} f(b_{r,i}) \geqslant 0$ denote the fixed sum of all other firms' budgets at $i$. Isolating firms $p$ and $q$,
\[
h_{p,i}(\mathbf{b}) = h_{p,i}(x,y;S) = \frac{f(x)}{f(x)+f(y)+\delta'}, \qquad \delta' := \delta + S > 0,
\]
so that the marginal return for firm $p$ is
\begin{equation}
\frac{\partial h_{p,i}(\mathbf b)}{\partial b_{p,i}} = \frac{\partial h_{p,i}}{\partial x}(x,y;S) = \frac{f'(x)\big(f(y)+\delta'\big)}{D(x,y)^2}, \qquad D(x,y) := f(x)+f(y)+\delta'
\label{eq:hpi-deriv}
\end{equation}

and for firm $q$, 

\begin{equation}
\frac{\partial h_{q,i}(\mathbf b)}{\partial b_{q,i}} = \frac{\partial h_{q,i}}{\partial y}(x,y;S) = \frac{f'(y)\big(f(x)+\delta'\big)}{D(x,y)^2}
\label{eq:hqi-deriv}
\end{equation}

Since the denominators of \eqref{eq:hpi-deriv} and \eqref{eq:hqi-deriv} are identical, this simplifies to comparing 
\[
\frac{f'(x)}{f(x)+\delta'}  \text{ and } \frac{f'(y)}{f(y)+\delta'}
\]

Let $F(t) := f'(t)/(f(t)+\delta')$. Then
\[
F'(t) = \frac{f''(t)\big(f(t)+\delta'\big) - f'(t)^2}{\big(f(t)+\delta'\big)^2}.
\]
Since $f$ is concave and increasing, $f'(t) > 0$ and $f''(t) < 0$, and $\delta' > 0$; thus $F'(t) < 0$, so $F$ is strictly decreasing. Hence for $x > y$, $F(y) > F(x)$, which gives

\begin{equation}
\frac{\partial h_{p,i}}{\partial b_{p,i}} \;<\; \frac{\partial h_{q,i}}{\partial b_{q,i}}, \qquad \text{for } b_{p,i} > b_{q,i}.
\tag{$\star$}
\label{eq:MD}
\end{equation}

\emph{Step 2: KKT conditions.} Since the utility is strictly concave in its own strategy, KKT conditions are necessary and sufficient. Firm $s$ solves
\[
\text{Maximize } u_s(\mathbf{b}) = \sum_i M_i h_{s,i}(\mathbf{b}) - \sum_i b_{s,i}, \qquad \text{subject to } \mathbf{b}_s \geqslant \mathbf{0},\ \sum_i b_{s,i} \leqslant C.
\]
Let $\mu_{s,i}, \lambda_s$ be the Lagrange multipliers for the two constraints, with Lagrangian
\[
\mathcal{L}(\mathbf{b}, \lambda_s, \mu_{s,i}) = u_s(\mathbf{b}) - \lambda_s\Big(\sum_i b_{s,i} - C\Big) - \sum_i \mu_{s,i}(-b_{s,i}).
\]
The KKT conditions are:
\begin{align}
&\text{Stationarity:} & M_i \frac{\partial h_{s,i}(\mathbf{b})}{\partial b_{s,i}} &= 1 + \lambda_s - \mu_{s,i}, \tag{S}\label{eq:S}\\
&\text{Dual feasibility:} & \mu_{s,i} &\geqslant 0, \quad \lambda_s \geqslant 0, \tag{D}\label{eq:D}\\
&\text{Complementary slackness:} & \mu_{s,i}b_{s,i} &= 0, \quad \lambda_s\Big(C - \sum_i b_{s,i}\Big) = 0. \tag{CS}\label{eq:CS}
\end{align}

\emph{Step 3: ruling out asymmetric equilibria.} Suppose $\mathbf{b}^\star$ is an asymmetric PSNE, i.e.\ $\mathbf{b}_p^\star \neq \mathbf{b}_q^\star$ for some two firms $p,q \in \mathscr{M}$.

Suppose $b^\star_{p,i_0} > b^\star_{q,i_0} \geqslant 0$ at some vertex $i_0$. Since $b^\star_{p,i_0} > 0$, \eqref{eq:CS} gives $\mu^\star_{p,i_0} = 0$, and \eqref{eq:S} gives
\begin{equation}
M_{i_0} \frac{\partial h_{p,i_0}}{\partial b_{p,i_0}}(\mathbf{b}^\star) = 1 + \lambda_p.
\label{eq:c1}
\end{equation}
For firm $q$, \eqref{eq:CS} gives $\mu_{q,i_0} \geqslant 0$, so \eqref{eq:S} gives
\begin{equation}
M_{i_0} \frac{\partial h_{q,i_0}}{\partial b_{q,i_0}}(\mathbf{b}^\star) = 1 + \lambda_q - \mu_{q,i_0} \leqslant 1 + \lambda_q.
\label{eq:c2}
\end{equation}
By the marginal dominance inequality \eqref{eq:MD} at $i_0$ (the other $m-2$ firms do not matter),
\begin{equation}
\frac{\partial h_{p,i_0}}{\partial b_{p,i_0}}(\mathbf{b}^\star) < \frac{\partial h_{q,i_0}}{\partial b_{q,i_0}}(\mathbf{b}^\star).
\label{eq:c3}
\end{equation}
Combining \eqref{eq:c1}, \eqref{eq:c2}, \eqref{eq:c3},
\begin{equation}
1 + \lambda_p < 1 + \lambda_q \implies \lambda_p < \lambda_q.
\label{eq:c4}
\end{equation}
That is: whoever bids strictly more at some vertex has a strictly smaller budget multiplier.

\smallskip
\noindent \textit{Case A: There exist two vertices $ i, j$ with $b^\star_{p,i} > b^\star_{q,i}$ and $b^\star_{q,j} > b^\star_{p,j}$}. This case arises whenever the two firms have equal total spend, i.e.\ $\sum_i b^\star_{p,i} = \sum_i b^\star_{q,i}$.  
Applying \eqref{eq:c4} at vertex $i$ gives $\lambda_p < \lambda_q$; applying it at vertex $j$ (roles reversed) gives $\lambda_q < \lambda_p$. Contradiction.

\smallskip
\noindent\textit{Case B: $b^\star_{p,i} \geqslant b^\star_{q,i}\ \forall i$, with strict inequality at some $i_0$}.
Then $\sum_i b^\star_{p,i} > \sum_i b^\star_{q,i}$ (unequal total spend). At $i_0$, \eqref{eq:c4} gives $\lambda_p < \lambda_q$.
\begin{itemize}
    \item If $\sum_i b^\star_{p,i} = C$, then $\sum_i b^\star_{q,i} < C$, so \eqref{eq:CS} gives $\lambda_q = 0$. Combined with $\lambda_p < \lambda_q = 0$, this forces $\lambda_p < 0$, violating \eqref{eq:D}.
    \item If $\sum_i b^\star_{p,i} < C$, then \eqref{eq:CS} gives $\lambda_p = 0$. Then $\lambda_p < \lambda_q$ forces $\lambda_q > 0$, which by \eqref{eq:CS} forces $\sum_i b^\star_{q,i} = C$. But this implies $\sum_i b^\star_{q,i} > \sum_i b^\star_{p,i}$, contradicting the case hypothesis.
\end{itemize}

\noindent
Hence, in every case, no asymmetric NE is possible. Therefore every NE is symmetric.

\medskip
\noindent\emph{Part (II): the symmetric NE is unique.}

At the symmetric profile $b_{s,i}=t_i$, let $a(t_i):=\partial h_{s,i}/\partial b_{s,i}$ be the marginal-return function. By the chain rule,
\begin{equation}
a'(t_i) = \left.\frac{\partial^2 h_{s,i}}{\partial b_{s,i}^2}\right|_{\mathrm{symm}} + (m-1)\left.\frac{\partial^2 h_{s,i}}{\partial b_{r,i}\, \partial b_{s,i}}\right|_{\mathrm{symm}}.
\label{eq:aprime}
\end{equation}
Writing $D = \sum_s f_s + \delta$ with $f_s := f(b_{s,i})$, direct differentiation of $h_{s,i} = f_s/D$ gives
\[
\frac{\partial^2 h_{s,i}}{\partial b_{s,i}^2} = \frac{(D - f_s)}{D^2}\left(f_s'' - \frac{2f_s'^2}{D}\right) < 0, \qquad \text{since } f_s'' < 0,\ f_s' > 0,\ D - f_s > 0.
\]
Evaluated at the symmetric point ($D = m f(t_i) + \delta$),
\begin{equation}
\left.\frac{\partial^2 h_{s,i}}{\partial b_{s,i}^2}\right|_{\mathrm{symm}} = \frac{\big((m-1)f(t_i)+\delta\big)}{D^2}\left(f''(t_i) - \frac{2f'(t_i)^2}{D}\right) < 0.
\label{eq:own2nd}
\end{equation}
Similarly, for $r \neq s$,
\[
\frac{\partial^2 h_{s,i}}{\partial b_{r,i}\, \partial b_{s,i}} = -f'(b_{s,i}) f'(b_{r,i}) \frac{D - 2f(b_{r,i})}{D^3},
\]
which at the symmetric point gives
\begin{equation}
\left.\frac{\partial^2 h_{s,i}}{\partial b_{r,i}\, \partial b_{s,i}}\right|_{\mathrm{symm}} = -\big(f'(t_i)\big)^2\, \frac{(m-2)f(t_i)+\delta}{D^3} < 0.
\label{eq:cross2nd}
\end{equation}
Substituting \eqref{eq:own2nd} and \eqref{eq:cross2nd} into \eqref{eq:aprime} gives $a'(t_i) < 0$; hence $a$ is strictly decreasing in $t_i$, hence injective.

Rewriting the KKT conditions for a symmetric equilibrium profile $\mathbf{t}^\star = (t_i^\star)_{i \in \mathscr{N}}$ with common multipliers $\lambda \geqslant 0$ and $\mu_i \geqslant 0$ (shared across firms by symmetry) :
\[
M_i\, a(t_i^\star) = 1 + \lambda - \mu_i 
\]
If $t_i^\star > 0$, then by complementary slackness, $\mu_i = 0$, the KKT becomes $M_i\, a(t_i^\star) = \lambda + 1 =: \Gamma$. If $t_i^\star = 0$, then $\mu_i \geqslant 0$ and we have $M_i\, a(0) = (\lambda+1) - \mu_i \leqslant \Gamma$.

Since $a$ is continuous and strictly decreasing, so is $M_i\, a(\cdot)$ for each fixed $M_i > 0$, and it is therefore invertible. Define
\begin{equation}
t_i^\star = \beta_i(\Gamma) := \max\left\{ a^{-1}\left(\frac{\Gamma}{M_i}\right),\ 0 \right\}.
\label{eq:beta-rcsf}
\end{equation}
Then $\beta_i(\Gamma)$ is single-valued and (weakly, strictly wherever positive) decreasing in $\Gamma$; hence for any $\Gamma > 0$ the budget $t_i^\star$ is unique. Next we prove that this $\Gamma$ is also unique for a given game instance.

Now consider the aggregate budget $S(\Gamma) := \sum_i \beta_i(\Gamma)$, and it is also continuous and strictly decreasing on the region where it is positive, with $S(\Gamma)\to0$ as $\Gamma\to\infty$. Since the total budget constraint is $\sum_i t_i^\star \leqslant C$, two cases arise considering the smallest possible $\Gamma = 1$:

\begin{enumerate}
    \item[(1)] $S(1) \leqslant C$: the unconstrained budget at the lowest possible marginal threshold $\Gamma = 1$ already satisfies the constraint, thus the budget multiplier is inactive, i.e. $\lambda = 0$. Thus there is a unique $\Gamma^\star = 1$, and $t_i^\star = \beta_i(1)$ is the unique solution.
    \item[(2)] $S(1) > C$: the budget constraint is active, $\lambda > 0$. Since $\lim_{\Gamma\to\infty}S(\Gamma)=0$, there exists a finite $\Gamma_1>1$ such that $S(\Gamma_1)<C$. Since $S$ is continuous on $(0,\infty)$, it is continuous on the closed interval $[1,\Gamma_1]$ with $S(1)>C>S(\Gamma_1)$. The Intermediate Value Theorem then implies that there exists a $\Gamma^\star > 1$ with $S(\Gamma^\star) = C$, and strict monotonicity implies that the $\Gamma^\star$ is unique as a strictly monotone continuous function crosses any given level exactly once. Thus $t_i^\star = \beta_i(\Gamma^\star)$.
\end{enumerate}
Thus in both the cases, there is a unique $\Gamma^\star \geqslant 1$ satisfying the budget constraint and hence at most one symmetric profile $\mathbf{t}^\star = (\beta_i(\Gamma^\star))_{i \in \mathscr{N}}$ satisfies the KKT system, so the symmetric equilibrium is unique. Together with \textbf{Part (I)}, the game admits a unique Nash equilibrium, and it is symmetric.
\hfill$\square$

\subsection{Proof of \texorpdfstring{\Cref{thm:RCSF_impossibility}}{Theorem 3}}
\label{app:RCSF_impossibility}

\emph{Step 1 (Reduction to symmetric profiles).}
The utility of a firm $s$ is $u_s(\mathbf{b}) = \sum_i M_i h_{s,i}(\mathbf b) - \sum_i b_{s,i}$ \eqref{eq:utility} .
Let $h_{s,i}(\mathbf{b}) = f(b_{s,i})/(F_i+\delta)$ with $F_i = \sum_r f(b_{r,i})$. Since $\sum_sf(b_{s,i})/(F_i+\delta)=F_i/(F_i+\delta)$ for every $i$, utilitarian welfare $W_1$ depends on $\mathbf{b}$ only through the aggregates $F_i(\mathbf{b}):=\sum_sf(b_{s,i})$ and $B_i(\mathbf{b}):=\sum_sb_{s,i}$:
\[
W_1(\mathbf{b}) = \frac1m\sum_s u_s(\mathbf{b}) = \frac1m\left[\sum_iM_i\frac{F_i(\mathbf{b})}{F_i(\mathbf{b})+\delta} - \sum_iB_i(\mathbf{b})\right].
\]
For $\mathbf{b}\in\mathcal{B}$, let $\bar{\mathbf{b}}$ be its \emph{symmetrization}, $\bar b_{s,i}:=\frac1m\sum_rb_{r,i}$ for every $s$. Then $\bar{\mathbf{b}}\in\mathcal{B}$, since $\sum_i \bar b_{s,i} = \frac1m\sum_r\sum_i b_{r,i}\leqslant C$, and symmetrization preserves aggregate spend exactly, $B_i(\bar{\mathbf{b}})=B_i(\mathbf{b})$. By concavity of $f$ (condition \ref{cond1} of \Cref{def:regular_rcsf}), Jensen's inequality gives $f(\bar b_{s,i})\geqslant\frac1m\sum_rf(b_{r,i})=F_i(\mathbf{b})/m$ for every $s$, so $F_i(\bar{\mathbf{b}})=\sum_sf(\bar b_{s,i})\geqslant F_i(\mathbf{b})$: symmetrization weakly increases the aggregate score. Since $t\mapsto t/(t+\delta)$ is strictly increasing, $F_i(\bar{\mathbf{b}})\geqslant F_i(\mathbf{b})$ together with $B_i(\bar{\mathbf{b}})=B_i(\mathbf{b})$ give $W_1(\bar{\mathbf{b}})\geqslant W_1(\mathbf{b})$: symmetrization never decreases average welfare. Because $\mathcal{B}$ is compact and $W_1$ continuous, a maximizer of $W_1$ over $\mathcal{B}$ exists; its symmetrization is itself a maximizer, and is symmetric. Consequently, if a welfare-maximizing profile exists, there exists one on the symmetric slice, and it suffices to compare $\mathbf{b}^\star$ which is symmetric by \Cref{lem:rcsf-unique-symmetric}, against a welfare-maximizing profile restricted to symmetric profiles.

\medskip
\noindent\emph{Step 2 (The two KKT systems).}
Restricted to a symmetric profile $b_{s,i}=t_i$, firm $s$'s stationarity condition is
\[
M_ia(t_i)-1=\lambda_s-\mu_{s,i}, \qquad \mu_{s,i}\geqslant0,\ \mu_{s,i}t_i=0,\ \lambda_s\geqslant0,\ \lambda_s\bigl(C-\textstyle\sum_it_i\bigr)=0,
\]
while the planner, maximizing $U:=mW_1=\sum_ru_r$, has
\[
M_i\psi(t_i)-1=\hat\lambda-\hat\mu_{i}, \qquad \hat\mu_{i}\geqslant0,\ \hat\mu_{i}t_i=0,\ \hat\lambda\geqslant0,\ \hat\lambda\bigl(C-\textstyle\sum_it_i\bigr)=0,
\]
where $a$ and $\psi$ are as in \Cref{def:wedge}. (The planner faces one budget constraint per firm; at a symmetric profile these carry a common multiplier, which we write as $\hat\lambda$ after rescaling by $m$.) A symmetric profile is simultaneously a Nash equilibrium and a welfare optimum only if some choice of multipliers satisfies both systems at once.

\medskip
\noindent\emph{Step 3 (The wedge is strictly positive).}
Writing $D=\sum_rf_r+\delta$ with $f_r := f(b_{r,i})$, direct differentiation of $h_{s,i}=f_s/D$ gives $\partial h_{s,i}/\partial b_{s,i} = f_s'(D-f_s)/D^2$ and, for $r\neq s$, $\partial h_{r,i}/\partial b_{s,i} = -f_r f_s'/D^2$. At the symmetric profile, where $f_r = f(t_i)$ for every $r$ and $D = mf(t_i)+\delta$,
\begin{equation}
\label{eq:rcsf-a-psi}
a(t_i) = \frac{f'(t_i)\bigl[(m-1)f(t_i)+\delta\bigr]}{\bigl(mf(t_i)+\delta\bigr)^2}, \quad d(t_i) = -\frac{f'(t_i)\,f(t_i)}{\bigl(mf(t_i)+\delta\bigr)^2}, \quad \psi(t_i) = \frac{\delta\, f'(t_i)}{\bigl(mf(t_i)+\delta\bigr)^2}.
\end{equation}
The $\delta$ term cancels in $w = a-\psi$ and the $(m-1)f(t_i)$ term survives:
\begin{equation}
\label{eq:wedge_RSCF}
w(t_i) = \frac{(m-1)\, f(t_i)\, f'(t_i)}{\bigl(mf(t_i)+\delta\bigr)^2} \;>\; 0 \qquad \text{for every } t_i>0,\ m\geqslant 2,
\end{equation}
since $f$ is increasing with $f(0)=0$, so $f(t_i)>0$ for $t_i>0$, and $f'(t_i)>0$.

\medskip
\noindent\emph{Step 4 (Failure of welfare optimality).}
On the symmetric slice evaluated at $b_{s,i}=t_i$, $W_1$ reduces to $\sum_i M_if(t_i)/(mf(t_i)+\delta) - \sum_it_i$. Each summand composes the concave increasing map $x\mapsto x/(mx+\delta)$ with the concave map $f$, hence is concave, so the planner's restricted problem is a concave maximization subject to a linear constraint and KKT stationarity is necessary \emph{and} sufficient for optimality. 

At any vertex $i$ with $t_i>0$, complementary slackness gives $\mu_{s,i}=\hat\mu_{i}=0$, so the two stationarity conditions of Step 2 reduce to
\begin{equation}
\label{eq:reduced-stationarity}
M_ia(t_i) = 1+\lambda_s, \qquad M_i\psi(t_i) = 1+\hat\lambda,
\end{equation}
with $\lambda_s,\hat\lambda\geqslant0$ shared across vertices. This holds regardless of whether the respective budget constraints bind, i.e.\ regardless of whether $\lambda_s,\hat\lambda$ are zero or strictly positive.

\emph{Case (i): heterogeneous centralities.} Suppose $\mathbf{b}^\star = \{t_i^\star\}_{i\in \mathscr N}$ is active at two vertices $i,j$ with $M_i\neq M_j$, i.e.  $t_i^\star,t_j^\star>0$. If $\mathbf{b}^\star$ also satisfied the planner's condition in \eqref{eq:reduced-stationarity}, dividing the two equations at each vertex would give
$a(t_i^\star)/\psi(t_i^\star) = (1+\lambda_s)/(1+\hat\lambda) = a(t_j^\star)/\psi(t_j^\star)$.
By \eqref{eq:rcsf-a-psi}, $a(t_i)/\psi(t_i) = \big[(m-1)f(t_i)+\delta\big]/\delta$, which is strictly increasing in $t_i$ because $f$ is strictly increasing. Hence $t_i^\star=t_j^\star$, and substituting back into \eqref{eq:reduced-stationarity}, using that $a$ is strictly decreasing and therefore injective (see the proof of \Cref{lem:rcsf-unique-symmetric}), gives $M_i=(1+\lambda_s)/a(t_i^\star)=(1+\lambda_s)/a(t_j^\star)=M_j$, contradicting $M_i\neq M_j$. So $\mathbf{b}^\star$ fails the planner's stationarity condition for every $\hat\lambda\geqslant0$, and by the concavity noted above it does not maximize $W_1$ even on the symmetric slice. Such instances exist. Take $m=2$, $f(b)=1-e^{-2b}$, $\delta=1.2$ (so $\delta>m-1$ and the CSF is a Regular RCSF), $C=0.5$, and the network of Table~\ref{tab:E}, whose centralities $M=(0.810,1.210,1.131,0.958,0.891)$ are pairwise distinct. The unique equilibrium is $\mathbf t^\star=(0.0468,0.1235,0.1095,0.0769,0.0636)$, active at all five consumers, with $W_1(\mathbf b^\star)=0.0994$, whereas $\max_{\mathcal B}W_1=0.1053$, attained at $\mathbf t^{\mathrm{opt}}=(0.0371,0.0957,0.0852,0.0604,0.0501)$. They are also the typical case, because equal centralities require a very special network. These instances alone therefore prove the theorem.

\emph{Case (ii): homogeneous centrality.} Suppose instead that all vertices carry the same centrality $M_i = M$. 
\begin{lemma}
\label{lem:sumM}
For every row-stochastic $\mathscr{E}$ and every $\alpha \in (0,1)$, $\sum_{i \in \mathscr{N}} M_i = n$ and $M_i\geqslant 1-\alpha$ for every $i\in\mathscr N$. 
\end{lemma}
\begin{proof}
Row-stochasticity gives $\mathscr{E}\mathbf{1} = \mathbf{1}$, hence $(I-\alpha\mathscr{E})^{-1}\mathbf{1} = \sum_{k\geqslant 0}\alpha^k\mathscr{E}^k\mathbf{1} = \bigl(\sum_{k \geqslant 0}\alpha^k\bigr)\mathbf{1} = (1-\alpha)^{-1}\mathbf{1}$. Therefore $\sum_i M_i = (1-\alpha)\,\mathbf{1}^\top(I-\alpha\mathscr{E})^{-1}\mathbf{1} = \mathbf{1}^\top\mathbf{1} = n$.
For the lower bound, note that every term $\alpha^k\mathscr E^k$ of the Neumann series has non-negative entries, so $(I-\alpha\mathscr E)^{-1}\geqslant I$ entrywise. Each column sum of $(I-\alpha\mathscr E)^{-1}$ is therefore at least $1$, and multiplying by $1-\alpha$ gives $M_i\geqslant 1-\alpha$.
\end{proof}

The above lemma forces $M_i=1$ for every $i$ (e.g. uniform complete graph). Then \eqref{eq:reduced-stationarity} does not by itself determine the relation between the firm and planner budget multipliers $\lambda_s$ and $\hat\lambda$, so the wedge could in principle be absorbed by their difference. We rule this out by exhibiting an instance whose optimum is interior $(t_i^{\mathrm{opt}}>0)$ and leaves the budget unexhausted $(\sum_it_i^{\mathrm{opt}}<C)$, which forces the budget multipliers $\hat\lambda=0$ and $\hat \mu_i = 0$. The planner's stationarity condition is therefore $M_i\psi(t_i)=1$, which reduces to $\psi(t)=1$ as $M_i = 1$ and all vertices hence have the same budget allocation $t_i = t$.

Take the saturating exponential generator $f(t)=1-e^{-\beta t}$ and choose parameters with
\begin{equation}
\label{eq:caseii-params}
m-1 < \delta < \beta, \qquad\text{and}\qquad n\,t^{\mathrm{opt}} < C,
\end{equation}
where $t^{\mathrm{opt}}$ is the symmetric welfare optimal budget which is same for all vertices $i$ (as defined later). The first chain of inequalities is consistent: $\delta>m-1$ makes the CSF a Regular RCSF, and $\delta<\beta=f'(0)$ is compatible with it. Now $\psi(t)=\delta f'(t)/(mf(t)+\delta)^2$ is continuous and strictly positive, with $\psi(0)=f'(0)/\delta=\beta/\delta>1$, and $\psi(t)\to0$ as $t \to\infty$ (if $f$ is bounded, concavity forces $f'\to0$; if $f$ is unbounded, $mf(t)+\delta\to\infty$). which implies $\psi(t) <1$ for all sufficiently large $t$. By the intermediate value theorem, there is some interior critical point $t^{\mathrm{opt}}\in(0,\infty)$ satisfying $\psi(t^{\mathrm{opt}})=1$, and $t^{\mathrm{opt}}\to0$ as $\beta\downarrow\delta$, so the second condition in \eqref{eq:caseii-params} can always be met by taking $\beta$ close enough to $\delta$. 
Since the welfare objective restricted to the symmetric slice is concave, the solution $\mathbf b^{\mathrm{opt}} := (t^{\mathrm{opt}},\ldots,t^{\mathrm{opt}})$ is the average welfare maximizer on that slice. By Step~1, symmetrization does not decrease average welfare, hence this symmetric planner optimum is also a maximizer of $W_1$ over the full feasible set $\mathcal B$.

We now show that this welfare optimum cannot simultaneously be a Nash equilibrium. Suppose, to the contrary, that $\mathbf b^\star=\mathbf b^{\mathrm{opt}}$ were a Nash equilibrium. Since $\sum_i t_i^{\mathrm{opt}} = n t^{\mathrm{opt}} < C$, the firm's budget is slack, so $\lambda_s=0$; and since $t^{\mathrm{opt}}>0$ the lower-bound multipliers vanish. Then \eqref{eq:reduced-stationarity} gives $a(t^{\mathrm{opt}})=1$ and $\psi(t^{\mathrm{opt}})=1$. Subtracting the two equations gives $a(t^{\mathrm{opt}})-\psi(t^{\mathrm{opt}}) = w(t^{\mathrm{opt}}) =0$, contradicting Step~3. So $\mathbf b^{\mathrm{opt}}$ is not a Nash equilibrium, and by uniqueness $\mathbf b^\star\neq\mathbf b^{\mathrm{opt}}$. Since $\mathbf{b}^{\mathrm{opt}}$ maximizes $W_1$ over $\mathcal{B}$, $\mathbf{b}^\star$ does not.

In either case $\mathbf{b}^\star$ fails to maximize $W_1$, which proves the theorem.
\hfill$\square$

\subsection{Proof of \texorpdfstring{\Cref{prop:alignment}}{Proposition 1}}
\label{app:alignment}

Let $(\lambda_s, \mu_{s,i})$ denote the multipliers of firm $s$'s best-response problem
$\max_{\mathbf{b}_s \in \mathcal{B}_s} u_s(\mathbf{b}_s, \mathbf{b}^{\mathrm{opt}}_{-s})$ and
$(\hat\lambda, \hat\mu_i)$ those of the planner's problem, both evaluated at the profile $\mathbf{b}^{\mathrm{opt}}$. Because all constraints are linear, the KKT conditions are necessary at both the equilibrium and the optimum. (As established in the proof of \Cref{thm:RCSF_impossibility} (\cref{app:RCSF_impossibility}), the planner's budget multipliers are equal at a symmetric profile and the KKT conditions are the same as in Step~2).

\emph{(i) Budget exhaustion suffices.} Substituting $\psi=ca$ into the planner's stationarity condition $M_i\psi(t_i^{\mathrm{opt}})-1=\hat\lambda-\hat\mu_{i}$ and adding and subtracting $c$ on the left gives
\[
c\big(M_ia(t_i^{\mathrm{opt}})-1\big)+(c-1) = \hat\lambda-\hat\mu_{i},
\]
so, dividing by $c>0$,
\begin{equation}
\label{eq:relabel}
M_ia(t_i^{\mathrm{opt}})-1 \;=\; \underbrace{\frac{\hat\lambda + (1-c)}{c}}_{=:\ \lambda_s} \;-\; \underbrace{\frac{\hat\mu_{i}}{c}}_{=:\ \mu_{s,i}}.
\end{equation}
This is exactly firm $s$'s stationarity condition $M_ia(t_i^{\mathrm{opt}})-1=\lambda_s-\mu_{s,i}$, with the multipliers relabelled. We check the remaining KKT requirements for the relabelled pair. Non-negativity holds: $\mu_{s,i}=\hat\mu_{i}/c\geqslant0$ since $c>0$, and $\lambda_s = (\hat\lambda+1-c)/c > 0$ since $\hat\lambda \geqslant 0$ and $c<1$. Complementary slackness on the lower bound is inherited, since $\mu_{s,i}t_i^{\mathrm{opt}} = \hat\mu_i t_i^{\mathrm{opt}}/c = 0$. The only condition left is complementary slackness on the budget, $\lambda_s\bigl(C-\sum_i t^{\mathrm{opt}}_{i}\bigr)=0$. Because $c<1$ forces $\lambda_s>0$ strictly, this holds precisely because the budget is exhausted by hypothesis, i.e. $\sum_i t^{\mathrm{opt}}_{i}=C$. So $\mathbf{b}^{\mathrm{opt}}$ satisfies every firm's KKT system, and concavity of $u_s$ in $\mathbf{b}_s$ makes these conditions sufficient, so
$\mathbf{b}^{\mathrm{opt}}$ is a best response for every firm, i.e.\ a Nash equilibrium.

\emph{(ii) Budget exhaustion is necessary when the optimum is nonzero.} Suppose
$\mathbf{b}^{\mathrm{opt}} \neq \mathbf{0}$ is a Nash equilibrium and, for contradiction, that
$\sum_i t^{\mathrm{opt}}_i < C$. Complementary slackness on the budget then forces
$\lambda_s = 0$ in every firm's system and $\hat\lambda = 0$ in the planner's. Since
$\mathbf{b}^{\mathrm{opt}} \neq \mathbf{0}$, there is a vertex $j$ with $t^{\mathrm{opt}}_j > 0$,
and complementary slackness on the lower bound gives $\mu_{s,j} = \hat\mu_j = 0$. The two
stationarity conditions at $j$ therefore reduce to
$M_j\, a(t^{\mathrm{opt}}_j) = 1$ and $M_j\, \psi(t^{\mathrm{opt}}_j) = 1$, implying
$\psi(t^{\mathrm{opt}}_j) = a(t^{\mathrm{opt}}_j)$. By \eqref{eq:proportional} this forces
$c = 1$, contradicting $c < 1$.

The hypothesis $\mathbf{b}^{\mathrm{opt}} \neq \mathbf{0}$ is needed because the all-zero profile is not ruled out by the proportional-marginals condition alone: if $t^{\mathrm{opt}}_i = 0$ at every vertex, the two stationarity conditions for the firm and planner become $M_i a(0) - 1 = -\mu_{s,i} \implies \mu_{s,i} = 1 - M_i a(0)$ and $M_i \psi(0) - 1 = -\hat\mu_i \implies \hat\mu_i = 1 - M_i c a(0)$ respectively. So whenever $M_i a(0) \leqslant 1$ for every $i$, both multipliers $\mu_{s,i}$ and $\hat\mu_i$ are non-negative and both complementary-slackness conditions $\mu_{s,i}t_i^{\mathrm{opt}} = \hat\mu_i t_i^{\mathrm{opt}} = 0$ hold trivially. The all-zero profile can then be a Nash
equilibrium and a welfare optimum simultaneously with non-exhausted budgets, so no contradiction arises.
\hfill$\square$

\subsection{Proof of \texorpdfstring{\Cref{lem:isoelastic}}{Lemma 5}}
\label{app:isoelastic}

Dividing numerator and denominator of \eqref{eq:c_PRCSF} by $f(B_i)$ gives
\begin{equation}
\label{eq:c-elasticity}
c(B_i) = \frac{m\,e(B_i)}{e(B_i)+m-1}.
\end{equation}
The map $x\mapsto mx/(x+m-1)$ is strictly increasing on $x>0$, hence injective, so $c(B_i)$ is constant in $B_i$ if and only if $e(B_i)$ is constant. For finding a function with constant elasticity $\eta$, i.e. $e(B_i)\equiv\eta$, solving the differential equation $f'(B_i)/f(B_i)=\eta/B_i$ gives the general solution $f(B_i)=\kappa B_i^\eta$; the requirements that $f$ be increasing with $f(0)=0$ force $\eta>0$, and the requirement in \Cref{def:srcsf} that $f$ be concave forces $\eta\leqslant1$, since $f''(B_i)=\kappa\eta(\eta-1)B_i^{\eta-2}$. Conversely $f(B_i)=\kappa B_i^\eta$ has $e(B_i)\equiv\eta$ identically. 
\hfill$\square$

\subsection{Proof of \texorpdfstring{\Cref{prop:power-align}}{Proposition 2}}
\label{app:power-align}

For the Power RCSF, the utility of a firm $s$ is concave in $\mathbf b_s$. This can be verified by the second derivative of firm $s$'s utility with respect to $b_{s,i}$, which is $M_i\kappa(\eta-1)B_i^{\eta-3}
\bigl(2\sum_{r\neq s}b_{r,i}+\eta b_{s,i}\bigr)<0$ for $\eta\in(0,1)$ and $B_i>0$. 

Since $f(B_i)=\kappa B_i^\eta$ is iso-elastic, i.e. $e(B_i)\equiv\eta$, by \Cref{lem:isoelastic}, the Power RCSF has proportional marginals. Hence, \eqref{eq:c-elasticity} gives $c=m\eta/(\eta+m-1)$, a constant.  Since $\eta<1$ we have $\eta(m-1)<m-1$, i.e.\ $m\eta<m+\eta-1$, so $c<1$; and $c>0$ since $m\geqslant2$ and $\eta>0$. Explicitly, evaluating \eqref{eq:srcsf-a-psi} at $B_i=mt_i$ gives $a(t_i) = \kappa m^{\eta-2}(m+\eta-1)\,t_i^{\eta-1}$ and $\psi(t_i)=\kappa\eta\,(mt_i)^{\eta-1}$.
\hfill$\square$

\subsection{Proof of \texorpdfstring{\Cref{prop:budget_bind_cond}}{Proposition 3}}
\label{app:budget_bind_cond}

\ref{bb:i} For the power RCSF, $\sum_s h_{s,i}=\kappa B_i^\eta$, so utilitarian welfare $W_1$ depends on $\mathbf{b}$ only through the total budget $B_i=\sum_sb_{s,i}$:
\[
W_1(\mathbf{b}) = \frac{1}{m}\sum_s u_s(\mathbf{b}) = \frac{1}{m} \Big[\kappa\sum_iM_iB_i^{\eta}-\sum_iB_i \Big],
\]
and the attainable totals over $\mathcal{B}$ are exactly $\{\mathbf{B}\geqslant\mathbf 0:\sum_iB_i\leqslant mC\}$ where any such $\mathbf B$ is realized by the symmetric split $t_{i}=B_i/m$. This objective is separable and strictly concave in $\mathbf B$, with unconstrained maximizer given by solving the planner's unconstrained first-order condition $M_i \psi(t_i) = M_i\kappa\eta B_i^{\eta-1}=1$, giving $B_i^{\mathrm{opt}}=(\kappa\eta M_i)^{\theta}$.
Hence at the welfare optimum, the full budget is exhausted if and only if the unconstrained maximizer requires at least the available aggregate budget, i.e., $\sum_i B_i^{\mathrm{opt}}\geqslant mC \implies \sum_{i}\bigl(\kappa\eta M_i\bigr)^{\theta}  \geqslant mC$ which is \eqref{eq:budget_bind_cond}.
In this case the constrained welfare optimum satisfies
$\sum_i B_i=mC$, which at a symmetric allocation is equivalent to $\sum_i t_{i} = C$ for every firm $s$.

\ref{bb:ii} Substituting the largest admissible $\kappa=(mC)^{-\eta}$ into \eqref{eq:budget_bind_cond} gives
$\eta^{\theta}(mC)^{-\eta\theta}\sum_iM_i^{\theta}\geqslant mC$. Since $1+\eta\theta = 1 + \eta/(1-\eta) = \theta$, this rearranges to $\eta^{\theta}\sum_iM_i^{\theta}\geqslant (mC)^{\theta}$, and taking $\theta^{\text{th}}$ roots gives $\eta\|M\|_\theta\geqslant mC$. Any smaller $\kappa$ makes the left side of \eqref{eq:budget_bind_cond} smaller, so this is necessary for every admissible $\kappa$.

\ref{bb:iii} For $\theta\geqslant1$ the $\ell^\theta$ norm is dominated by the $\ell^1$ norm, so $\|M\|_{\theta}\leqslant\sum_iM_i=n$ by \Cref{lem:sumM}. With $\eta<1$ this gives $\eta\|M\|_\theta< n \leqslant mC$, contradicting \ref{bb:ii}. It remains to explain why this rules out alignment. By part \ref{bb:i}, the symmetric welfare-maximizing profile then leaves every budget unexhausted. Moreover, since $M_i \geqslant 1-\alpha > 0$ (\cref{lem:sumM}) and the planner's marginal return $M_i\kappa\eta B_i^{\eta-1}$ grows without bound as $B_i\to0$, moving a small amount of budget to any vertex with $B_i=0$ strictly increases $W_1$, so $t_i^{\mathrm{opt}}>0$ on every vertex $i$, i.e. $\mathbf{b}^{\mathrm{opt}} \neq 0$. \Cref{prop:alignment}(ii) therefore applies, and it shows that this profile cannot be a Nash equilibrium.
\hfill$\square$

\subsection{Proof of \texorpdfstring{\Cref{prop:tax-alignment}}{Proposition 4}}
\label{app:tax-alignment}

Under the taxed utilities, the firm's first-order condition (FOC) becomes
\begin{equation}
\label{eq:firm-foc-tau}
M_ia(t_i) - (1+\tau) = \lambda_s - \mu_{s,i}.
\end{equation}
whereas the planner's FOC remains unchanged:
\begin{equation}
\label{eq:planner-foc}
M_i\psi(t_i) - 1 = \hat\lambda - \hat\mu_{i}.
\end{equation}
Let $\mathbf{b}^{\mathrm{opt}}$ be a welfare-maximizing symmetric profile with common vertex values
$b^{\mathrm{opt}}_{s,i} = t^{\mathrm{opt}}_i$ for every firm $s \in \mathscr{M}$. Substituting $\psi=ca$ in the planner's FOC \eqref{eq:planner-foc} and dividing by $c$,
\[
M_ia(t_i^{\mathrm{opt}}) = \frac{1+\hat\lambda-\hat\mu_i}{c} = \frac1c + \frac{\hat\lambda}{c} - \frac{\hat\mu_i}{c} = (1+\tau) + \lambda_s - \mu_{s,i},
\qquad \lambda_s := \frac{\hat\lambda}{c},\quad \mu_{s,i}:=\frac{\hat\mu_i}{c},
\]
using $1+\tau=1/c$. We obtain exactly the firm's FOC \eqref{eq:firm-foc-tau} with the relabeled multipliers $\lambda_s$ and $\mu_{s,i}$. These multipliers are non-negative because $c>0$, and both complementary slackness conditions are inherited directly from the planner's, since $\mu_{s,i}t_i^{\mathrm{opt}}=\hat\mu_it_i^{\mathrm{opt}}/c=0$ and $\lambda_s(C-\sum_it_i^{\mathrm{opt}})=\hat\lambda(C-\sum_it_i^{\mathrm{opt}})/c=0$. Since $\widetilde{u}_s$ is concave in $\mathbf{b}_s$, KKT is necessary and sufficient, so $\mathbf{b}^{\mathrm{opt}}$ is a best response for every firm.
\hfill$\square$

\subsection{Proof of \texorpdfstring{\Cref{lem:power-unique-sym}}{Lemma 8}: uniqueness and symmetry for the Power RCSF}
\label{app:power-unique}

\emph{Part I: The equilibrium is always symmetric.}\\
Throughout, firms face the common unit charge $1+\tau\geqslant1$, so utilities are $\widetilde u_s$ as in \eqref{eq:pigouvian_utility}. 
For the Power RCSF, 
\[
\frac{\partial^2h_{s,i}}{\partial b_{s,i}^2} = \kappa(\eta-1)B_i^{\eta-3}\bigl[2B_i+(\eta-2)b_{s,i}\bigr] = \kappa(\eta-1)B_i^{\eta-2}\bigl[2+(\eta-2) \frac{b_{s,i}}{B_i}\bigr]
\]
Since $\eta-2\in(-2,-1)$ and $\frac{b_{s,i}}{B_i}\leqslant1$, the bracket is at least $2+(\eta-2)=\eta>0$, and $\kappa(\eta-1)<0$, so the second derivative is strictly negative everywhere. Hence $\widetilde u_s$ is strictly concave in $\mathbf b_s$ for any fixed $\mathbf b_{-s}$. Hence the game admits a pure strategy symmetric equilibrium because it is a symmetric game with compact and convex strategy spaces and continuous and strictly-concave utility function \citep{Cheng2004}. Now it remains to show that the game doesn't have any asymmetric equilibria. 

This proof follows similar arguments as in the proof of \cref{lem:rcsf-unique-symmetric} in \cref{app:RRCSF_sym_unique} establishing the uniqueness and symmetry of the equilibrium under the Regular RCSF. For the Power RCSF, we derive the same \emph{marginal dominance (MD) inequality} as follows:

Fix a vertex $i$ and firms $p,q$, where $b_{p,i} = x$, $b_{q,i} = y$ with $0 \leqslant y < x \leqslant C$ and $S=\sum_{r\neq p,q}b_{r,i}$, so $B_i=x+y+S$. By \eqref{eq:power-derivs}, $\partial h_{p,i}/\partial x = \kappa B_i^{\eta-2}[B_i-(1-\eta)x]$ and $\partial h_{q,i}/\partial y = \kappa B_i^{\eta-2}[B_i-(1-\eta)y]$. Subtracting the two, we get
\[
\frac{\partial h_{p,i}}{\partial x} - \frac{\partial h_{q,i}}{\partial y}
= \kappa B_i^{\eta-2}(1-\eta)(y-x) \;<\; 0 ,
\]
since $\kappa B_i^{\eta-2}>0$ and $1-\eta>0$. Thus, 
\[
\frac{\partial h_{p,i}}{\partial b_{p,i}} \;<\; \frac{\partial h_{q,i}}{\partial b_{q,i}}, \qquad \text{for } b_{p,i} > b_{q,i}.
\]
which is exactly \eqref{eq:MD} for the Power RCSF.

Since the taxed utility \eqref{eq:pigouvian_utility} is strictly concave in its own strategy, KKT conditions are necessary and sufficient. Since the rebate $\rho_s$ does not depend on $\mathbf b_s$, the KKT conditions are:
\begin{align}
&\text{Stationarity:} & M_i \frac{\partial h_{s,i}(\mathbf{b})}{\partial b_{s,i}} &= 1 + \tau + \lambda_s - \mu_{s,i}, \tag{S}\label{eq:S_2}\\
&\text{Dual feasibility:} & \mu_{s,i} &\geqslant 0, \quad \lambda_s \geqslant 0, \tag{D}\label{eq:D_2}\\
&\text{Complementary slackness:} & \mu_{s,i}b_{s,i} &= 0, \quad \lambda_s\Big(C - \sum_i b_{s,i}\Big) = 0. \tag{CS}\label{eq:CS_2}
\end{align}

Now we can write the same arguments as in \cref{app:RRCSF_sym_unique} and produce the same contradictions on the budget multipliers in order to rule out all asymmetric equilibria as those arguments use the \emph{MD inequality} \eqref{eq:MD} which has been derived here too. 

Suppose $\mathbf{b}^\star$ is an asymmetric PSNE, i.e.\ $\mathbf{b}_p^\star \neq \mathbf{b}_q^\star$ for some two firms $p,q \in \mathscr{M}$. 
Suppose $b^\star_{p,i_0} > b^\star_{q,i_0} \geqslant 0$ at some vertex $i_0$. Since $b^\star_{p,i_0} > 0$, \eqref{eq:CS_2} gives $\mu^\star_{p,i_0} = 0$, and \eqref{eq:S} gives
\begin{equation}
M_{i_0} \frac{\partial h_{p,i_0}}{\partial b_{p,i_0}}(\mathbf{b}^\star) = 1 + \tau + \lambda_p.
\label{eq:c1p}
\end{equation}
For firm $q$, \eqref{eq:CS_2} gives $\mu_{q,i_0} \geqslant 0$, so \eqref{eq:S_2} gives
\begin{equation}
M_{i_0} \frac{\partial h_{q,i_0}}{\partial b_{q,i_0}}(\mathbf{b}^\star) = 1 + \tau + \lambda_q - \mu_{q,i_0} \leqslant 1 + \tau + \lambda_q.
\label{eq:c2p}
\end{equation}
By the marginal dominance inequality \eqref{eq:MD} at $i_0$ (the other $m-2$ firms do not matter),
\begin{equation}
\frac{\partial h_{p,i_0}}{\partial b_{p,i_0}}(\mathbf{b}^\star) < \frac{\partial h_{q,i_0}}{\partial b_{q,i_0}}(\mathbf{b}^\star).
\label{eq:c3p}
\end{equation}
Combining \eqref{eq:c1p}, \eqref{eq:c2p}, \eqref{eq:c3p},
\begin{equation}
1 + \tau + \lambda_p < 1 + \tau + \lambda_q \implies \lambda_p < \lambda_q.
\label{eq:c4p}
\end{equation}
That is: whoever bids strictly more at some vertex has a strictly smaller budget multiplier.

\smallskip
\noindent \textit{Case A: There exist two vertices $ i, j$ with $b^\star_{p,i} > b^\star_{q,i}$ and $b^\star_{q,j} > b^\star_{p,j}$}. This case arises whenever the two firms have equal total spend, i.e.\ $\sum_i b^\star_{p,i} = \sum_i b^\star_{q,i}$.  
Applying \eqref{eq:c4p} at vertex $i$ gives $\lambda_p < \lambda_q$; applying it at vertex $j$ (roles reversed) gives $\lambda_q < \lambda_p$. Contradiction.

\smallskip
\noindent\textit{Case B: $b^\star_{p,i} \geqslant b^\star_{q,i}\ \forall i$, with strict inequality at some $i_0$}.
Then $\sum_i b^\star_{p,i} > \sum_i b^\star_{q,i}$ (unequal total spend). At $i_0$, \eqref{eq:c4p} gives $\lambda_p < \lambda_q$.
\begin{itemize}
    \item If $\sum_i b^\star_{p,i} = C$, then $\sum_i b^\star_{q,i} < C$, so \eqref{eq:CS_2} gives $\lambda_q = 0$. Combined with $\lambda_p < \lambda_q = 0$, this forces $\lambda_p < 0$, violating \eqref{eq:D_2}.
    \item If $\sum_i b^\star_{p,i} < C$, then \eqref{eq:CS_2} gives $\lambda_p = 0$. Then $\lambda_p < \lambda_q$ forces $\lambda_q > 0$, which by \eqref{eq:CS_2} forces $\sum_i b^\star_{q,i} = C$. But this implies $\sum_i b^\star_{q,i} > \sum_i b^\star_{p,i}$, contradicting the case hypothesis.
\end{itemize}

\noindent
Hence, in every case, no asymmetric NE is possible. Therefore every NE is symmetric.

\medskip
\noindent\emph{Part II: uniqueness of the symmetric equilibrium.} \\
At a symmetric profile with $B_i=mt_i$, \eqref{eq:power-derivs} gives the closed form for the private marginal return
\begin{equation}
a(t_i) = \kappa\,m^{\eta-2}(m+\eta-1)\,t_i^{\eta-1} = K\,t_i^{\eta-1}, 
\label{eq:Phi-closed-form}
\end{equation}
where $K = \kappa\,m^{\eta-2}(m+\eta-1) > 0$. So $a'(t_i) = K(\eta-1)t_i^{\eta-2}<0$ since $\eta < 1$ and hence $a(t_i)$ is strictly decreasing on $t_i \in (0, \infty)$ with $a(t_i)\to+\infty$ as $t_i\to0^+$.

Rewriting the KKT conditions for a symmetric equilibrium profile $\mathbf{t}^\star = (t_i^\star)_{i \in \mathscr{N}}$ with common multipliers $\lambda \geqslant 0$ and $\mu_i \geqslant 0$ (shared across firms by symmetry) :
\[
M_i\, a(t_i^\star) = 1 + \tau + \lambda - \mu_i 
\]
If $t_i^\star > 0$, then by complementary slackness, $\mu_i = 0$, the KKT becomes $M_i\, a(t_i^\star) = \lambda + 1 + \tau =: \Gamma$. If $t_i^\star = 0$, then $\mu_i \geqslant 0$ and we have $M_i\, a(0^+) = (\lambda+1+\tau) - \mu_i \leqslant \Gamma$. 
Note that since $a(t)$ is only defined for $t>0$, we write $a(0^+) := \lim_{t_i^\star \downarrow 0} a(t_i^\star) = +\infty$. Hence for any vertex with $M_i>0$, $M_i a(0^+)=+\infty$ and at the boundary solution $t_i^\star=0$, the KKT condition which requires $M_i a(0^+) \leqslant \Gamma$ cannot hold for any finite $\Gamma$. Thus the symmetric solution $t_i^\star > 0$ always. This simplifies the argument relative to the RCSF case (\Cref{app:RRCSF_sym_unique}), where $a(0)$ is finite and corner solutions ($t_i^\star=0$) are possible.

Since $a$ is continuous and strictly decreasing, so is $M_i\, a(\cdot)$ for each fixed $M_i > 0$, and it is therefore invertible. Let $\theta := 1/(1-\eta)$. Inverting \eqref{eq:Phi-closed-form} to obtain the symmetric equilibrium solution $t_i^\star$,
\[
t_i^\star = \beta_i(\Gamma) := a^{-1}\Bigl(\frac{\Gamma}{M_i}\Bigr) = \Bigl(\frac{M_iK}{\Gamma}\Bigr)^{\theta}, \qquad \Gamma>0,
\]
which is single-valued and strictly decreasing in $\Gamma$ because $\theta>0$. This proves that for each $\Gamma>0$, the budget $t_i^\star$ is unique.
Consequently the aggregate demand for budget, $S(\Gamma)=\sum_i\beta_i(\Gamma)$ is continuous and strictly decreasing on all of $\Gamma \in (0,\infty)$, with $S(\Gamma)\to0$ as $\Gamma\to\infty$. Since the total budget constraint is $\sum_i t_i^\star \leqslant C$, two cases arise considering the smallest possible $\Gamma = 1 + \tau$:

\begin{enumerate}
    \item[(1)] $S(1 + \tau) \leqslant C$: the unconstrained budget at the lowest possible marginal threshold $\Gamma = 1 + \tau$ already satisfies the constraint, thus the budget multiplier is inactive, i.e. $\lambda = 0$. Thus there is a unique $\Gamma^\star = 1 + \tau$, and $t_i^\star = \beta_i(1 + \tau)$ is the unique solution.
    \item[(2)] $S(1 + \tau) > C$: the budget constraint is active, $\lambda > 0$. Since $\lim_{\Gamma\to\infty}S(\Gamma)=0$, there exists a finite $\Gamma_1>1 + \tau$ such that $S(\Gamma_1)<C$. Since $S$ is continuous on $(0,\infty)$, it is continuous on the closed interval $[1 + \tau,\Gamma_1]$ with $S(1 + \tau)>C>S(\Gamma_1)$. The Intermediate Value Theorem then implies that there exists a $\Gamma^\star > 1 + \tau$ with $S(\Gamma^\star) = C$, and strict monotonicity implies that the $\Gamma^\star$ is unique as a strictly monotone continuous function crosses any given level exactly once. Thus $t_i^\star = \beta_i(\Gamma^\star)$.
\end{enumerate}

The two cases are mutually exclusive and exhaustive, determined by the sign of $S(1 + \tau)-C$, and each yields at most one candidate. In both, $S(\Gamma^\star)=\min\{S(1+\tau),C\}$. Combined with \textit{Part I}, the equilibrium is unique and symmetric.
\hfill$\square$

\subsection{Proof of \texorpdfstring{\Cref{prop:unbounded}}{Proposition 5}}
\label{app:unbounded}

Fix $i$ and let $\mathbf{b}^{(B_i)}\in\mathcal{B}$ place $b_{s,i}=B_i/m$ at $i$ for every firm $s$, distributing each firm's remaining budget $C-B_i/m$ over the other $n-1\geqslant1$ vertices; this is feasible for every $B_i\in(0,mC]$. Differentiating \eqref{eq:power_rcsf} twice gives
\begin{equation}
\label{eq:power-second-derivs}
\frac{\partial^2 h_{s,i}}{\partial b_{s,i}^2} = \kappa(\eta-1)B_i^{\eta-3}\bigl[2B_i+(\eta-2)b_{s,i}\bigr],
\qquad
\frac{\partial^2 h_{s,i}}{\partial b_{r,i}\,\partial b_{s,i}} = \kappa(\eta-1)B_i^{\eta-3}\bigl[B_i+(\eta-2)b_{s,i}\bigr].
\end{equation}
At $\mathbf{b}^{(B_i)}$, where $b_{s,i}=B_i/m$, the diagonal entry of block $H_i$ has magnitude:
\[
|(H_i)_{ss}| = M_i\kappa(1-\eta)\frac{2m+\eta-2}{m}B_i^{\eta-2}
\]
and $2m+\eta-2>0$ for $m\geqslant2$ and $\eta\in(0,1)$. Since $\|H_i\|_{\mathrm{op}}\geqslant|\mathbf{e}_s^\top H_i\mathbf{e}_s|=|(H_i)_{ss}|$ for the $s$th standard basis vector $\mathbf e_s$, we get
\[
\|H_i\|_{\mathrm{op}}\geqslant M_i\kappa(1-\eta)\,\tfrac{2m+\eta-2}{m}\,B_i^{\eta-2},
\]
and this operator norm diverges as $B_i\to0^+$ because $\eta-2<0$. By the reduction of SB property to vertex-wise blocks $H_i$ in \cref{app:reduction}, the unboundedness of $\|H_i\|_{\mathrm{op}}$ implies that the bounded-operator-norm condition in \Cref{def:SB} cannot hold for any finite $L$.
\hfill$\square$

\subsection{Proof of \texorpdfstring{\Cref{lem:explicit-floor}}{Lemma 7}}
\label{app:explicit-floor}

Let
\[
  \bar\varepsilon_1 := \Bigl(\frac{K\min_iM_i}{1+\tau}\Bigr)^{\theta},
  \qquad
  \bar\varepsilon_2 := C\,\frac{\min_iM_i^{\theta}}{\sum_jM_j^{\theta}},
\]
so that $\bar\varepsilon = \min\{\bar\varepsilon_1,\bar\varepsilon_2\}$ by definition. 
Recall from \Cref{lem:power-unique-sym} that $t_i^\star=\beta_i(\Gamma^\star)$, where
$\beta_i(\Gamma):=(M_iK/\Gamma)^{\theta}$, and $\Gamma^\star$ is the unique value with
$S(\Gamma^\star)=\min\{S(1+\tau),C\}$, $S(\Gamma):=\sum_i\beta_i(\Gamma)$. 
We show $\min_i t_i^\star = \bar\varepsilon$ by identifying, in each of the two cases of $S(\Gamma)$ in the proof of \Cref{lem:power-unique-sym} (\cref{app:power-unique}), which of $\bar\varepsilon_1,\bar\varepsilon_2$ equals $\min_i t_i^\star$ exactly, and then verifying that this
value is always the smaller of the two, so that taking the minimum of the two constant expressions recovers
the correct one in either case.

Since $\beta_i(\Gamma)=(K/\Gamma)^{\theta}M_i^{\theta}$ for every $i$, the factor
$(K/\Gamma)^{\theta}$ cancels in the ratio $\beta_i(\Gamma)/S(\Gamma)=M_i^{\theta}/\sum_jM_j^{\theta}$,
which is therefore independent of $\Gamma$; multiplying through by $S(\Gamma)$ and taking the minimum
over $i$ gives, for every $\Gamma>0$,
\begin{equation}
\label{eq:min-beta-ratio}
  \min_i\beta_i(\Gamma) \;=\; S(\Gamma)\,\frac{\min_iM_i^{\theta}}{\sum_jM_j^{\theta}}.
\end{equation}

\emph{Case (i): $S(1+\tau)\leqslant C$.} Here $\Gamma^\star=1+\tau$, so applying
\eqref{eq:min-beta-ratio} with $S(\Gamma) = (K/\Gamma)^{\theta}\sum_jM_j^{\theta}$ at $\Gamma=1+\tau$,
\[
  \min_it_i^\star \;=\; \min_i\beta_i(1+\tau)
  \;=\; \Bigl(\frac{K}{1+\tau}\Bigr)^{\theta}\min_iM_i^{\theta} \\
  \;=\; \Bigl(\frac{K\min_iM_i}{1+\tau}\Bigr)^{\theta} \;=\; \bar\varepsilon_1.
\]
Because $\theta=1/(1-\eta)>0$, the map $x\mapsto x^{\theta}$
is increasing on $x\geqslant 0$ which implies $\min_i M_i^\theta = (\min_i  M_i)^\theta$.

We check that $\bar\varepsilon_1\leqslant\bar\varepsilon_2$ in this regime. Writing $\bar\varepsilon_1 = (K/(1+\tau))^{\theta}\min_iM_i^{\theta}$ and using
$S(1+\tau)=(K/(1+\tau))^{\theta}\sum_jM_j^{\theta}\leqslant C$, we get
$(K/(1+\tau))^{\theta}\leqslant C/\sum_jM_j^{\theta}$. Multiplying both sides by
$\min_iM_i^{\theta}\geqslant 0$ preserves the inequality:
\[
  \bar\varepsilon_1 \;=\; \Bigl(\frac{K}{1+\tau}\Bigr)^{\theta}\min_iM_i^{\theta}
  \;\leqslant\; C\,\frac{\min_iM_i^{\theta}}{\sum_jM_j^{\theta}} \;=\; \bar\varepsilon_2.
\]
Hence $\min\{\bar\varepsilon_1,\bar\varepsilon_2\}=\bar\varepsilon_1=\min_it_i^\star$ throughout this regime.

\emph{Case (ii): $S(1+\tau)>C$.} Here $\Gamma^\star$ is the value with $S(\Gamma^\star)=C$, so applying
\eqref{eq:min-beta-ratio} at $\Gamma=\Gamma^\star$,
\[
  \min_it_i^\star \;=\; \min_i\beta_i(\Gamma^\star)
  \;=\; S(\Gamma^\star)\,\frac{\min_iM_i^{\theta}}{\sum_jM_j^{\theta}}
  \;=\; C\,\frac{\min_iM_i^{\theta}}{\sum_jM_j^{\theta}} \;=\; \bar\varepsilon_2.
\]
We check that $\bar\varepsilon_2<\bar\varepsilon_1$ in this regime. From $S(1+\tau)>C$, i.e.
$(K/(1+\tau))^{\theta}\sum_jM_j^{\theta}>C$, we get $(K/(1+\tau))^{\theta}>C/\sum_jM_j^{\theta}$. Multiplying
both sides of this inequality by $\min_iM_i^{\theta}>0$ (since every vertex $i$ has $M_i\geqslant 1-\alpha >0$ (\cref{lem:sumM})) gives
\[
  \bar\varepsilon_1 \;=\; \Bigl(\frac{K}{1+\tau}\Bigr)^{\theta}\min_iM_i^{\theta}
  \;>\; C\,\frac{\min_iM_i^{\theta}}{\sum_jM_j^{\theta}} \;=\; \bar\varepsilon_2.
\]
Hence $\min\{\bar\varepsilon_1,\bar\varepsilon_2\}=\bar\varepsilon_2=\min_it_i^\star$ throughout this regime
as well.

Since Cases (i) and (ii) are exhaustive and in each of them $\min_it_i^\star$ coincides with the smaller
of $\bar\varepsilon_1,\bar\varepsilon_2$, we conclude that, regardless of which case actually holds,
\[
  \min_it_i^\star \;=\; \min\{\bar\varepsilon_1,\bar\varepsilon_2\} \;=\; \bar\varepsilon.
\]
Moreover $\bar\varepsilon_1,\bar\varepsilon_2>0$,
since $K,\min_iM_i,C>0$, so $\bar\varepsilon=\min\{\bar\varepsilon_1,\bar\varepsilon_2\}>0$.

Now let $\varepsilon$ satisfy $0<\varepsilon<\bar\varepsilon$. Since $\bar\varepsilon=\min_it_i^\star$,
every coordinate satisfies $t_i^\star\geqslant\bar\varepsilon>\varepsilon$, so the floor is inactive at the equilibrium for every such $\varepsilon$. 
Finally, $\min_iM_i^{\theta}/\sum_jM_j^{\theta}\leqslant1/n$ (from \cref{lem:sumM}, the smallest of $n$ non-negative numbers never exceeds their average, so $\min_iM_i^\theta\leqslant\frac1n\sum_jM_j^\theta$), so $\bar\varepsilon\leqslant C/n$ and $n\varepsilon<n\bar\varepsilon\leqslant C$. Hence
$\mathcal{B}^{\varepsilon}$ is non-empty.
\hfill$\square$

\subsection{Why is a participation floor needed?}
\label{app:floor-why}

\Cref{prop:unbounded} is a statement about the Power RCSF family \eqref{def:power_rcsf} and the SB-based convergence argument used here, rather than a general impossibility result for two reasons: the SB property (\Cref{def:SB}) is sufficient but not necessary for BRD convergence, and \Cref{lem:isoelastic} characterizes the alignment condition only within the SRCSF class \eqref{eq:SRCSF}.

In practice, there are three ways to address the unbounded curvature as $B_i\to0$. First, one could modify the
CSF near zero: since the equilibrium and welfare optimum lie strictly above $\bar\varepsilon$ (\Cref{lem:explicit-floor}), one could choose $B_0<m\bar\varepsilon$, retain the iso-elastic form on $[B_0,mC]$, and replace $f(B_i)$ on $[0,B_0)$ by a function with bounded second derivative, as is common for smoothing a contest success function near zero
\citet{ghosh2023best,elkind2024continuous}.
Second, one could retain the Power RCSF and replace the SB-based convergence argument with a method that
tolerates unbounded curvature. 
Third, one could restrict the strategy space so that $B_i$ is bounded away from zero, which is what the participation floor does.

Within the SRCSF class and the SB-based BRD convergence framework, exact iso-elasticity on the whole feasible set and globally bounded curvature cannot hold together, so one of the three must be chosen. We choose the floor because it leaves the Power RCSF unchanged and, for $\varepsilon<\bar\varepsilon$, it is inactive at the solution. Because the corrective charge aligns the firms' and the planner's KKT systems whether or not budgets are exhausted (\Cref{prop:tax-alignment}), the choice of $\varepsilon$ need not depend on which case of \eqref{eq:epsbar} applies. The floor's only role is to restore the SB property, and \Cref{lem:explicit-floor} confirms that it does so without disturbing the solution.

\subsection{Proof of \texorpdfstring{\Cref{lem:power-SB}}{Lemma 9}: the Power RCSF satisfies SB on the floored set}
\label{app:power-SB}

By \Cref{app:reduction} it suffices to establish \eqref{eq:blockwise-SB} for each vertex block $H_i$, and the linear cost term (with or without the charge) does not affect $H_i$.

\medskip
\emph{Structure.} Direct differentiation of the Power RCSF \eqref{eq:power_rcsf} gives (from \eqref{eq:power-second-derivs}), 
\[
 (H_i)_{ss}=M_i\kappa(\eta-1)B_i^{\eta-3}\,[2B_i+(\eta-2)b_{s,i}]=c_0\,(v_s+B_i) 
\]
\[
(H_i)_{sr}=M_i\kappa(\eta-1)B_i^{\eta-3}\,[B_i+(\eta-2)b_{s,i}]=c_0v_s \;\; \text{for } r \neq s
\]
where
$c_0:=M_i\kappa(\eta-1)B_i^{\eta-3}<0$ and $v_s:=B_i+(\eta-2)b_{s,i}$. Hence the vertex-wise Hessian block can be written in the matrix form as $H_i=c_0(\mathbf{v}\mathbf{1}^\top+B_iI_m)$. 

\medskip
\emph{Strong concavity.} $\tfrac12(H_i+H_i^\top)=c_0 \mathsf M$ with $\mathsf M:=B_iI_m+\tfrac12(\mathbf{v}\mathbf{1}^\top+\mathbf{1}\mathbf{v}^\top)$. Since $c_0<0$, it
suffices to show $\mathsf M\succeq\chi(m, \eta) B_i\,I_m$. 
The symmetric matrix $\tfrac12(\mathbf v\mathbf1^\top+\mathbf1 \mathbf v^\top)$ has rank at
most two (since Rank$(\mathbf v\mathbf1^\top)$ = Rank$(\mathbf1 \mathbf v^\top) = 1$), so it vanishes on the orthogonal complement of $\operatorname{span}\{\mathbf 1,\mathbf v\}$. Hence it suffices to analyze its restriction to $\operatorname{span}\{\mathbf 1,\mathbf v\}$ on which it has eigenvalues
$\mu_\pm=\tfrac12\big(\mathbf1^\top \mathbf v\pm\sqrt m\,\|\mathbf v\|_2\big)$ and its remaining eigenvalues are zero, with multiplicity at least $m-2$. By
Cauchy--Schwarz inequality, $\mathbf1^\top \mathbf v \leqslant \sqrt{m} \;\|\mathbf v \|$, which implies $\mu_-\leqslant 0$, so the minimum eigenvalue of matrix $\mathsf M$ is $\lambda_{\min}(\mathsf M)=B_i+\mu_-$. Next we have to find a lower bound on $\mu_-$ so that we can bound $\lambda_{\min}(\mathsf M)$. 

\[
\mu_-=\tfrac12\big(\mathbf1^\top \mathbf v - \sqrt m\,\|\mathbf v\|_2\big)
\]
 
Now $\mathbf1^\top \mathbf v=\sum_s\big[B_i+(\eta-2)b_{s,i}\big]=(m+\eta-2)B_i$, and,
using $\sum_s b_{s,i}^2\leqslant B_i^2$, which holds for any nonnegative vector summing to $B_i$,
\[
m\,\|\mathbf v\|_2^2
= m\Big[(m+2\eta-4)\,B_i^2+(2-\eta)^2\sum_s b_{s,i}^2\Big]
\;\leqslant\; m\,(m-2\eta+\eta^2)\,B_i^2 .
\]
Substituting in the minimum eigenvalue of $\mathsf M$:
\[
\lambda_{\min}(\mathsf M)
= B_i+\tfrac12\Big[(m+\eta-2)B_i-\sqrt m\,\|\mathbf v\|_2\Big]
\;\geqslant\;
\frac{m+\eta-\sqrt{m(m-2\eta+\eta^2)}}{2}\;B_i
\;=\;\chi(m,\eta) \,B_i .
\]
By rationalizing the denominator of $\chi(m, \eta)$, we get
\[
\chi(m, \eta) = \;\frac{\eta\,\big(4m-(m-1)\eta\big)}
{2\,\big[\,m+\eta+\sqrt{m(m-2\eta+\eta^2)}\,\big]}
\;>\;0 ,
\]
since $(m+\eta)^2-m(m-2\eta+\eta^2)=\eta\big(4m-(m-1)\eta\big)>0$ for
$\eta\in(0,1)$ and $m\geqslant2$ (as $(m-1)\eta<m-1<4m$).  Hence, $\mathsf M\succeq\chi(m,\eta) B_i\,I_m$ and multiplying by $c_0=-M_i\kappa(1-\eta)B_i^{\eta-3}<0$ and using $B_i\leqslant mC\Rightarrow B_i^{\eta-2}\geqslant(mC)^{\eta-2}$ (since $\eta-2<0$ reverses the inequality), gives $\tfrac12(H_i+H_i^\top)\preceq-\widetilde{\lambda} I_m$ for every $i$  and hence in \eqref{eq:modified-SB}, $\tfrac12\bigl(\widetilde{G}(\mathbf b)+\widetilde{G}(\mathbf b)^\top\bigr)+\widetilde{\lambda} I_{mn}\preceq0$ where
\[
 \widetilde{\lambda} := \kappa(1-\eta)\chi(m,\eta)(mC)^{\eta-2}\min_iM_i \;>\; 0
\]
Since the strong-concavity bound uses only the upper bound $B_i\leqslant mC$ and holds on the whole unrestricted set, so the participation floor doesn't play any role here.

\medskip
\emph{Bounded operator norm.} For $0\leqslant b_{s,i}\leqslant B_i$ and
$\eta-2\in(-2,-1)$ we have $|2B_i+(\eta-2)b_{s,i}|\leqslant 2B_i$ and
$|B_i+(\eta-2)b_{s,i}|\leqslant B_i$, so every entry of $H_i$ has magnitude at most
$M_i\kappa(1-\eta)B_i^{\eta-3}\cdot2B_i
=2M_i\kappa(1-\eta)B_i^{\eta-2}
\leqslant 2M_i\kappa(1-\eta)(m\varepsilon)^{\eta-2}$,
using $B_i\geqslant m\varepsilon$. Hence
$\|H_i\|_{\mathrm{op}}\leqslant\|H_i\|_F\leqslant m\cdot2M_i\kappa(1-\eta)
(m\varepsilon)^{\eta-2}\leqslant 2m\,\kappa\,(1-\eta)\,(m\varepsilon)^{\eta-2}\,\max_{i}M_i :=\widetilde{L}$, and in \eqref{eq:modified-SB}, $\|\widetilde G\|_{\mathrm{op}}\leqslant \widetilde{L}$. This bound uses the lower bound $B_i\geqslant m\varepsilon$ and is therefore the step at which the participation floor is essential.
\hfill$\square$

\end{document}